\documentclass[apj,twocolumn]{openjournal}
\usepackage{amsmath}
\usepackage{booktabs}
\usepackage{multirow}
\usepackage{color}
\usepackage{soul}
\usepackage{threeparttable}
\usepackage{float}
\usepackage{graphicx}
\usepackage{CJK}
\usepackage{xspace}
\usepackage{afterpage}
\usepackage{placeins}
\usepackage[breaklinks,colorlinks,citecolor=blue,urlcolor=blue,linkcolor=blue,filecolor=blue]{hyperref}

\shorttitle{All the Little Things: JWST dwarf galaxies at high redshift}
\shortauthors{Naidu \& Matthee et al.}

\usepackage[export]{adjustbox}

\newcommand{\mstar}{\ensuremath{\log(M_{\rm{\star}}/M_{\rm{\odot}})}}
\newcommand{\orcidauthor}[3]{\author{\href{http://orcid.org/#1}{#2$^{#3}$}}}

\begin{document}

\title{\vspace{-0.8cm}All the Little Things in Abell 2744: $>1000$ Gravitationally Lensed Dwarf Galaxies at $\MakeLowercase{z}=0-9$ from JWST NIRCam Grism Spectroscopy\vspace{-1.6cm}}

\orcidauthor{0000-0003-3729-1684}{Rohan P. Naidu}{1,*,\dagger,\ddagger}
\orcidauthor{0000-0002-1655-5604}{Jorryt Matthee}{2,*, \dagger}
\orcidauthor{0000-0001-5346-6048}{Ivan Kramarenko}{2}
\orcidauthor{0000-0001-8928-4465}{Andrea Weibel}{3}
\orcidauthor{0000-0003-2680-005X}{Gabriel Brammer}{4,5}
\orcidauthor{0000-0001-5851-6649}{Pascal A.\ Oesch}{3,4,5}
\orcidauthor{}{Peter Lechner}{2}
\orcidauthor{0000-0001-6278-032X}{Lukas J. Furtak}{6}
\orcidauthor{0000-0003-1408-7373}{Claudia Di Cesare}{2,7}
\orcidauthor{0000-0001-5586-6950}{Alberto Torralba}{8,9}
\orcidauthor{0009-0007-1062-0253}{Gauri Kotiwale}{2}
\orcidauthor{0000-0001-5063-8254}{Rachel Bezanson}{10}
\orcidauthor{0000-0002-4989-2471}{Rychard J. Bouwens}{11}
\orcidauthor{0000-0002-0572-8012}{Vedant~Chandra}{12}
\orcidauthor{0000-0001-7940-1816}{Ad\'{e}la\"{i}de Claeyssens}{13}
\orcidauthor{0000-0002-9708-9958}{A. Lola Danhaive}{14,15}
\orcidauthor{0000-0002-2139-7145}{Anna Frebel}{1}
\orcidauthor{0000-0002-2380-9801}{Anna de Graaff}{16}
\orcidauthor{0000-0002-5612-3427}{Jenny E. Greene}{17}
\orcidauthor{0000-0002-9389-7413}{Kasper~E.~Heintz}{3,4,5}
\orcidauthor{0000-0002-4863-8842}{Alexander P. Ji}{18,19}
\orcidauthor{0000-0001-9044-1747}{Daichi Kashino}{20}
\orcidauthor{0000-0003-1561-3814}{Harley Katz}{18,19}
\orcidauthor{0000-0002-2057-5376}{Ivo Labbe}{21}
\orcidauthor{0000-0001-6755-1315}{Joel Leja}{22,23,24}
\orcidauthor{0000-0002-0682-3310}{Yijia Li}{22,24}
\orcidauthor{0000-0003-0695-4414}{Michael V. Maseda}{25}
\orcidauthor{0000-0001-5492-1049}{Johan Richard}{26}
\orcidauthor{0000-0003-4702-7561}{Irene Shivaei}{27}
\orcidauthor{0000-0003-3769-9559}{Robert A.\ Simcoe}{1}
\orcidauthor{0000-0001-8823-4845}{David Sobral}{28,29}
\orcidauthor{0000-0002-1714-1905}{Katherine A. Suess}{30}
\orcidauthor{0000-0002-8224-4505}{Sandro Tacchella}{14,15}
\orcidauthor{0000-0003-2919-7495}{Christina C.\ Williams}{31}

\affiliation{$^1$MIT Kavli Institute for Astrophysics and Space Research, 70 Vassar Street, Cambridge, MA 02139, USA}
\affiliation{$^2$Institute of Science and Technology Austria (ISTA), Am Campus 1, 3400 Klosterneuburg, Austria}
\affiliation{$^3$Department of Astronomy, University of Geneva, Chemin Pegasi 51, 1290 Versoix, Switzerland}
\affiliation{$^4$Cosmic Dawn Center (DAWN), Copenhagen, Denmark}
\affiliation{$^5$Niels Bohr Institute, University of Copenhagen, Jagtvej 128, K{\o}benhavn N, DK-2200, Denmark}
\affiliation{$^6$Department of Physics, Ben-Gurion University of the Negev, P.O. Box 653, Be'er-Sheva 84105, Israel}
\affiliation{$^7$INAF/Osservatorio Astronomico di Roma, Via di Frascati 33, 00078 Monte Porzio Catone, Italy}
\affiliation{$^8$Observatori Astron\`omic de la Universitat de Val\`encia, Ed. Instituts d'Investigaci\'o, Parc Cient\'ific. C/ Catedr\'atico Jos\'e Beltr\'an, n2, 46980 Paterna, Valencia, Spain}
\affiliation{$^9$Departament d'Astronomia i Astrof\'isica, Universitat de Val\`encia, 46100 Burjassot, Spain}
\affiliation{$^{10}$Department of Physics and Astronomy and PITT PACC, University of Pittsburgh, Pittsburgh, PA 15260, USA}
\affiliation{$^{11}$Leiden Observatory, Leiden University, NL-2300 RA Leiden, Netherlands}
\affiliation{$^{12}$Center for Astrophysics $\mid$ Harvard \& Smithsonian, 60 Garden St, Cambridge, MA 02138, USA}
\affiliation{$^{13}$The Oskar Klein Centre, Department of Astronomy, Stockholm University, AlbaNova, SE-10691 Stockholm, Sweden}
\affiliation{$^{14}$Kavli Institute for Cosmology, University of Cambridge, Madingley Road, Cambridge, CB3 0HA, UK}
\affiliation{$^{15}$Cavendish Laboratory, University of Cambridge, 19 JJ Thomson Avenue, Cambridge, CB3 0HE, UK}
\affiliation{$^{16}$Max-Planck-Institut f\"ur Astronomie, K\"onigstuhl 17, D-69117, Heidelberg, Germany}
\affiliation{$^{18}$Department of Astronomy \& Astrophysics, University of Chicago, 5640 S Ellis Avenue, Chicago, IL 60637, USA}
\affiliation{$^{19}$Kavli Institute for Cosmological Physics, The University of Chicago, 5640 South Ellis Avenue, Chicago, IL 60637, USA}
\affiliation{$^{20}$National Astronomical Observatory of Japan, 2-21-1 Osawa, Mitaka, Tokyo 181-8588, Japan}
\affiliation{$^{21}$Centre for Astrophysics and Supercomputing, Swinburne University of Technology, Melbourne, VIC 3122, Australia}
\affiliation{$^{22}$Department of Astronomy \& Astrophysics, The Pennsylvania State University, University Park, PA 16802, USA}
\affiliation{$^{23}$Institute for Computational \& Data Sciences, The Pennsylvania State University, University Park, PA 16802, USA}
\affiliation{$^{24}$Institute for Gravitation and the Cosmos, The Pennsylvania State University, University Park, PA 16802, USA}
\affiliation{$^{25}$Department of Astronomy, University of Wisconsin-Madison, 475 N. Charter St., Madison, WI 53706 USA}
\affiliation{$^{26}$Univ Lyon, Univ Lyon1, ENS de Lyon, CNRS, Centre de Recherche Astrophysique de Lyon UMR5574, Saint-Genis-Laval, France}
\affiliation{$^{27}$Centro de Astrobiolog\'{i}a (CAB), CSIC-INTA, Carretera de Ajalvir km 4, Torrej\'{o}n de Ardoz, 28850, Madrid, Spain}
\affiliation{$^{28}$Departamento de F\'isica, Faculdade de Ci\`encias, Universidade de Lisboa, Edif\'icio C8, Campo Grande, PT1749-016 Lisbon, Portugal}
\affiliation{$^{29}$BNP Paribas Corporate \& Institutional Banking, Torre Ocidente Rua Galileu Galilei, 1500-392 Lisbon, Portugal}
\affiliation{$^{30}$Department for Astrophysical \& Planetary Science, University of Colorado, Boulder, CO 80309, USA}
\affiliation{$^{31}$NSF National Optical-Infrared Astronomy Research Laboratory, 950 North Cherry Avenue, Tucson, AZ 85719, USA}

\thanks{$^*$E-mail: \href{mailto:rnaidu@mit.edu}{rnaidu@mit.edu}, \href{mailto:jorryt.matthee@ist.ac.at}{jorryt.matthee@ist.ac.at}}
\thanks{$\dagger$ These authors are the PIs of ALT (JWST Program \#3516) and contributed equally to this work.}
\thanks{$\ddagger$ NASA Hubble Fellow}

\begin{abstract}
Dwarf galaxies hold the key to crucial frontiers of astrophysics, however, their faintness renders spectroscopy challenging. Here we present the JWST Cycle 2 survey, All the Little Things (ALT, PID 3516), which is designed to seek late-forming Pop III stars and the drivers of reionization at $z\sim6$-$7$. ALT has acquired the deepest NIRCam grism spectroscopy yet (7-27 hr), at JWST's most sensitive wavelengths (3-4 $\mu$m), covering the powerful lensing cluster Abell 2744. Over the same 30 arcmin$^{2}$, ALT's ultra-deep F070W+F090W imaging ($\sim30$ mag, 5$\sigma$) enables selection of very faint sources at $z>6$. We demonstrate the success of ALT's novel ``butterfly" mosaic to solve spectral confusion and contamination, and introduce the ``Allegro" method for emission line identification. By collecting spectra for every source in the field of view, ALT has measured precise ($\mathcal{R}\sim1600$) redshifts for 1630 sources at $z=0.2$-$8.5$. This includes one of the largest samples of distant dwarf galaxies: [1015, 475, 50] sources less massive than the SMC, Fornax, and Sculptor with $\log(M_{*}/M_{\rm{\odot}})<$[8.5, 7.5, 6.5]. We showcase ALT's discovery space with: (i) spatially resolved spectra of lensed clumps in galaxies as faint as $M_{\rm{UV}}\sim-15$; (ii) large-scale clustering -- overdensities at $z$=[2.50, 2.58, 3.97, 4.30, 5.66, 5.77, 6.33] hosting massive galaxies with striking Balmer breaks; (iii) small-scale clustering -- a system of satellites around a Milky Way analog at $z\sim6$; (iv) spectroscopically confirmed multiple images that help constrain the lensing model underlying all science in this legacy field; (v) sensitive star-formation maps based on dust-insensitive tracers such as Pa$\alpha$; (vi) direct spectroscopic discovery of rare sources such as AGN with ionized outflows. These results provide a powerful proof of concept for how grism surveys maximize the potential of strong lensing fields.
\end{abstract}

\section{Introduction}
\label{sec:intro}

The hallmark of $\Lambda$CDM cosmology is hierarchical structure formation, wherein galaxies like our Milky Way continually assimilate smaller dwarf galaxies \citep[e.g.,][]{White91}. Directly observing these dwarf galaxies in the early Universe holds the key to unraveling the two great transformations wrought on the cosmos by the very first generations of stars: the synthesis of the chemical elements \citep[e.g.][]{Tinsley80}, and the epoch of Hydrogen reionization \citep{Loeb&Barkana01}. 

Metal-free stars (so-called Population III [Pop III] stars) forming out of pristine gas after the Big Bang are predicted to have forged heavier elements in their cores for the first time, setting off the chemical evolution of the universe that would one day seed life on Earth \citep[e.g.,][]{Patridge67}. While directly witnessing the light from the very first Pop III stars at $z\approx15-20$ may require a 100m telescope in space \citep[e.g.,][]{Schauer20}, \textit{late-forming} Pop III stars may be observable well into the epoch of reionization. Pristine pockets of dense gas may persist even amidst subsequent generations of stars due to e.g., inhomogeneous metal mixing where the ejecta from supernovae never pollute these pockets, due to re-accretion of pristine gas that was ejected into the halo, and by remaining shielded on the peripheries of galaxies growing inside-out \citep[e.g.,][]{Tornatore07,Katz23, Venditti23, Venditti24}, though note the galaxies in which this is possible and survival redshifts vary widely across these models. Late-blooming low-mass galaxies in under-dense voids that take longer to collapse compared to denser regions of the Universe may also host such pristine pockets \citep[e.g.,][]{Pallottini14, Xu16, Jaacks19}. The chances of pollution are a function of star-formation activity and feedback strength, and are likely lower in dwarf galaxies that typically display low star-formation efficiency \citep[e.g.,][]{Ma16}.

While JWST \citep[][]{Gardner23} is yet to detect unambiguous Pop III signatures such as bright HeII emission \citep[e.g.,][]{Schaerer03} combined with the absence of metals \citep[e.g.,][]{Wang24HeII,Maiolino24}, current models and their expectations suggest clear search strategies. While HeII emission is faint and may be powered by other sources such as AGNs or Pop II stars \citep[e.g.][]{Plat2019}, all hydrodynamical simulations that include Pop III models underscore the importance of spatially resolved observations to identify metal-poor pockets \citep[e.g.][]{Pallottini15,Venditti23}. This is a natural fit for gravitationally lensed fields, where JWST's exquisite spatial resolution is paired with galaxies that are magnified into arcs and resolved into individual star-forming complexes \citep[e.g.,][]{Welch22JW, Hsiao23, Rigby23templates, Rivera-Thorsen24, Adamo24}. Another challenge is demonstrating that a pocket in an arc is indeed metal-poor -- a variety of spectral diagnostics have been proposed \citep[e.g.,][]{Nakajima22}, of which the most efficient for a blind search across large samples is [OIII]/H$\beta$ owing to the strength of these lines and the resulting stringent limits on the metallicity that are practically feasible. Indeed, the most metal-poor object detected to date with JWST (LAP1 at $z=6.64$, 12 + log(O/H)$<6$) was identified behind the MACS J0416 cluster, with NIRSpec/IFU spectrosocopy revealing clumps with extremely low [OIII]/H$\beta$ line ratios \citep{Vanzella23}. Given that [OIII] and H$\beta$ suffer from enormous sensitivity loss at $z\gtrsim9$, and the higher the redshift the better the chances of pristine pockets, Pop III surveys using these lines are most efficiently directed at $z<9$, and due to JWST's sensitivity function are most effective at $z\sim7$, right in the heart of the epoch of reionization. 

Cosmic reionization is another potential legacy of Pop III stars and their immediate descendants. This was the last cosmological phase transition where neutral gas pervading the Universe was heated into a hot, ionized state \citep[e.g.,][]{Loeb&Barkana01}. Even with the advent of JWST, the protagonists of reionization remain a subject of debate due to uncertainties in galaxies' emissivities \citep[e.g.,][]{Munoz24,Simmonds24}. The high numbers of UV-faint active galactic nuclei (AGNs) being discovered have defied extrapolations of quasar luminosity functions \citep[e.g.,][]{Harikane23, Maiolino23, Matthee24, Greene24, Kocevski24,Taylor24}. These abundant sources have breathed new life into proposals for AGN-driven reionization \citep[e.g.,][]{Giallongo15, Madau15, Finkelstein19, Dayal24,Madau24, Grazian24}. However, what fraction of ionizing photons is escaping the dusty tori and interstellar media of their host galaxies to make it to the intergalactic medium remains to be seen.

Reionization by ``oligarchs", i.e., relatively rare, bright star-forming galaxies ($L\gtrsim 0.05 L^{*}$, $M_{\rm{UV}}<-18$; e.g., \citealt{Sharma16,Naidu20, MN22}) is buoyed by the lack of evolution in the bright end of the UV LF \citep[e.g.,][]{Bouwens23} that may be due to ejection of dust \citep[e.g.,][]{Ferrara23}, coupled with ionizing photon production at par or exceeding pre-JWST models \citep[e.g.,][]{Matthee23,Simmonds24,Pahl24}. The rising and bursty star-formation histories (SFHs) inferred for such galaxies \citep[e.g.,][]{Cole23, Pizatti24, Endsley24}, is exactly what is required for the escape of ionizing photons \citep[e.g.,][]{Ma16,Ma20,NM22, Choustikov24,Flury24}. This is further supported by the detailed inventory of photons in the bubble around the luminous COLA1 galaxy -- a confirmed, powerful luminous leaker that dominates its local ionizing budget \citep{Torralba-Torregrosa24, Matthee18, MasonGronke20}. 

The key missing item in JWST's reionization budget is the emissivity from faint sources ($<0.02 L^{*}$, $M_{\rm{UV}}>-16.5$) that have so far been poorly explored. Due to their sheer numbers and predicted ionizing photon production efficiencies \citep[e.g.][]{Saxena24}, these sources have been classically viewed as the primary agents of reionization \citep[e.g.,][]{Yajima11, Wise14, Paardekooper15, Kimm17, Finkelstein19}. Observations remain sparse due to the large integration times required -- e.g., \citealt{Atek24} recently found four such galaxies to be strong ionizers. However, based on SED modeling, \cite{Endsley23} show indications of a larger scatter in emission line equivalent widths (and therefore ionizing photon production efficiencies) towards fainter galaxies. Clearly, larger spectroscopic statistics are beneficial. Balmer lines offer the most direct, efficient window into the ionizing power of galaxies -- JWST NIRCam is able to survey the strongest of these, H$\alpha$, out to $z\approx7$ and H$\beta$ out to $z\approx9$.

To seek extremely metal-poor pockets and reveal the protagonists of reionization, we designed the ``All the Little Things" (ALT) NIRCam grism survey over the Abell 2744 strong lensing cluster. We were inspired by the remarkable success of early NIRCam grism programs, which have returned the largest samples of reionization-era galaxies across all spectroscopic programs to date \citep[][]{Kashino23, Oesch23, Wang23} and have already yielded candidate metal-poor galaxies in blank fields \citep{Matthee23}. By collecting spatially resolved spectra for every single galaxy in the field of view, ALT is designed to maximize the chances of detecting metal-poor clumps. Pushing to much deeper flux limits than Cycle 1 surveys, we designed ALT to gather the statistical samples required to inventory the ionizing properties of faint galaxies. Simultaneously, the vast amounts of spectroscopic redshifts would benefit numerous science studies in this legacy field.

The outline for this paper follows. In \S\ref{sec:data} we motivate our survey design -- why we chose the Abell 2744 cluster (\S\ref{sec:publicdata}), what considerations informed our grism strategy (\S\ref{sec:rationale}), the development and validation of a ``butterfly" mosaicking strategy to solve confusion and contamination (\S\ref{sec:butterfly}), the implementation of our observations (\S\ref{sec:implement}), and data reduction (\S\ref{sec:datareduction}). In \S\ref{sec:methods} we discuss how we translate the reduced data into science-ready products by creating photometric catalogs (\S\ref{sec:photometry}), estimating photometric redshifts (\S\ref{sec:photozs}), identifying emission lines in the grism data via two independent methods -- \textit{Allegro} (\S\ref{sec:allegro}) and \textit{grizli} (\S\ref{sec:grizli}), and modeling the 27-band SEDs to derive physical parameters (\S\ref{sec:sed}). We describe the data release accompanying this paper and validate our reported redshifts in \S\ref{sec:DR1}. In \S\ref{sec:results} we present first science results that showcase ALT's capabilities on highly magnified arcs (\S\ref{sec:arcs}), multiply-imaged systems (\S\ref{sec:lensing}), large-scale environments (\S\ref{sec:envlarge}), small-scale environments (\S\ref{sec:envsmall}), clustering statistics (\S\ref{sec:whygrism}), obscured star-formation (\S\ref{sec:paschen}), and the serendipitous discovery of rare sources such as Little Red Dots (LRDs; \S\ref{sec:lineprofiles}). Throughout this work, we adopt a \citet[][]{Planck18} cosmology. We often reference $L^{*}$, the characteristic luminosity in Schechter function parametrizations of luminosity functions as per \citet[][]{Bouwens21}. Magnitudes are in the AB system \citep[e.g.,][]{Oke83}. For summary statistics, we typically report medians with errors on the median from bootstrapping (16$^{\rm{th}}$ and 84$^{\rm{th}}$ percentiles).

\section{Data}
\label{sec:data}
\subsection{Public Data}
\label{sec:publicdata}
The field around the Abell 2744 \citep[e.g.,][]{Abell89} lensing cluster is emerging as one of the key extra-galactic survey fields in the JWST era. Previously, the main core of the A2744 cluster was part of the Hubble Frontier Fields program and received ample attention including HST and Spitzer imaging \citep{Lotz17, Steinhardt20}, HST grism \citep{Treu15}, MUSE \citep{Richard21}, and the GLASS NIRISS ERS program \citep{Treu22}. Flanking fields have been observed by various parallel programs, most well known of which are the GLASS-NIRCam parallels that led to the early JWST discovery of luminous $z>10$ galaxies \citep[e.g.,][]{Castellano22, Naidu22}. A key distinguishing feature of A2744 is the wealth of immediately public data beginning with GLASS and the Cycle 1 UNCOVER survey \citep{Bezanson22}.

\begin{figure*}[htb]
    \centering
    \includegraphics[width=0.95\linewidth]{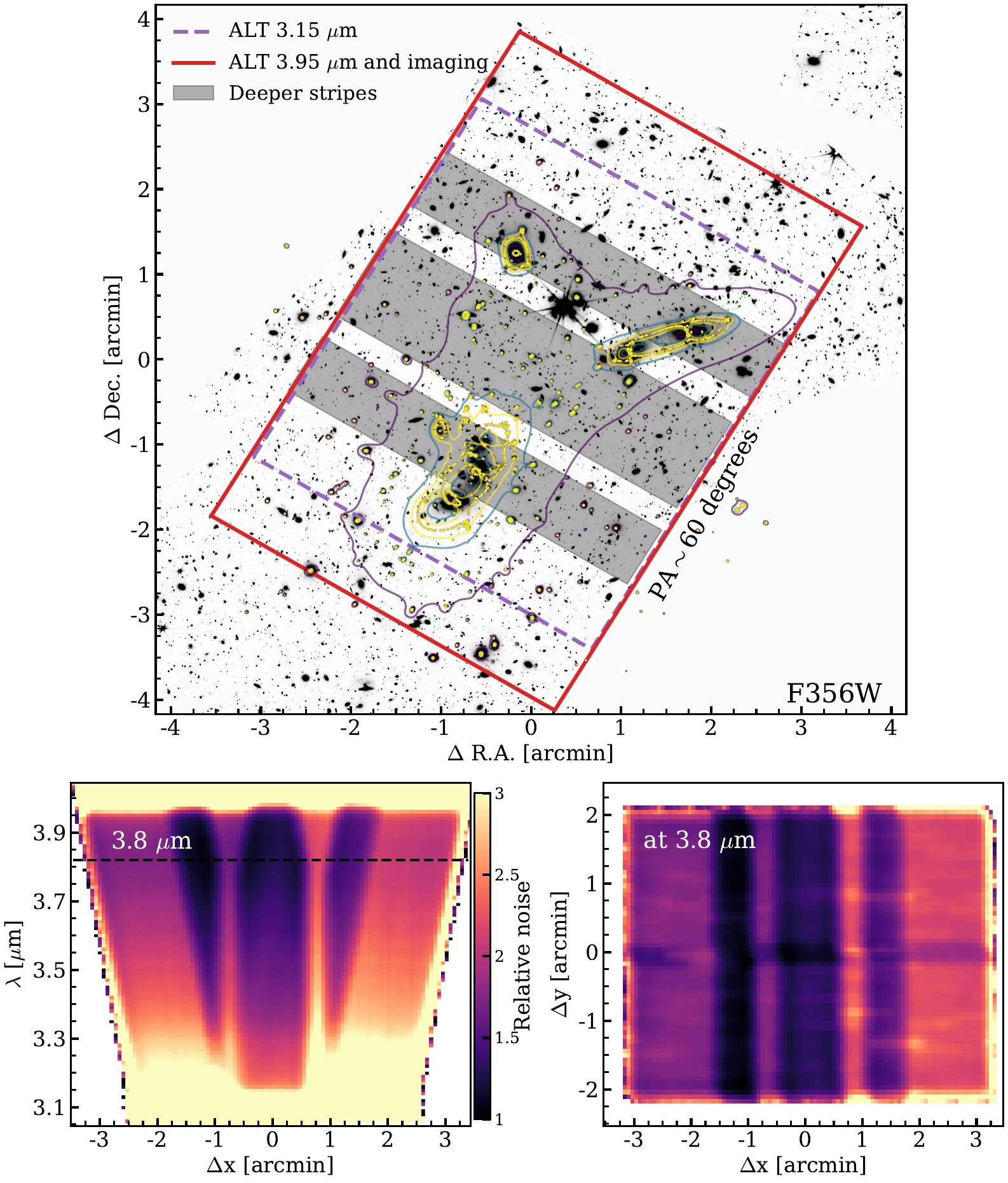}
    \caption{\textbf{The {\bf The ALT survey footprint around the Abell 2744 lensing cluster}.} Top: Maximum on sky field of view of the ALT NIRCam WFSS data (red lines; at the undeviated wavelength, $\lambda=3.95 \mu$m) illustrated on top of a background F356W image (combining ALT and UNCOVER data). The dashed purple line shows the minimum field of view at 3.15 $\mu$m. Contours highlight the increasing magnification closer to the three main cores of the cluster (at $z=6$ from the model by \citealt{Furtak23}). The center of the footprint is at $\alpha=$0:14:18.0, $\delta=-$30:22:47.3 (J2000). Bottom, left: Wavelength dependence of the spectroscopic data and the total field of view (along the main axis of the mosaic, it is constant along the minor axis). Our data are most sensitive in the deep stripes and around 3.8 $\mu$m. Bottom, right: Spatial variation of the sensitivity of the grism data, shown at a wavelength of 3.8 $\mu$m. The deep stripes and mosaic pattern, as well as the difference in sensitivity between module A and B data can clearly be seen. One can also notice the slight decrease in sensitivity near bright objects.}
    \label{fig:RADEC}
\end{figure*}

\begin{figure*}
    \centering
    \includegraphics[width=0.95\linewidth]{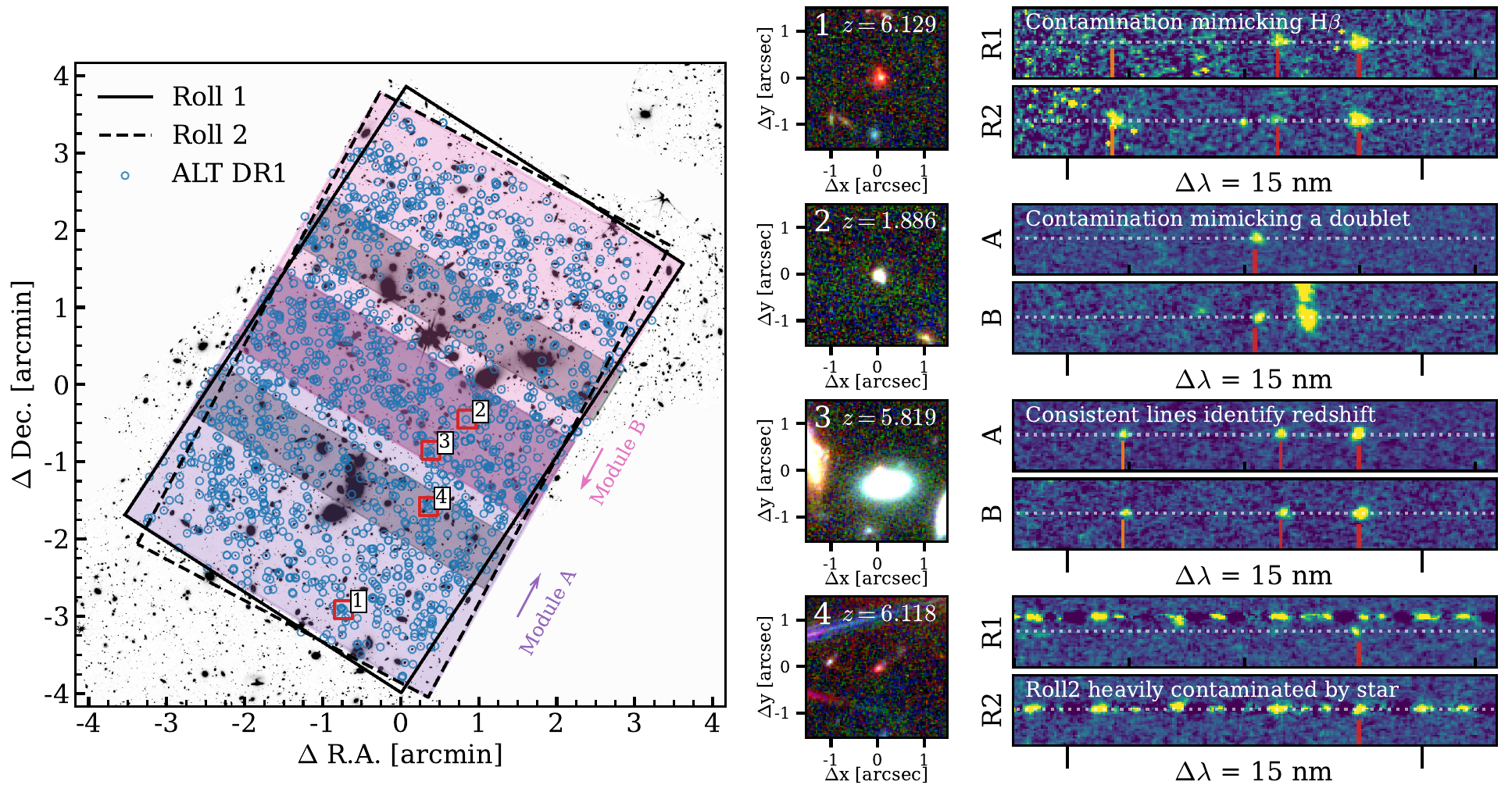}
    \caption{\textbf{Illustration of how the mosaic design facilitates the analysis of the grism data}. \textbf{Left}: ALT field of view in the two rolls (solid and dashed black line) on top of the same background F356W image as in Figure $\ref{fig:RADEC}$. Blue circles show the galaxies for which the ALT data measured a spectroscopic redshift. The three shaded regions highlight the stripes observed twice (either twice with module A, or once with each module, or twice with module B, from left to right respectively). We highlight the positions of four example objects with red squares. \textbf{Right:} NIRCam false colour stamps (based on F115W/F200W/F356W imaging) and 2D grism emission-line spectra for the four objects highlighted in the left panel. The grism spectra span $\approx15$ nm in the observed frame. For objects 1 and 4, we show 2D spectra separated by Roll 1 vs. Roll 2, while for objects 2 and 3 we separate Module A vs. Module B. These spectra illustrate how observations of the same source across different rolls and modules helps in disentangling contamination and confusion.} 
    \label{fig:rolls}
\end{figure*}

As discussed in \cite{Bezanson22}, A2744 is a merging galaxy cluster. Its structure and lensing properties are extremely well-characterized as a result of more than a decade of detailed modeling efforts \citep[e.g.,][]{Merten11, Richard14, Wang15, Jauzac15, Kawamata16}. Among well-studied lensing fields (e.g., those included in the HFF), it has the largest area with a magnification above 2 (17 arcmin$^2$). Mosaicking such an area with NIRCam has several convenient advantages: 1) efficient use of parallels, 2) a well-matched NIRSpec field of view for follow-up with flexibility in position angles, 3) a large intermediate lensing regime with mean magnification $\mu>2$, where lensing models are more reliable than in the highly magnified regimes. Exploiting these efficiencies, the UNCOVER survey \citep[][]{Bezanson22, Price24} acquired deep NIRCam imaging in all broadbands except F070W and F090W, along with follow-up prism spectroscopy. 

Cycle 2 saw the award of four survey programs that included A2744 as a target: ALT (this paper), MegaScience \citep[][]{Suess24}, and two other NIRCam grism programs over lensing fields: \#2883 (MAGNIF, PI: Sun), \#3538 (PI: Iani). MegaScience is a particularly important addition, as it surveyed the cluster in \textit{all} the NIRCam medium-bands. ALT and MegaScience also completed the broad-band filter-set with F070W and F090W imaging. We also incorporate additional data from a transient follow-up program that observed NIRCam parallels \#2756, PI: Chen \citep[e.g.][]{Chen22} and the pure-parallel NIRCam imaging program \#3990, PI: Morishita. Combining these programs, imaging is now available in all the 20 NIRCam medium and broad bands across the ALT footprint (see Figure $\ref{fig:RADEC}$). 

\subsection{ALT NIRCam imaging and  grism observations: rationale}
\label{sec:rationale}
The NIRCam grism is performing exceptionally well \citep[][]{Sun23,Rieke23}, yielding spectroscopic identifications of hundreds of galaxies beyond redshift $z>6$ \citep[e.g.,][]{Kashino23, Meyer24, CoveloPaz24}. The large samples have led to the serendipitous identification and characterisation of rare, genre-defining sources such as the ``Little Red Dots" \citep{Matthee24}. These grism surveys are greatly facilitated by a 20-40 \% higher throughput than expected \citep{Rigby22}, as well as the simplicity with which continuum contamination can be removed thanks to the relatively high spectral resolution of $R\approx1600$ \citep{Kashino23}, $\approx10\times$ higher than the HST grisms \citep{Brammer12}.

The first deep grism data from the EIGER program \citep[][]{Kashino23} identified galaxies through their [OIII] doublet emission at $z\sim6$ down to magnitudes as faint as M$_{\rm UV}\approx-17$ \citep{Matthee23}, interestingly including galaxies with [OIII]/H$\beta \lesssim3$ implying relatively low metallicites (with oxygen abundances of 1/50 Z$_{\odot}$) and including spatially resolved galaxies with varying [OIII]/H$\beta$ line ratios. Obtaining such data in a field targeting a strong lensing cluster was a logical next step. We set out to do this in the A2744 field primarily because of the relatively large existing NIRCam mosaic from the UNCOVER program \citep{Bezanson22}, which enables an efficient mosaicking survey design to optimize the grism analysis.

We chose to observe with the F356W filter primarily because this is the broad-band filter in which the grism is most sensitive (as WFSS data are background-limited) and it has the largest fraction of the field of view with complete wavelength coverage ($\lambda=3.15-3.95 \mu$m). The redshift $z\approx6$ is where JWST can most strongly constrain the [OIII]/H$\beta$ ratio that is the most sensitive metallicity indicator in the extremely low metallicity regime \citep[e.g.][]{Nakajima22}. Additionally, the H$\alpha$ flux (needed to estimate the nebular attenuation) could be inferred from its impact on the F410M+F444W photometry from the UNCOVER data\footnote{Note that our ability to infer the H$\alpha$ flux from $z\approx6$ galaxies is now much easier with the acquisition of additional medium-band photometry from the simultaneously executed Cycle 2 MegaScience program \citep{Suess24}.}. As the background level is minimal at 3.5 micron \citep{Rigby23}, swapping F356W for a medium-band filter (such as F335M or F360M) instead would yield $<\sqrt{2}$ sensitivity increase at the expense of losing more than half of the covered volume and wavelength coverage of identified galaxies. We also note that the wider LW filter (F322W2) would result in much higher backgrounds, as well as a smaller fraction of the field of view with complete wavelength coverage.  Simultaneous to our grism observations, our program acquired imaging in the F070W and F090W filters with the NIRCam SW channel over the same footprint. This extends the UNCOVER wavelength coverage blueward and adds dropout photometric bands for exactly the redshifts of interest probed by [OIII]+H$\beta$.

\subsection{ALT NIRCam imaging and grism observations: motivation and validation of a ``butterfly" mosaic}
\label{sec:butterfly}

Early NIRCam WFSS analyses demonstrated that continuum light from neighbouring objects is not the catastrophic source of contamination it may seem at first glance. It can be subtracted relatively easily with simple methods such as a running median filter, adding only a little extra Poisson noise on the detector due to the high resolution of the NIRCam grism ($R\sim1600$) and the largely flat spectrum of most objects at $\lambda_{\rm obs}\approx3.5$ micron \citep[][]{Kashino23,Matthee23}.

Source confusion of detected emission-lines, however, can be a significant issue in case grism observations are taken only with a single dispersion direction. The nominally recommended way of addressing this is via multiple observations taken with grism dispersers that separate the direction of dispersion by 90 degrees, such as the grismR + grismC combination. However, due to the wavelength dependence of the field of view (in particular, due to parts of the spectra falling off the detector), and the fixed relative locations of modules A and B, efficient mosaic designs with this constraint are challenging. Furthermore, using grismC requires out-of-field images taken at 35'' offsets, and therefore separate visits, which result in enormous overheads without any gain in sensitivity (see discussion in \S2.1.1 of \citealt{Oesch23}).

The EIGER team developed a novel mosaic design where pointings are offset in the horizontal direction (using grism R alone), which produces survey areas with overlapping NIRCam modules, and yields Module A+B spectra that disperse in opposite directions (by 180 degrees) from each other \citep{Kashino23}. This solves the confusion issue, helps remove complex contamination and can accurately measure kinematics on $>100$ km s$^{-1}$ scales. Nonetheless, it can only be applied in $\approx15$ \% of the survey area where both NIRCam modules overlap. 

We further developed this idea by covering the deep field around A2744 with two sets of a 2x2 mosaic pattern (i.e., 8 visits/pointings in total). The two sets of mosaics are identical except they are observed with slightly different roll angles (V3-PA of 56.1 and 60.1 degrees). The four-degree difference was chosen as a compromise between a maximal field of view with full coverage and confusion-solving power. The ability to solve confusion with such small difference in dispersion angle depends on the specific position in the mosaic and on the wavelength. Generally, confusion can more easily be solved in the blue end of the spectrum and in the outskirts of the mosaic where the offset PAs produce more leverage. Towards the center, where this advantage diminishes, the joint module A + B coverage kicks in. To summarize, in this ``butterfly" mosaic,  contamination in the wings is solved with the four-degree PA difference, and in the center it is solved via the module overlap.

Figure $\ref{fig:rolls}$ shows this strategy in action. It depicts the maximum field of view of the grism data obtained with the two roll angles (left panel), and illustrates how the survey strategy helps to solve confusion and contamination issues (right panel). The first highlighted object 1 has only been covered by module A, and its Roll2 spectrum is consistent with [OIII] and H$\beta$ emission with a very low [OIII]/H$\beta$ ratio. However, no H$\beta$ is detected in the Roll1 spectrum, indicating that this emission line is not associated to this galaxy and contaminates the Roll2 spectrum. Objects 2 and 3 fall in the region with joint module A and B coverage. Emission-lines that originate from the objects in question are observed in identical positions in their module A and B spectra, respectively, while other emission-lines only appear in a single spectrum. Object 4 is unfortunately contaminated by the trace of a star in the Roll2 observation, but its emission-line can still be identified in the Roll1 data. 

\subsection{ALT NIRCam imaging and grism observations: implementation}
\label{sec:implement}

After deciding on the mosaic strategy, we set a target sensitivity to produce a mass-complete sample of main sequence galaxies with H$\beta$ detections (with a signal to noise of 10) at masses above $10^8$ M$_{\odot}$ at $z=6$. According to the JAGUAR simulations \citep{Williams18}, we found that this required a $10^{-18}$ erg s$^{-1}$ cm$^{-2}$ line-sensitivity limit (assuming a median magnification $\mu=2$). Of course, whether this limit indeed implies mass-completeness above this mass depends on the exact shape of the relation between H$\beta$ luminosity and mass (and its scatter), and is a subject of further study (Di Cesare et al. in prep). 

To achieve this sensitivity, we found that we required a typical exposure time of $\approx25$ ks. This is obtained by splitting each visit (out of the total of 8) in two sequences of WFSS observations in the F356W filter with 3 INTRAMODULEX primary dithers and 4 subpixel dithers (to optimize the image quality in the short-wavelength imaging data). By splitting in two sequences, we can change the filter used in the simultaneous SW imaging between F070W and F090W. For all WFSS observations, we used MEDIUM8 with 5 groups/integration as readout mode for an optimal compromise between data rate and ability to identify cosmic ray hits. The total WFSS exposure time is 12.4 ks per visit, yielding a typical total WFSS exposure time of 24.8 ks for the majority of the sky area after both rolls are observed. The exposure time is a factor two higher for three central stripes that cover the main regions with high magnification (see Figure $\ref{fig:RADEC})$. Direct and two out of field images in the F356W images were taken with exposure times 526s at the end of the WFSS exposure sequences, with SHALLOW4 readouts with 10 groups/integration. Coordinated parallel NIRISS imaging data are taken in the F090W and F277W filters, as those data were lacking in the respective parallel fields. The total allocation of telescope time was 47.7 hours. Observations were taken in 2023 on November 26-27, December 1, 5, 6 and 10.

\section{Methods}
\label{sec:methods}
\subsection{Data reduction} \label{sec:datareduction}
We process all publicly available NIRCam imaging in A2744 (see \S $\ref{sec:publicdata}$ for the list of programs) with the \texttt{grizli} software following the steps outlined in \citet{Valentino23}. Briefly, level-2 NIRCam exposures are retrieved from the Mikulski Archive for Space Telescopes (MAST), calibrated with `jwst\_0995.pmap', and then processed with custom algorithms that handle various artefacts such as snowballs, 1/f noise, wisps, and bad pixels. Images are aligned to stars from Gaia DR3 \citep{GaiaDR3} and drizzled to a shared pixel grid (40 mas for long-wavelength channel and 20 mas for short-wavelength channel images). These images are publicly available as v7.2 JWST mosaics from the DAWN JWST archive (DJA)\footnote{\url{https://dawn-cph.github.io/dja/index.html}}. Archival HST imaging is incorporated from the Complete Hubble Archive for Galaxy Evolution (CHArGE; \citealt{Kokorev22}). All the filters used in our analysis are listed in Table \ref{table:depth}.

\begin{figure*}
    \centering
    \includegraphics[width=18cm]{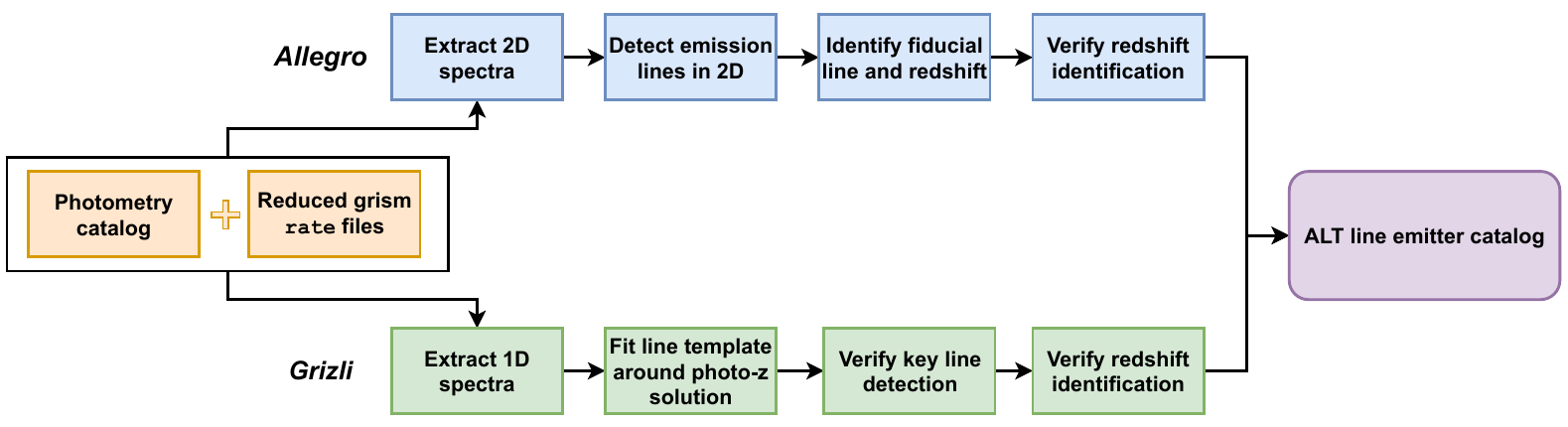}
    \caption{{\bf A schematic illustrating the procedure of constructing our ALT DR1 emission-line galaxy catalog.} Our starting points are the photometry catalog and the reduced grism images. Spectroscopic redshifts are measured with two independent and complementary approaches: {\it Allegro}, which detects emission-lines in 2D spectra and identifies a fiducial redshift using the photometric redshift and {\it Grizli}, which searches for spectroscopic redshift solutions around photometric redshift priors in optimally extracted 1D spectra. While {\it Grizli} is by designed for optimal sensitivity and exploitation of existing data, {\it Allegro} is designed for robustness to contamination, photometric redshift errors and complete identification of all lines.}
    \label{fig:flowchart}
\end{figure*}

For the grism data reduction as well as spectral extractions, we follow two independent, complementary approaches (see Figure \ref{fig:flowchart}). In the first approach we deploy the \texttt{grizli} pipeline end-to-end akin to FRESCO \citep[][]{Oesch23}, following similar reduction steps as above for the imaging. Astrometric alignment of individual grism exposures is performed using a catalog of bright sources (F356W$<$26 mag) extracted from the imaging mosaics. Sensitivity curves and dispersion functions for the grism images are drawn directly from CRDS \texttt{specwcs} reference files. To facilitate emission line searches, the grism images are median-filtered following \citet{Kashino23} to remove continuum. The filter is 71 pixels wide with a central hole of 10 pixels. To prevent over-subtraction around emission lines, the filtering is run in two steps where the second pass is run after masking S/N$>$3 pixels identified in the first pass. We emphasize that continuum spectra are also recoverable from the unfiltered data -- as shown in Figure \ref{fig:rolls}, our multi-roll strategy helps disambiguate overlap and contamination (see examples of clean continuum spectra with absorption features in Figure \ref{fig:lineprofiles}).

In our second approach (``Allegro"), the pre-processing steps of the NIRCam grism data reduction are the following. First, we acquire the {\sc RATE} files from MAST and then we use the {\sc jwst} pipeline\footnote{\url{https://github.com/spacetelescope/jwst}} (version 1.11.0; CRDS 11.16.20, pmap 1166) to assign a World Coordinate System (WCS) using {\sc assign-wcs}. We refine this solution by comparing the positions of sources in the direct F356W images to their positions in the GAIA-aligned Abell2744 image (see above). Offsets are typically small ($\approx0\farcs1$), but we correct for them by fitting coordinate transformation matrix (allowing for both a shift and a rotation). This procedure is applied for each visit individually. We then use {\sc Image2} to flat-field the data. The background subtraction is not trivial due to the high number of pixels that are affected by traces from sources (due to the high source density in the cluster field). We therefore estimate the background level by scaling the norm (i.e., the most common value) of object-free pixels to match a background template that we based on deep archival F356W GrismR data from the EIGER program. This background template is constructed by measuring the sigma-clipped median value along each column. We scale by matching the norm, rather than the mean, as it is less sensitive to the presence of sources. Similar to \cite{Kashino23}, we then split the grism data into a continuum and emission-line component. The continuum image is obtained by running a median filter along the dispersion direction with a kernel with 71 pixels width, and a central hole of 9 pixels.

The post-processing steps of the grism data reduction primarily involve the extraction of spectra. We extract spectra using the {\sc grismconf} package\footnote{\url{https://github.com/npirzkal/GRISMCONF}}, which predicts the (x,y)$_{\rm grism}$ coordinates on the grism images for each (RA,DEC,$\lambda$) value. We combine spectra using a three-times iterated 5 $\sigma$-clipped mean. We create combinations of all data, as well as separate combinations of data using only one of the NIRCam modules and the data from the individual roll angles. The flux calibration is applied in the spectral extraction step using the most up-to-date (V9) grism sensitivity curves. The sensitivity of the grism data depends on the wavelength and in the position of the source in the field of view, as illustrated in the bottom panel of Figure $\ref{fig:RADEC}$. The best sensitivity of $6\times10^{-19}$ erg s$^{-1}$ cm$^{-2}$ ($5\sigma$) is achieved at $\lambda\approx3.8$ micron, whereas the typical $5\sigma$ compact-source line-flux sensitivity is $8\times10^{-19}$ erg s$^{-1}$ cm$^{-2}$ and always better than $2\times10^{-18}$ erg s$^{-1}$ cm$^{-2}$.

\begin{deluxetable}{lrrrrrrrrrrr}
\tabletypesize{\footnotesize}
\tablecaption{5$\sigma$ depth of images used in this analysis \label{table:depth}}
\tablehead{
\colhead{Band} & \colhead{Depth (0\farcs16)} & \colhead{Depth (0\farcs08)}\\
\colhead{} & \colhead{[mag$_{\rm AB}$]} & \colhead{[mag$_{\rm AB}$]}
}
\startdata
\vspace{-0.3cm}\\
\multicolumn{3}{c}{HST (ACS, WFC3)}\\
$F$435$W$ & 28.91 & 29.87\\
$F$606$W$ & 28.62 & 29.62\\
$F$814$W$ & 28.52 & 29.51\\
$F$105$W$ & 27.63 & 28.63\\
$F$125$W$ & 27.73 & 28.76\\
$F$140$W$ & 29.06 & 30.11\\
$F$160$W$ & 27.45 & 28.44\\
\hline
\vspace{-0.2cm}\\
\multicolumn{3}{c}{JWST (NIRCam)}\\
$F$070$W$ & 28.91 & 29.89\\
$F$090$W$ & 29.36 & 30.34\\
$F$115$W$ & 29.02 & 30.05\\
$F$140$M$ & 28.03 & 29.04\\
$F$150$W$ & 29.12 & 30.14\\
$F$162$M$ & 28.11 & 29.11\\
$F$182$M$ & 28.48 & 29.50\\
$F$200$W$ & 29.21 & 30.26\\
$F$210$M$ & 28.37 & 29.37\\
$F$250$M$ & 28.11 & 29.11\\
$F$277$W$ & 29.64 & 30.66\\
$F$300$M$ & 28.63 & 29.64\\
$F$335$M$ & 28.65 & 29.70\\
$F$356$W$ & 29.82 & 30.86\\
$F$360$M$ & 28.72 & 29.72\\
$F$410$M$ & 29.02 & 29.99\\
$F$430$M$ & 27.98 & 28.98\\
$F$460$M$ & 27.75 & 28.72\\
$F$480$M$ & 27.91 & 28.87\\
$F$444$W$ & 29.44 & 30.43
\enddata
\tablecomments{Measured in 0\farcs16 radius and  0\farcs08 radius circular apertures. Notably, ALT adds $\approx29$ hrs in F070W and F090W while ``Medium Bands, Mega Science" \citep{Suess24} observed all the medium-bands to complete the NIRCam filter-set over Abell 2744 when combined with the Cycle 1 UNCOVER survey \citep{Bezanson22}. See \S\ref{sec:publicdata} for a full list of imaging programs. The average depth reported for F140W is higher than all the other WFC3 bands because the coverage in this filter is limited to deep images taken over the cluster core, whereas for the other filters the average includes shallower imaging taken across the ALT mosaic.}
\end{deluxetable}

\subsection{Photometric catalog}
\label{sec:photometry}
Starting from the reduced NIRCam + HST mosaics, we generate a photometric catalog largely following the methods outlined in \citet{Weibel24} as summarised below, and with a few custom steps adapted to the ALT data set and its specific application purposes. In order to detect potentially highly magnified objects near or behind the cluster we wish to remove cluster members before measuring photometric fluxes. To this end we run a simple median filter with a box size of $101\times101$ pixels ($4\farcs04$ $\times$ 4\farcs04) on all the images (similar to e.g., \citealt{Bouwens17b}). We use sigma-clipping to reduce over-subtraction around stars. Subtracting the median filter map reliably removes most of the light coming from bright foreground objects. We then run \texttt{SourceExtractor} in dual mode with an inverse-variance weighted stack of the median filter subtracted images in all the available long-wavelength channel (LW) filters (F250M, F277W, F300M, F335M, F356W, F360M, F410M, F430M, F444W, F460M and F480M) as the detection image. Fluxes are measured in circular apertures with a radius of 0\farcs16 on all images, PSF-matched to F444W. We use a low deblending parameter of $10^{-4}$ to ensure we detect faint sources whose light may still be contaminated by foreground objects even after the median filter subtraction. We note that a consequence of this deblending parameter is that some clumpy galaxies are broken up in multiple components with separate entries in our catalog. Each PSF-matched aperture flux is first scaled to the flux measured through a Kron-like aperture on a PSF-matched version of the detection image. Then, it is scaled to a total flux by dividing by the encircled energy of the Kron ellipse on the F444W PSF. 

To estimate the average depth of the images, we divide the flux images by the rms images to get effectively a ``signal-to-noise'' image. Then, we sum the pixels on that image through 0\farcs16 radius circular apertures in 5000 random positions across the field, selected to not be contaminated by light from any nearby sources according to the segmentation map from \texttt{SourceExtractor}. The scatter among these 5000 measurements provides an estimate of the ($1\sigma$) depth of each image for the given aperture size. We specify the average 5$\sigma$ depths in each filter measured within the ALT footprint in Table \ref{table:depth}. 

\subsection{Photometric redshifts}
\label{sec:photozs}

Photometric redshifts (photo-$z$s) play an important supporting role in the identification of emission lines in the grism data. We derive \texttt{eazy} \citep{Brammer08} photo-$z$s following \citet{Weibel24}. We adopt the \texttt{agn\_blue\_sfhz}\footnote{\url{https://github.com/gbrammer/eazy-photoz/tree/4747b5955ef95117615560c45b96e1d4487e6613/templates/sfhz}} template set which is comprised of thirteen \texttt{FSPS} templates \citep{FSPS1,FSPS2,FSPS3, FSPS4, python-FSPS}. A key feature of these templates is redshift-dependent star-formation histories that e.g., do not allow templates older than the age of the universe. In order to incorporate JWST's discoveries, two empirical templates are included that are fit to: i) a $z=8.5$ source from \citet{Carnall23} representative of the strong nebular emission line galaxies at these redshifts \citep[e.g.,][]{Matthee23, Meyer24,Boyett24}; ii) a $z=4.5$ source from \citet{Killi23} representative of the ``little red dots" thought to host broad-line AGN \citep[e.g.,][]{Matthee24, Kocevski24, Kokorev24} with ``V-shaped" SEDs that are blue in the rest-UV and red in the rest-optical \citep[e.g.,][]{Labbe23, Greene24}. Zero-point photometric offsets are allowed in the \texttt{eazy} fits and are found to be $<15\%$ for all bands, including the medium-bands and HST bands.

We test the photo-$z$s against literature spectroscopic redshifts, as well as ``blind" ALT grism redshifts  -- i.e., multi-line redshifts, or redshifts that use very weak photometric constraints such as non-detection below the implied Lyman limit (see the discussion of in \S\ref{sec:grism}). The redshift precision is $\langle |\Delta z|/(1+z) \rangle = 0.005$ with a standard deviation of 0.16, with 3$\%$ catastrophic outliers. These numbers are in excellent agreement with \citealt{Suess24}, who independently report similar precision and accuracy in their analysis of imaging in Abell 2744. We note that our comparison spectroscopic sample is dominated by galaxies with strong emission lines, which are easier to derive precise redshifts for, and so these numbers are optimistic relative to the entire galaxy population. The remarkable precision and accuracy is perhaps unsurprising given the complete NIRCam filter-set and archival HST imaging available. This allows us to adopt strict photo-$z$ priors to improve some of the grism emission line search methods discussed in \S\ref{sec:grism}.

\subsection{Emission-line galaxy identification in Grism data} \label{sec:grism}
In order to construct an exhaustive catalog of line-emitters, we undertook two complementary approaches to the grism data. These approaches are summarised in Figure $\ref{fig:flowchart}$. Both approaches start from the sources identified in the photometry catalog, but use the grism data in slightly different ways.

\subsubsection{Allegro}
\label{sec:allegro}
The {\it Allegro} approach (Kramarenko \& Matthee in prep.) is based upon the line-targeted search developed for EIGER (see \citealt{Kashino23}) and is conceptually similar to emission-line searches in narrow-band or IFU data \citep[e.g.][]{Ouchi08,Bacon16,Matthee16}. We extract 2D emission-line spectra from the continuum-subtracted grism data (\S\ref{sec:datareduction}) for each object in the photometric catalog (\S $\ref{sec:photometry}$). Emission lines are detected and identified in the stack of all data using {\sc Source-Extractor} \citep{Bertin96}. We also measure line fluxes in the combined data from each roll and module individually using {\sc Source-Extractor}'s dual-mode. These measurements enable us to easily add consistency requirements -- for example, a contaminant line will have very different flux and source centroids in data from different rolls and modules (as illustrated in Figure $\ref{fig:rolls}$).

We then use the catalogs of emission lines produced by {\sc Source-Extractor} to identify the highest S/N line within $\Delta y = \pm 3.5~$\,pixels of the trace. At this step, we only consider emission lines detected in the stacked 2D spectra with a $\mathrm{S/N}>10$. Additionally, we require that line fluxes measured in two different rolls, F$_{\mathrm{R1}}$ and F$_{\mathrm{R2}}$, are similar, i.e.: $0.5 < \mathrm{F}_{\mathrm{R1}}/\mathrm{F}_{\mathrm{R2}} < 2.0$. If for some galaxy the grism data are available in both NIRCam modules (A and B), we also consider emission lines with $0.5 < \mathrm{F}_{\mathrm{A}}/\mathrm{F}_{\mathrm{B}} < 2.0$ even if they are discarded based on the $\mathrm{F}_{\mathrm{R1}}/\mathrm{F}_{\mathrm{R2}}$ criterion.

We proceed to automatically assign a redshift to each galaxy as follows. For each line identified in the previous step, we test different redshift solutions by assuming that the line is either [OIII]$_{5008}$, H$\alpha$, [SIII]$_{9533}$, HeI$_{10833}$, Paschen-$\beta$, Paschen-$\alpha$ or Bracket-$\beta$. Taking advantage of the exquisite photometric data, we use the photometric redshifts (\S\ref{sec:photozs}) as priors and require $\Delta z/(1+z) < 0.15$. This usually narrows down the choice to one or two lines, for example [OIII]$_{5008}$ and H$\alpha$. For each of these tentative redshift solutions, we search for ``secondary'' emission lines in the stacked 2D spectra that would allow us to unambiguously tell the redshift of a galaxy (e.g., [OIII]$_{4960}$). Here, we consider all $\mathrm{S/N}>5$ emission lines that lie within $\Delta y = \pm 1.5~$pix of the highest S/N line and have a maximum velocity offset of $\Delta \upsilon = 500$~km~s$^{-1}$ with respect to their expected position in the spectrum. We also apply the same consistency checks as in the previous step. These criteria result in the detection of ``secondary'' lines in approximately 30~\% of the galaxies. For the rest of the sample, we adopt a redshift solution that yields the smallest $\Delta z/(1+z)$ with respect to the photometric redshift.

We then visually inspect all redshift solutions suggested by our algorithm, and particularly check whether any secondary line vetoes or confirms the redshift. For a small fraction of galaxies (6\%), we change the redshift based on a tentative ($\mathrm{S/N}<5$) detection of a secondary line, or when a suspected H$\alpha$-emitter is assigned a low-redshift solution due to photometry contaminated by a foreground galaxy.

The main strengths of {\it Allegro} are therefore that it is fast, flexible and enables a well-controlled and easily reproducible selection of line emitters. Completeness is high, since no line is ``left behind", i.e., all viable line-candidates with consistent positions along the trace are considered. A relative weakness of the method is that it does not reach optimal signal-to-noise as it does not consider the source morphology when identifying emission-lines (i.e., no optimal extraction is used), leading to a slightly lower sensitivity. Another weakness is that in grism datasets where multiple rolls are unavailable (the majority of the area currently covered by deep NIRCam grism surveys), the number of viable lines may become intractable. When supporting photometry is sparse and photo-$z$ constraints are weak (e.g., EIGER, \citealt{Kashino23}), assigning the single line candidates a rest-wavelength becomes more challenging. Extending \textit{Allegro} to address these situations will be tackled in a forthcoming paper (Kramarenko \& Matthee in prep.), which builds upon the experience of the ALT data reduction. For now, the ``butterfly mosaic" and exquisite photometry allow us to apply this method.

\subsubsection{grizli}
\label{sec:grizli}
The \texttt{grizli} approach is complementary -- here we take the photometric redshifts as our starting point instead of in the final step, and then seek emission lines at their expected observed wavelengths. In detail, optimally-extracted 1D spectra are produced for each roll based on segmentation masks of the objects derived from the detection image (NIRCam LW stack in \S\ref{sec:photometry}). Emission lines are detected and identified in these 1D spectra. We iteratively search for lines in windows of $\Delta z$ such that $\Delta z/(1+z) < [1\%, 5\%, 10\%, 25\%]$, where the search is stopped when at least one S/N$>5$ line is found. Pure emission-line templates scaled from \texttt{FSPS} are used for this search, and all expected lines at $3-4\mu$m are fit simultaneously. The iterative search leverages the high photo-$z$ precision available in the ALT survey field. Starting with tighter windows also protects against contaminating emission lines from interloper galaxies that may be picked up with higher S/N. Protection against contamination is necessary due to the depth of the ALT data that is sensitive to emission lines down to $\approx6\times10^{-19}$ erg s$^{-1}$ cm$^{-2}$, due to the complex morphology of extended lensed arcs that are dispersed over several rows, and due to residuals from median filtering to remove continuum (e.g., \#4 in Figure  \ref{fig:rolls}). Spurious lines \texttt{grizli} latches onto are apparent in visual inspections (Figure \ref{fig:rolls}) and in cross-checks with the other method. 

The relative strengths of the \texttt{grizli} approach are that it uses more sensitive optimal extractions \citep[e.g.,][]{Horne86} based on morphology priors, and that it uses more data for redshift fits (there is no S/N requirement on detected lines, S/N$>5$ is required only for the strongest line and fainter lines are also picked up). The main weakness is the total reliance on photometric redshifts. The filter coverage and depth of the Abell-2744 photometry represents the best-case scenario for JWST redshifts, and photo-$z$s are likely to be less certain elsewhere -- e.g., the quasar fields surveyed by the EIGER \citep{Kashino23} and ASPIRE \citep{Wang23} grism surveys have imaging in only three NIRCam bands. Furthermore, the selection function of the emission line catalog is not as well-controlled as it also depends on the accuracy, completeness, and precision of the photometric redshifts, which are non-trivial to characterize and challenging to combine across various surveys requiring detailed recovery and completeness simulations.

While the two methods are strictly independent, we did not run them independently in the process of creating our exhaustive line-emitter catalog. In practice, we implemented insights obtained through one method with the other. For example, initial [OIII] and HeI+Pa$\gamma$ (pseudo-)doublet searches where we could identify redshifts without photo-$z$s with {\it Allegro} quickly revealed that the photo-$z$s typically were very precise. This enabled the use of strong priors in the {\it grizli} method, which then succeeded in identifying large samples of single-line H$\alpha$ emission line galaxies that in turn inspired us to implement the highest-S/N-line identification approach that {\it Allegro} uses. 

\begin{deluxetable}{lrrrrrrrrrrr}
\tabletypesize{\footnotesize}
\tablecaption{F356W Grism Specs}
\tablehead{
\colhead{Property} & \colhead{Measurement}
}
\startdata
\vspace{-0.3cm}\\
Nominal Wavelength Range & 3-4 $\mu$m \\
Spectral Resolution & $R\approx1600$\\
Typical Exposure Time & 24.8 ks\\
Depth (5$\sigma$ averaged) & 8$\times10^{-19}$ erg s$^{-1}$ cm$^{-2}$\\
Depth (best) & 6$\times10^{-19}$ erg s$^{-1}$ cm$^{-2}$\\
Depth (worst) & 2$\times10^{-18}$ erg s$^{-1}$ cm$^{-2}$\\
\enddata
\tablecomments{In detail, the wavelength range varies as a function of mosaic position (see bottom panel of Figure \ref{fig:RADEC}). The best depth is achieved in the stripes highlighted in Figure \ref{fig:RADEC} which received $2\times$ the typical exposure time, whereas the sensitivity is lowest towards the edges covered by a single roll.}
\end{deluxetable}

\begin{figure*}
    \centering
    \begin{tabular}{cc}
    \includegraphics[width=0.49\linewidth]{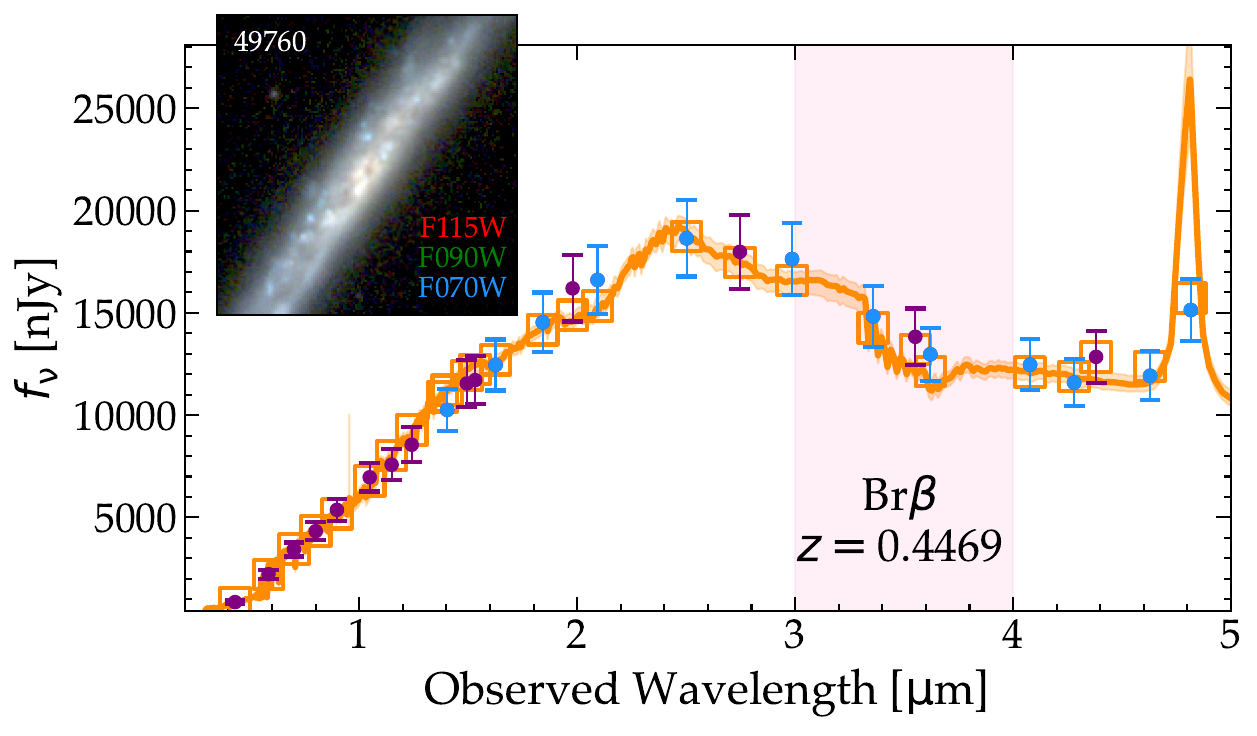} & \includegraphics[width=0.49\linewidth]{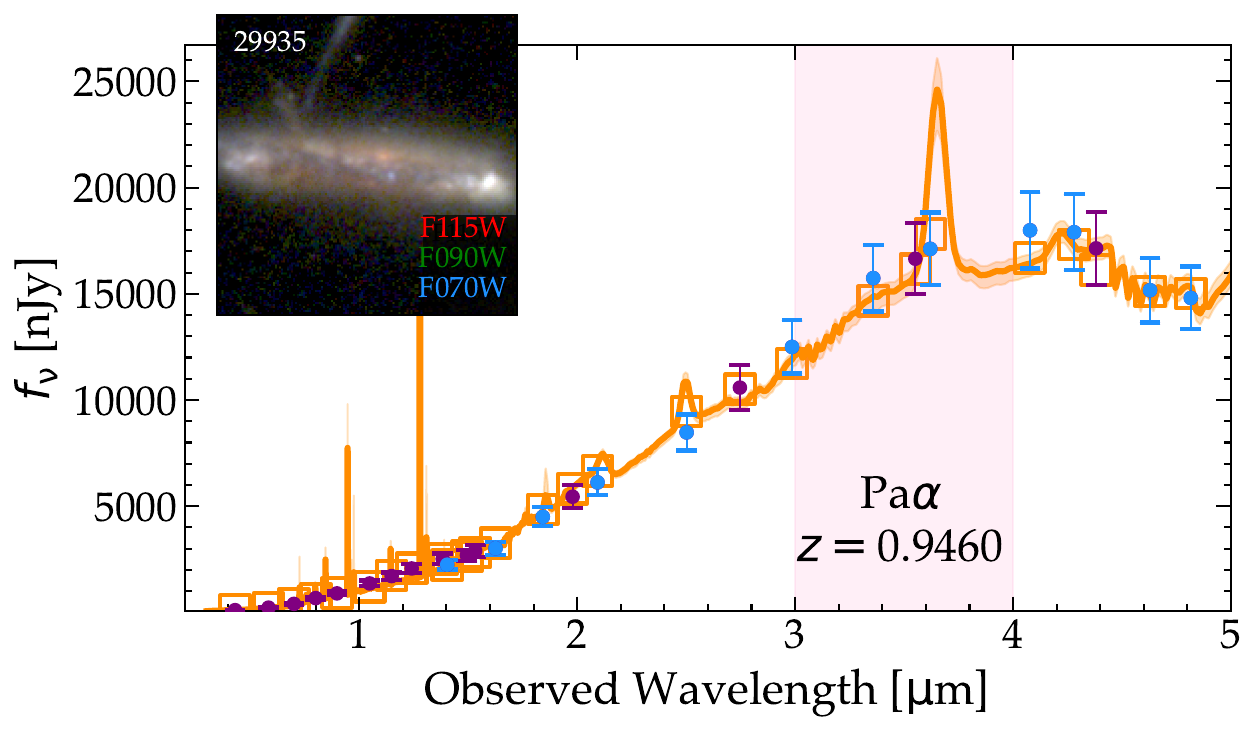} \\  
\includegraphics[width=0.49\linewidth]{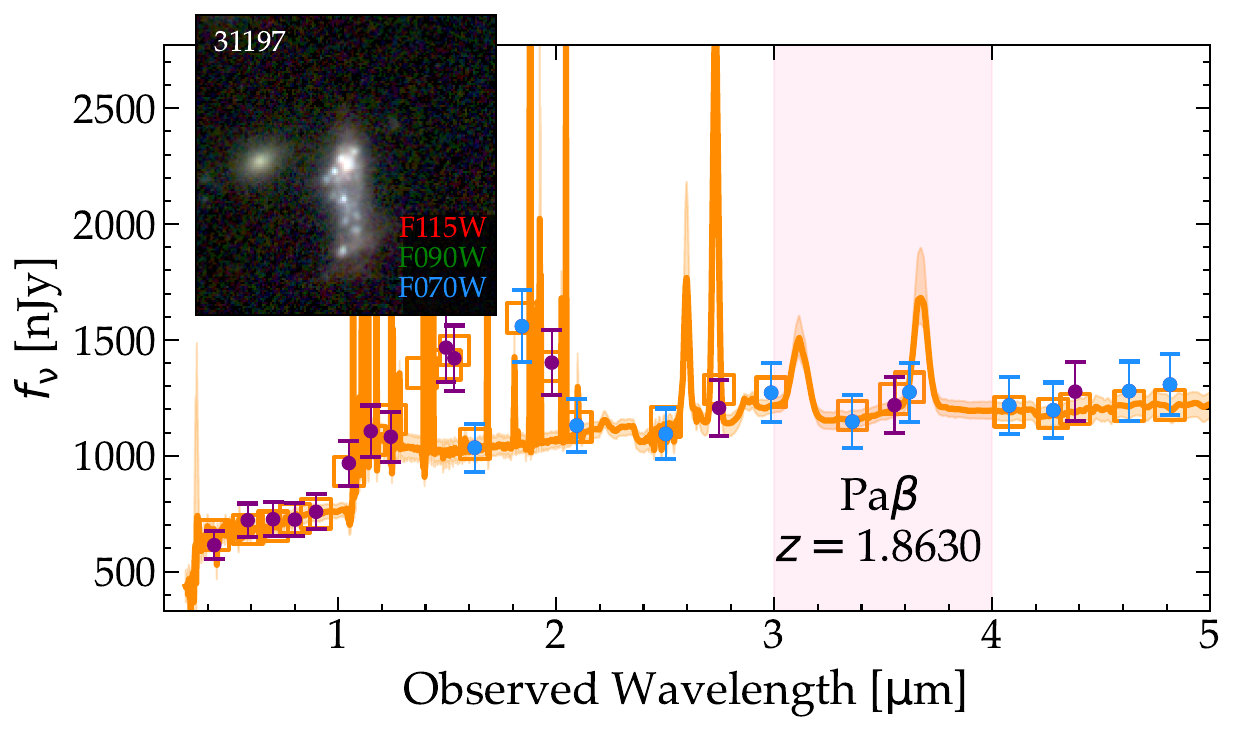} & \includegraphics[width=0.49\linewidth]{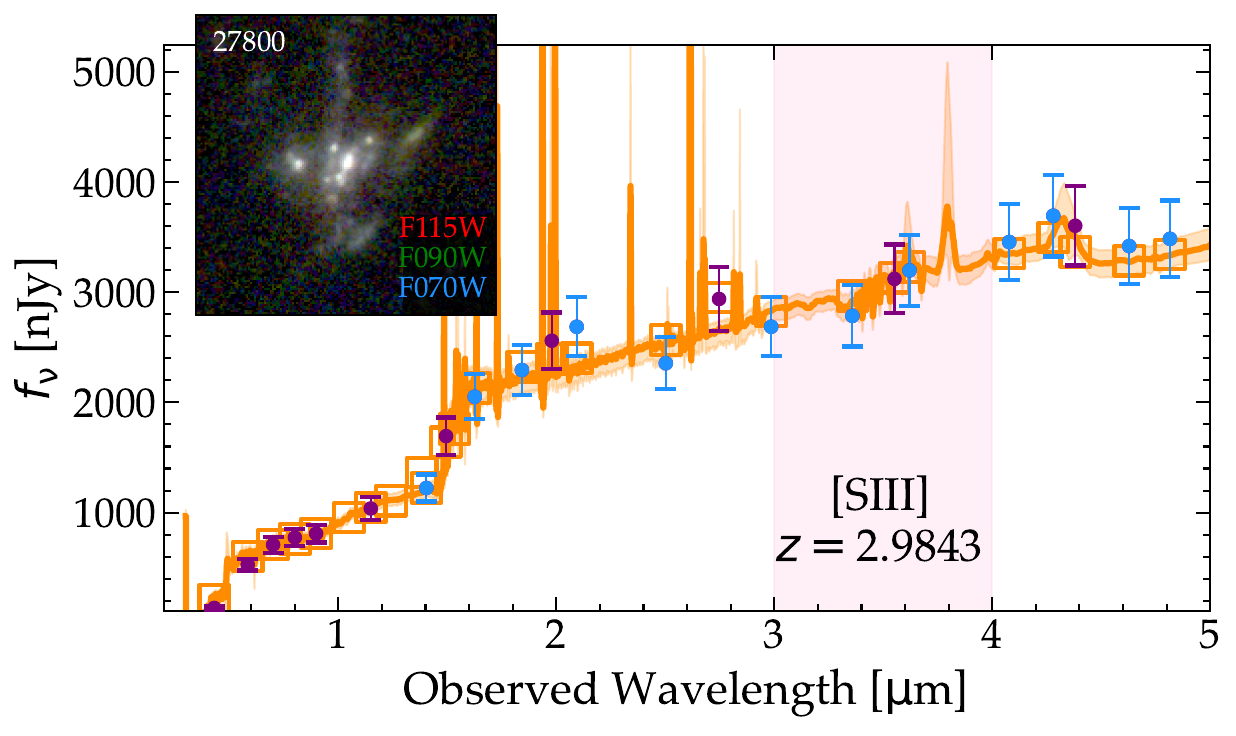} \\       
\includegraphics[width=0.49\linewidth]{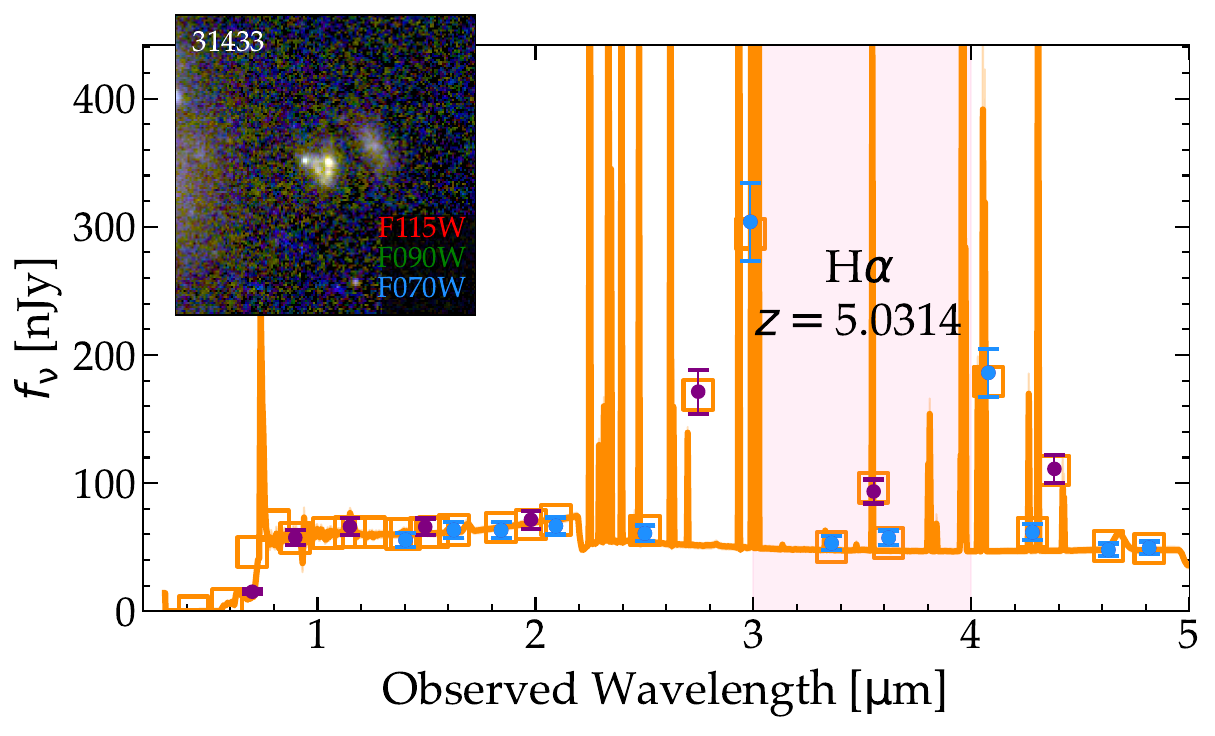} & \includegraphics[width=0.49\linewidth]{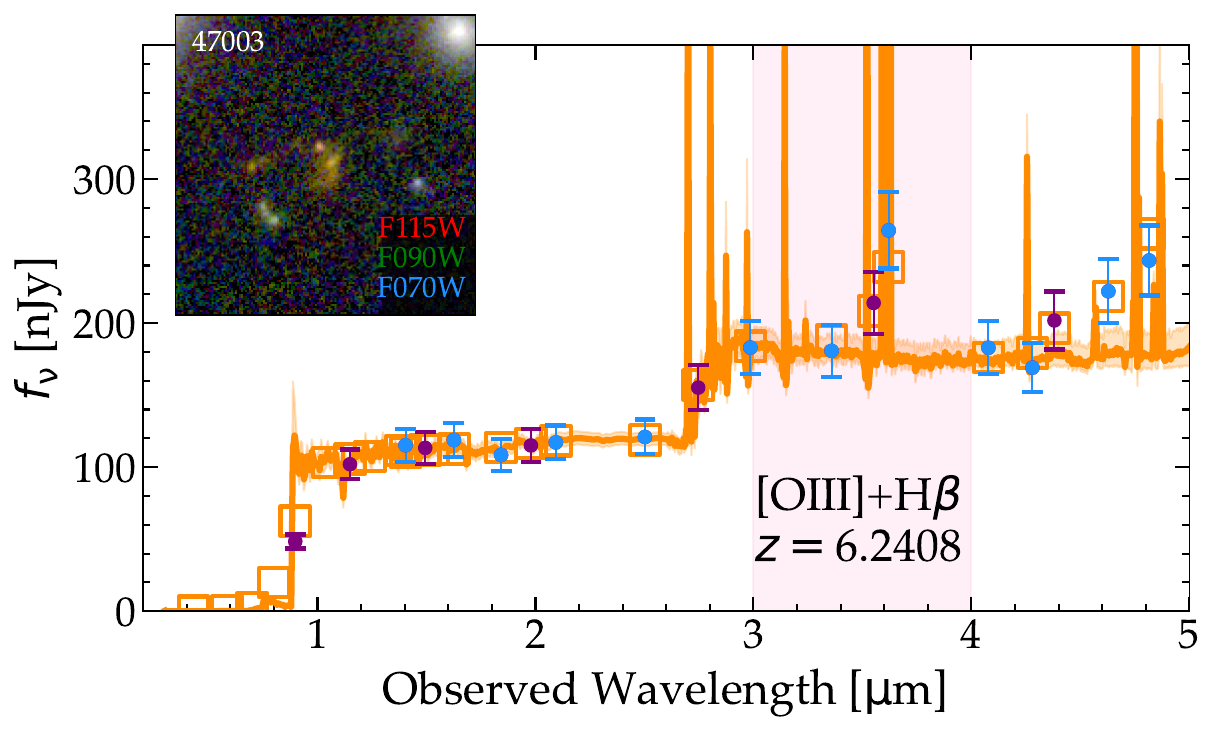} \\     
    
     \end{tabular}
     
    \caption{{\bf \texttt{Prospector} SEDs for representative galaxies in the various redshift regimes probed by ALT.} The SEDs (orange) are constrained by up to 27 JWST+HST broad-bands (purple) and medium-bands (blue). The wavelength range probed by ALT is highlighted in a pink band along with the primary spectral feature picked up by the survey. 100 draws from the posterior spectrum are summarized by their median as a solid orange line and the corresponding photometry is shown in orange squares. The area between the 16th and 84th percentiles of the draws is shaded, but is barely discernible at most wavelengths given the highly constraining data. Inset 3\farcs0$\times$3\farcs0 RGB images showcase the deep, high-resolution F070W and F090W images acquired by ALT that reveal rich substructure and the diversity of galaxies captured by the survey ranging from young, clumpy star-forming systems with extreme emission lines and Balmer jumps (bottom-left) to much more mature systems with older stellar populations, disks and dust lanes (top).}
    \label{fig:seds}
\end{figure*}

\subsection{SED modeling}
\label{sec:sed}

Physical parameters for all galaxies are estimated using the \texttt{Prospector} SED-fitting framework \citep{Leja17, Leja19, Johnson21} closely following the choices outlined in \citet[][]{Naidu22,Naidu22Schro,Tacchella22}. We use FSPS \citep{FSPS1,FSPS2,FSPS3} with the \texttt{MIST} stellar models \citep[][]{Choi17} and \texttt{MILES} spectral library \citep[][]{MILES2011} and assume a \citet{Chabrier03} IMF. Nebular emission is modeled with the \texttt{CLOUDY} \citep{Ferland17} grid described in \citet{Byler17}. The parameters we fit for include ten bins describing a non-parametric star-formation history, the total stellar mass, stellar and gas-phase metallicities, nebular emission parameters, and a flexible dust model \citep{KriekConroy13}. We adopt a ``bursty continuity" prior for the star-formation history \citep{Tacchella22} with the time bins logarithmically spaced up to a formation redshift of $z=20$. We hold the first two bins fixed at lookback times of 0-5 and 5-10 Myrs to capture bursts that power strong emission lines \citep[e.g.,][]{Whitler22} and that are expected to grow ubiquitous with increasing redshift. The large number of bins in the SFH are motivated by the rich data spanning 0.4-5.0 $\mu$m  (up to 27 bands total, see Table \ref{table:depth}).

We fit all available photometry and fix the redshift to the ALT grism redshift. Bands including and blueward of Ly$\alpha$ are masked (see Heintz et al. in prep., for modeling of DLA features at these wavelengths). Observed fluxes are corrected for magnification with an updated version of the lensing model from \citet[][]{Furtak23} presented in \citet[][]{Price24}\footnote{\url{https://jwst-uncover.github.io/DR4.html}}. This lensing model is now fully spectroscopically constrained using multiply imaged systems discovered using ALT and UNCOVER NIRSpec/Prism spectroscopy \citep{Price24} -- see \S\ref{sec:lensing} for more details. 

The highly constraining nature of the data and excellent quality of the fits is illustrated in Figure \ref{fig:seds}, where we show representative examples from different redshift slices. The spectroscopic redshifts combined with the complete NIRCam filter-set thoroughly pins down the detailed shape of the SEDs as well as excesses arising from emission lines (see also Fig. 4 of \citealt{Suess24}). The fine wavelength sampling constrains subtle features such as the 2.5$\mu$m dip in the Br$\beta$ emitter (top-left) and the Balmer jump in the H$\alpha$ emitter (bottom-left). These fits will be even further improved by incorporating ALT line-fluxes that will be included in our DR2.

\section{Data Release 1: The Catalog}
\label{sec:DR1}
In this paper, we publicly release the catalog with source coordinates and spectroscopic redshifts of the 1630 sources that we identified (DR1). Table $\ref{table:DR1}$ shows the example entries of the objects that are shown in this paper. In this section, we discuss general characteristics of the galaxy sample, such as the redshift distributions, the lines that we typically detect, how our catalog compares to other spectroscopic galaxy samples in the field, and how the spectroscopic redshifts compare to photometric redshifts.

\begin{deluxetable}{lrrr}
\tabletypesize{\footnotesize}
\tablecaption{Example entries of the ALT DR1 catalog \label{table:DR1}}
\tablehead{
\colhead{ID} & \colhead{R.A. (J2000)} & \colhead{Dec. (J2000)} & \colhead{$z_{\rm grism}$}\\
}
\startdata
18986 & 3.56053 & $-$30.41615 & 2.2909 \\
20720 & 3.61408 & $-$30.41045 & 6.3230 \\
23000 & 3.63696 & $-$30.40636 & 8.5116\\
26222 & 3.62512 & $-$30.40142 & 2.4941\\
26974 & 3.56893 & $-$30.40279 & 2.5835 \\
27075 & 3.54669 & $-$30.39973 & 0.7628\\
29936 & 3.61755 & $-$30.39501 & 4.5692 \\
39921 & 3.58247 & $-$30.37717 & 4.3055\\
41171 & 3.55831 & $-$30.37617 & 2.5780\\
42272 & 3.60451 & $-$30.38044 & 7.8771\\
42549 & 3.54281 & $-$30.38065 & 5.7918\\
45408 & 3.55918 & $-$30.38521 & 1.8854 \\
62975 & 3.56573 & $-$30.33597 & 3.9898 \\
69688 & 3.56943 & $-$30.34823 & 4.3069 \\
71962 & 3.57193 & $-$30.35168 & 4.3048 \\
80091 & 3.52965 & $-$30.36484 & 2.5034\\ 
\enddata
\tablecomments{This list includes various sources shown in this paper. The full catalog is available at \url{https://zenodo.org/records/13871850}.}
\end{deluxetable}

\begin{figure}
    \centering
    \includegraphics[width=0.99\linewidth]{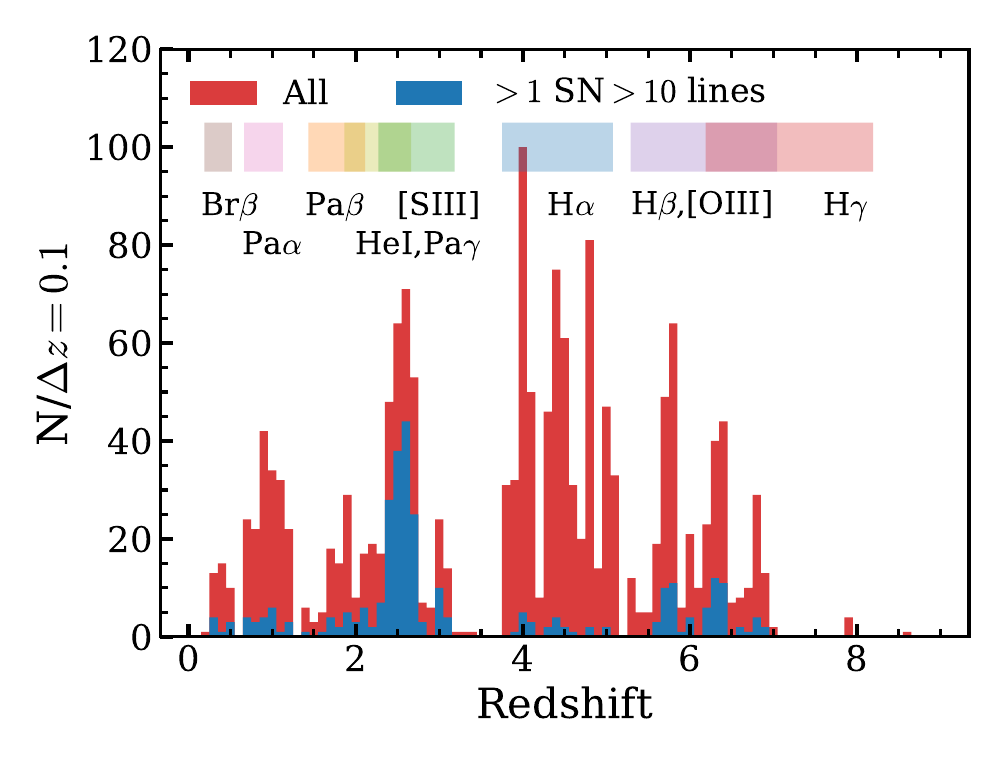} 
    \caption{{\bf Spectroscopic redshift distribution of the emission-line galaxies in ALT DR1.} The red histogram shows all redshifts, while the blue histogram show the galaxies for which more than one line is detected above S/N $>10$. This illustrates that the majority of pairs are HeI+Pa$\gamma$ at $z\sim2.5$ and the [OIII] doublet at $z\sim6$ and that the vast majority of H$\alpha$ emitters are single-line emitters in our data. We highlight the redshift ranges at which the primary lines are covered in the F356W grism spectra. }
    \label{fig:redshifts}
\end{figure}

\begin{figure*}
    \centering
    \begin{tabular}{cc}
    \includegraphics[width=0.44\linewidth]{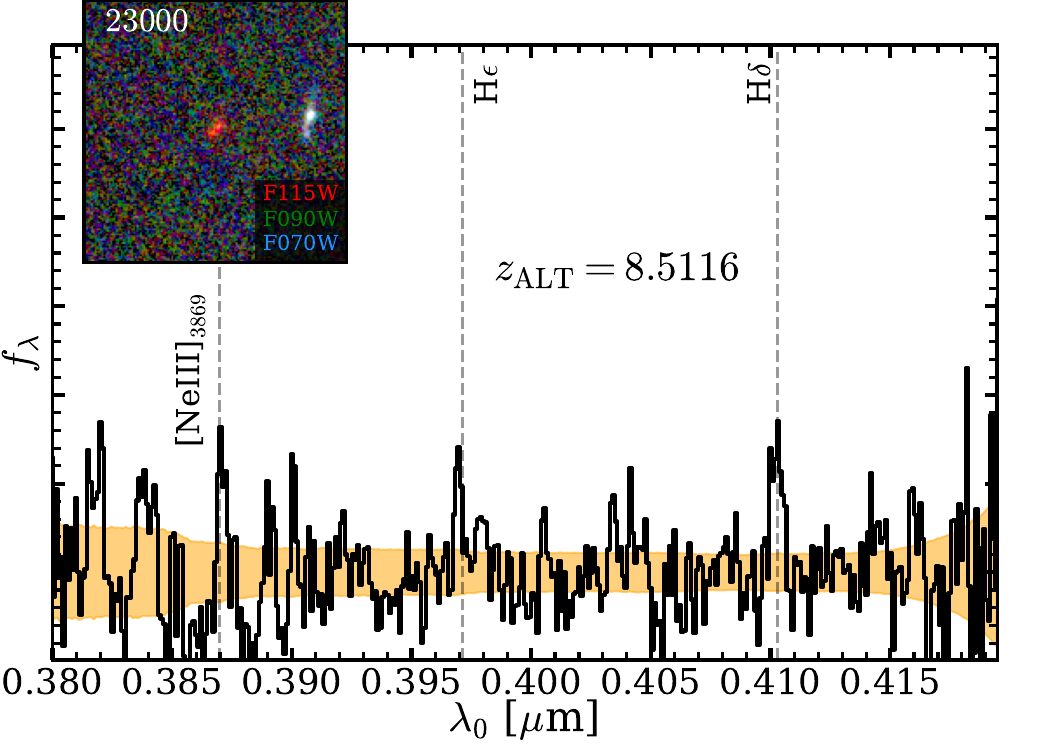} & \includegraphics[width=0.44\linewidth]{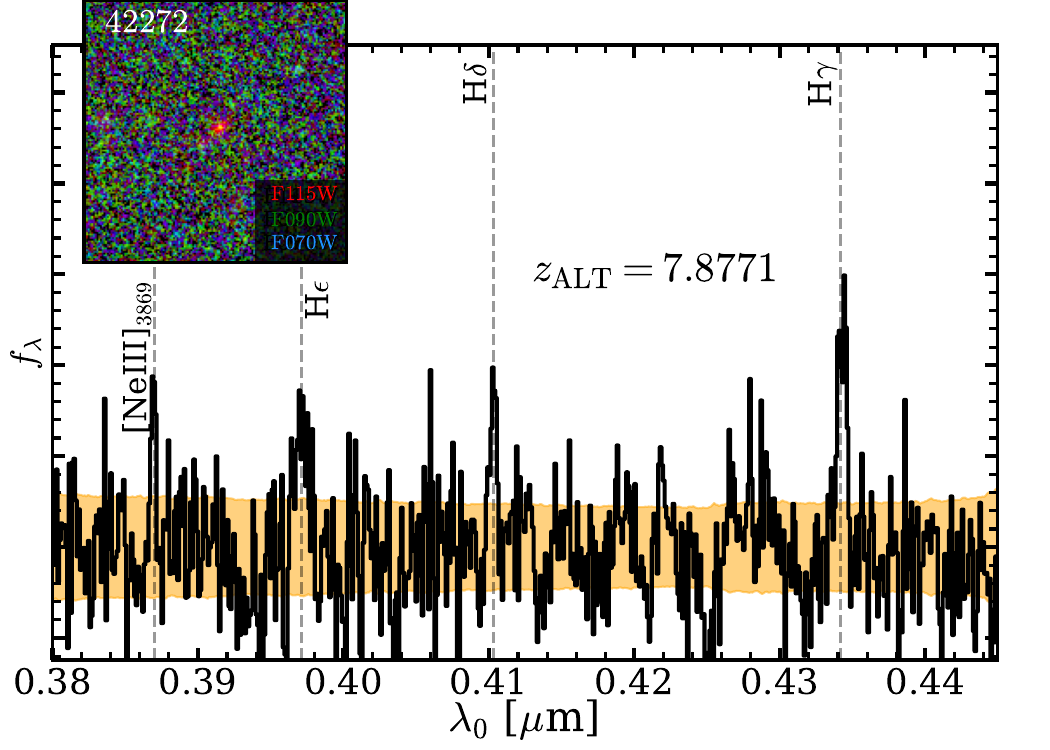}\\ \vspace{-0.15cm}\includegraphics[width=0.44\linewidth]{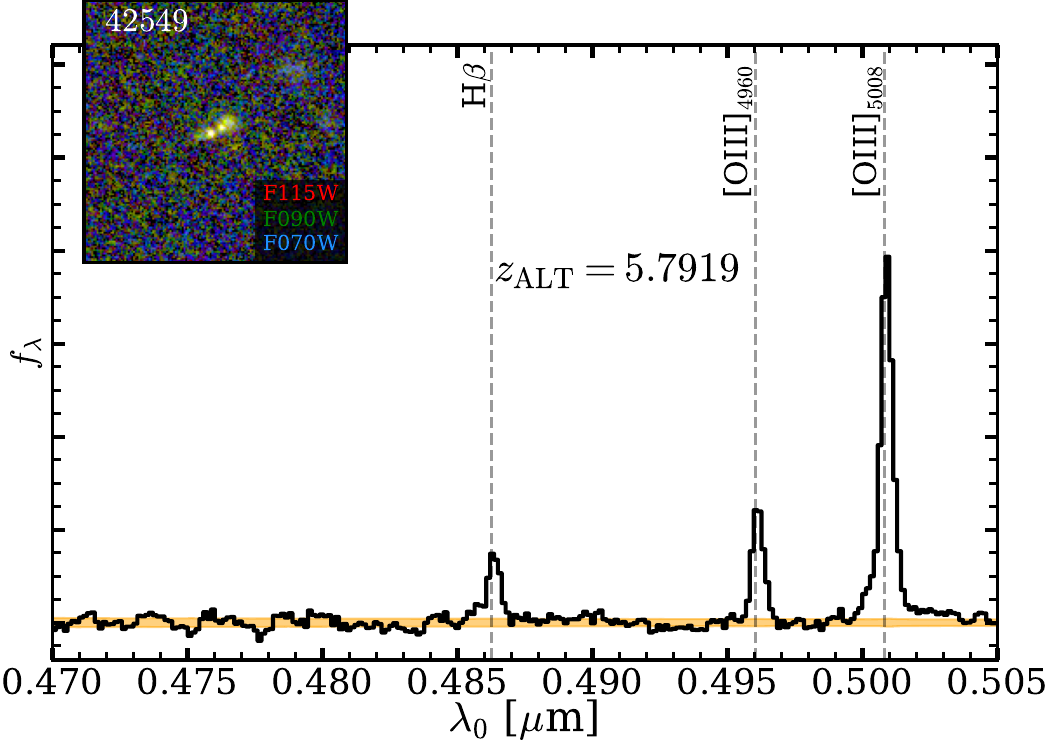}  &
\includegraphics[width=0.44\linewidth]{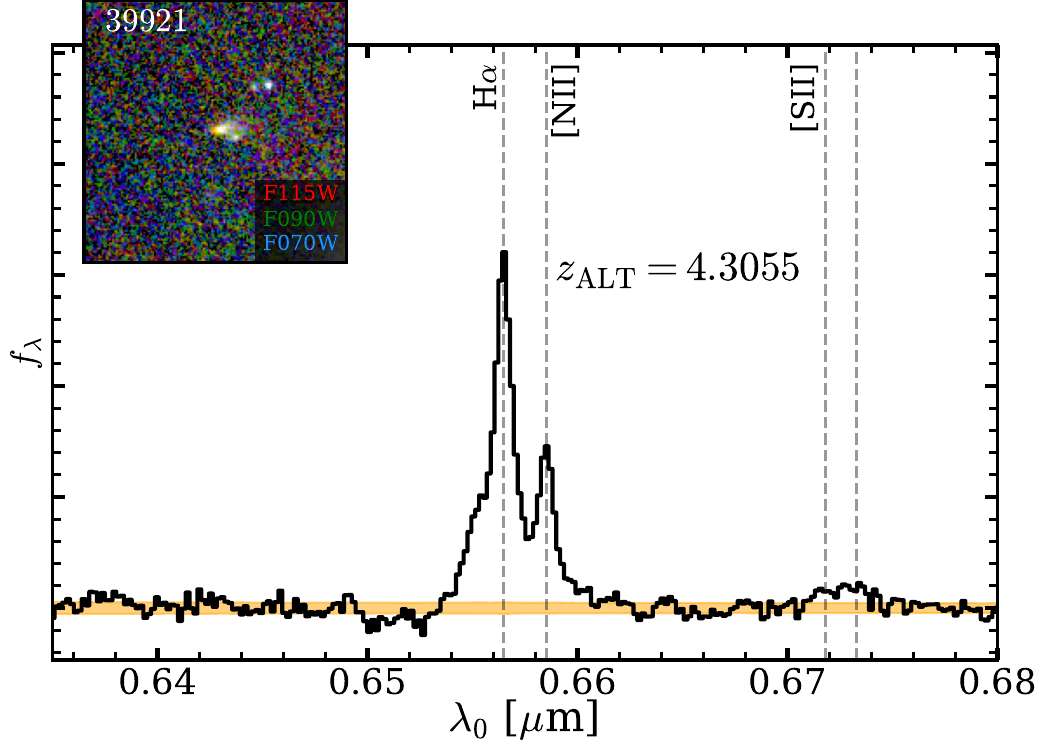} \\ \vspace{-0.15cm}
\includegraphics[width=0.44\linewidth]{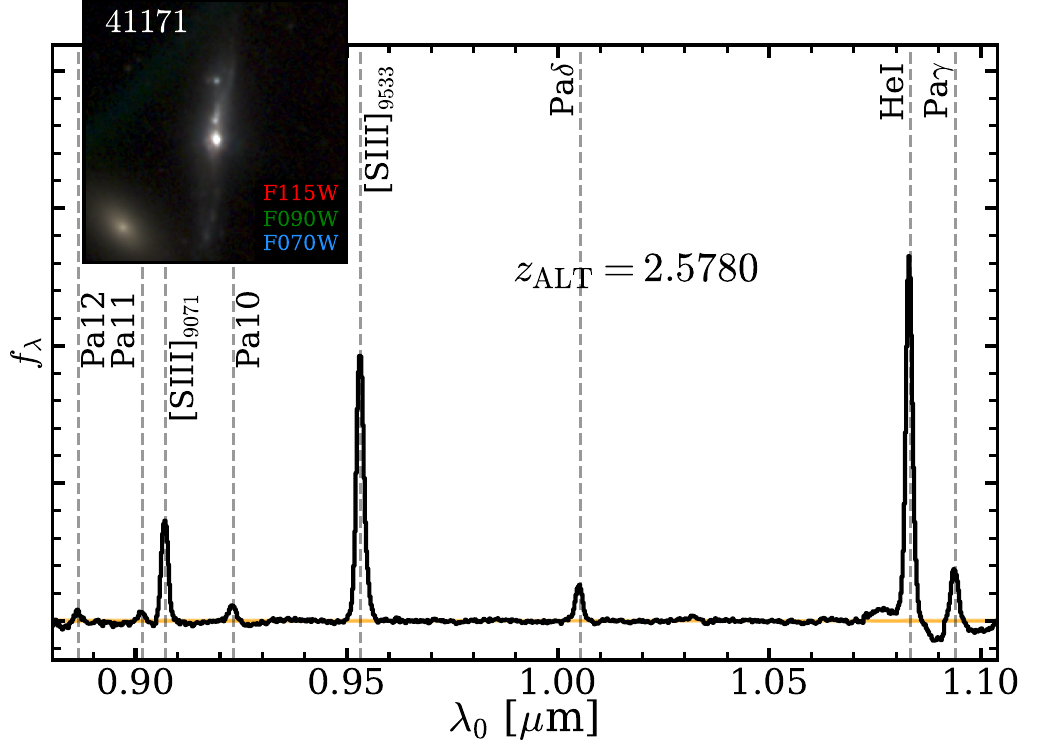} & \includegraphics[width=0.44\linewidth]{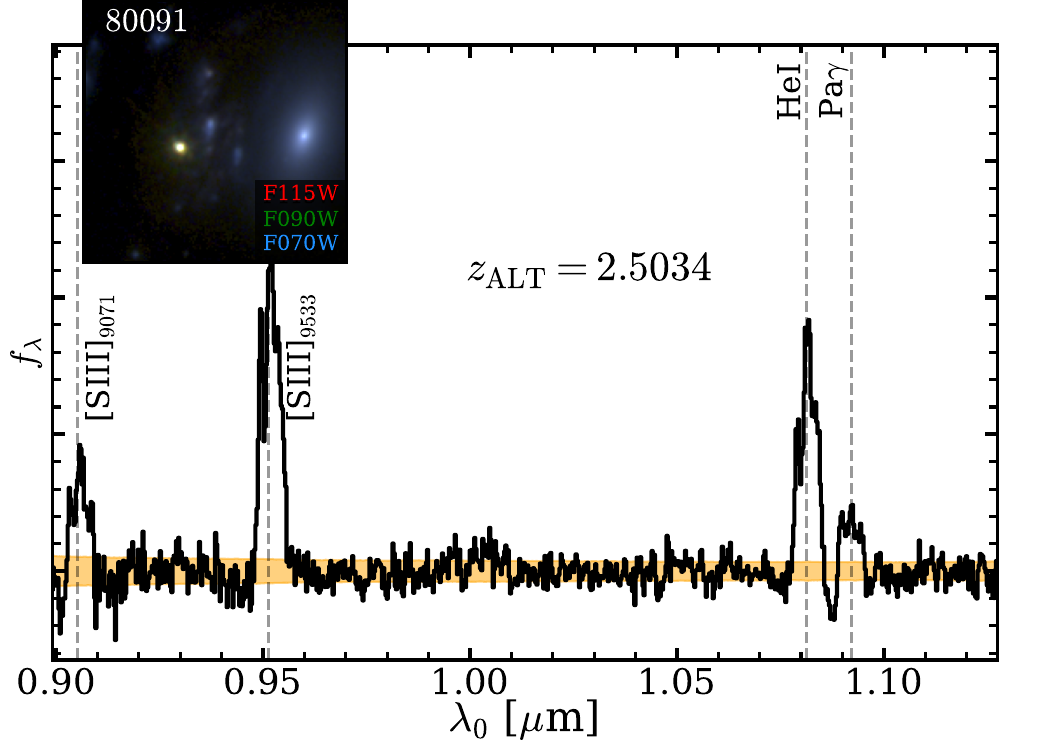}\\    \vspace{-0.15cm}   
\includegraphics[width=0.44\linewidth]{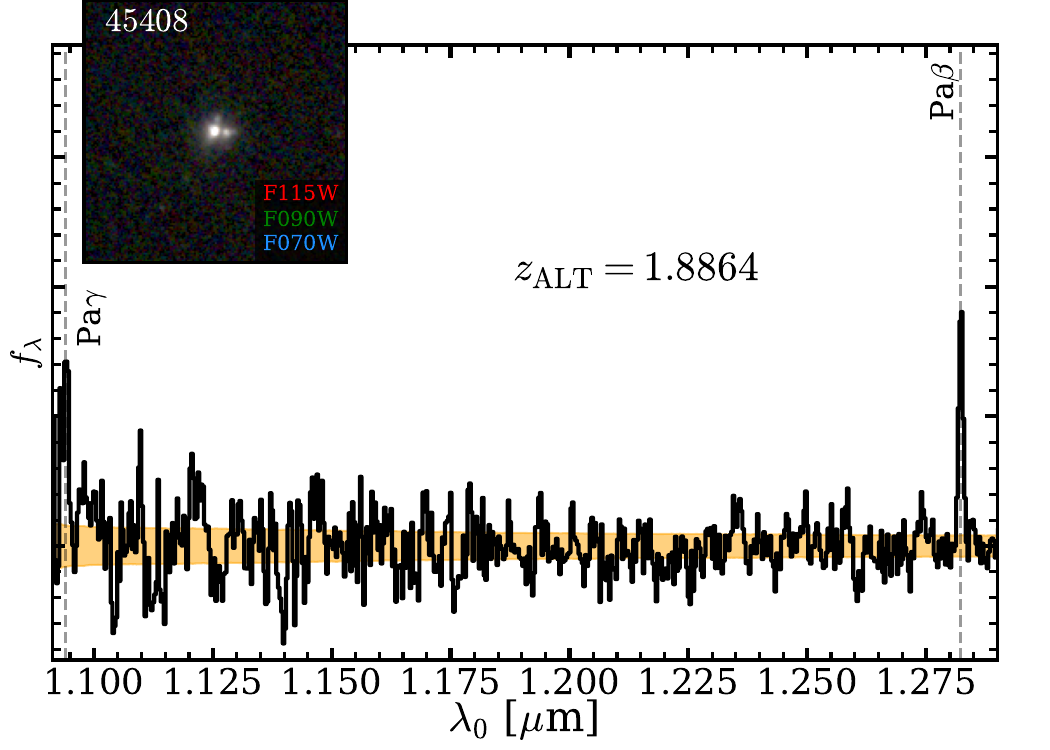} & \includegraphics[width=0.44\linewidth]{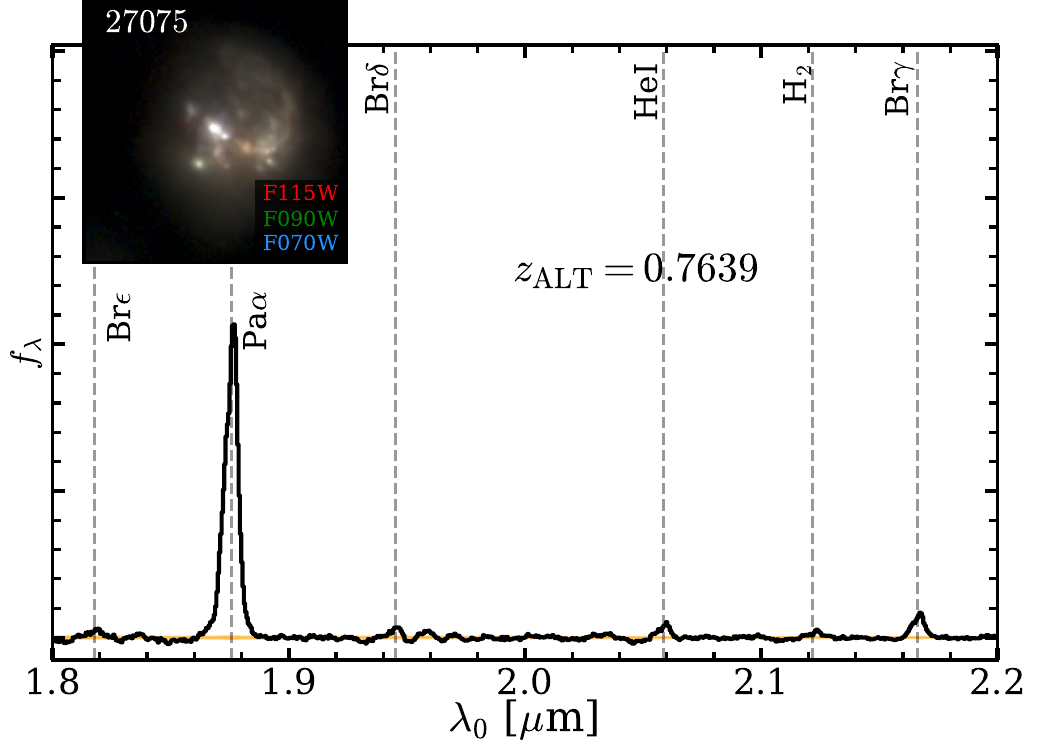}
     \end{tabular}
    \caption{{\bf F356W grism spectra showcasing the variety of emission-lines detected by ALT.} The orange shaded region shows the uncertainty on the flux. Inset $3\arcsec\times3\arcsec$ RGB images are comprised of F070W+F090W+F115W. See \S\ref{sec:numbers} for details on individual sources.}
    \label{fig:lines}
\end{figure*}

\subsection{The numbers}
\label{sec:numbers}
Figure $\ref{fig:redshifts}$ shows the distribution of the spectroscopically measured redshifts. This immediately shows the strong redshift clustering of our emission-line sample, with prominent redshift spikes at, for example, $z=2.6, 4.0, 4.3, 5.7$ (see \S\ref{sec:envlarge}), illustrating the high bias of early galaxies \citep[e.g.,][]{Fry96, Tegmark98, Desjacques18}. We highlight the primary emission-lines that are covered in our grism data (with F356W) at various redshifts. We also show which objects have multiple lines at S/N$>10$. For the majority of galaxies ($\approx74$ \%), we only detect a single emission-line whose {\it identity} strictly relies on additional photometric information, i.e., these are photo-$z$ identified, spectroscopically measured redshifts. For doublets, in particular [OIII]$_{4960, 5008}$ and HeI+Pa$\gamma$, the redshifts can be spectroscopically identified and measured. While the observed separations of the [OIII] doublet and the HeI+Pa$\gamma$ lines can be similar, we find that HeI is always brighter than Pa$\gamma$ (see also \citealt{Brinchmann23}), which is the opposite of the [OIII] doublet where the redder line is the brighter one. As there are no strong emission-lines in the rest-frame $\lambda=0.7$-0.9 micron range, our F356W GrismR survey has a lack of redshifts at $z\approx3.3$-3.8. A less prominent redshift valley is visible around $z\approx1.5$, where we lack strong lines between Pa$\alpha$ and Pa$\beta$ ($\approx1.3$-1.8 micron). Beyond $z>7$, H$\beta$ and the strong [OIII] doublet shift out of the filter, so spectroscopic redshifts rely on much weaker lines like H$\gamma$ and [OII], effectively limiting our survey to $z\approx7$ apart from a handful of exceptionally bright sources. 

Figure $\ref{fig:lines}$ showcases example spectra of the variety of galaxies in our catalog. This includes the most distant galaxy in our catalog at $z=8.51$ that was previously confirmed spectroscopically in \cite{Fujimoto23}, and a new spectroscopically confirmed galaxy at $z=7.877$ that likely belongs to a known over-density at that redshift in the field \citep{Morishita23}. These two sources highlight the capability of NIRCam grism spectroscopy to identify redshifts beyond the redshifts covered by the strong [OIII] doublet. In addition to the strong [OIII] doublet and H$\alpha$ line that have been the focus of the majority of NIRCam grism analyses so far, we also show galaxies at $z\sim1-2$ that display a suite of emission-lines from [SIII], HeI and various Paschen- and Bracket-series transitions. The spectra of ID 41171 and 27075 appear smoothed due to the spatial extent of these galaxies along the dispersion direction. ID 80091 is a clumpy galaxy, which is reflected by multiple peaks in the spectrum. ID 45408 is highlighted as it is an unusually young and star-bursting low mass galaxy at $z=1.88$ that shares various properties with the typical [OIII] emitter at $z\sim6$ in our sample.

\begin{figure*}
    \centering
    \includegraphics[width=0.9\linewidth]{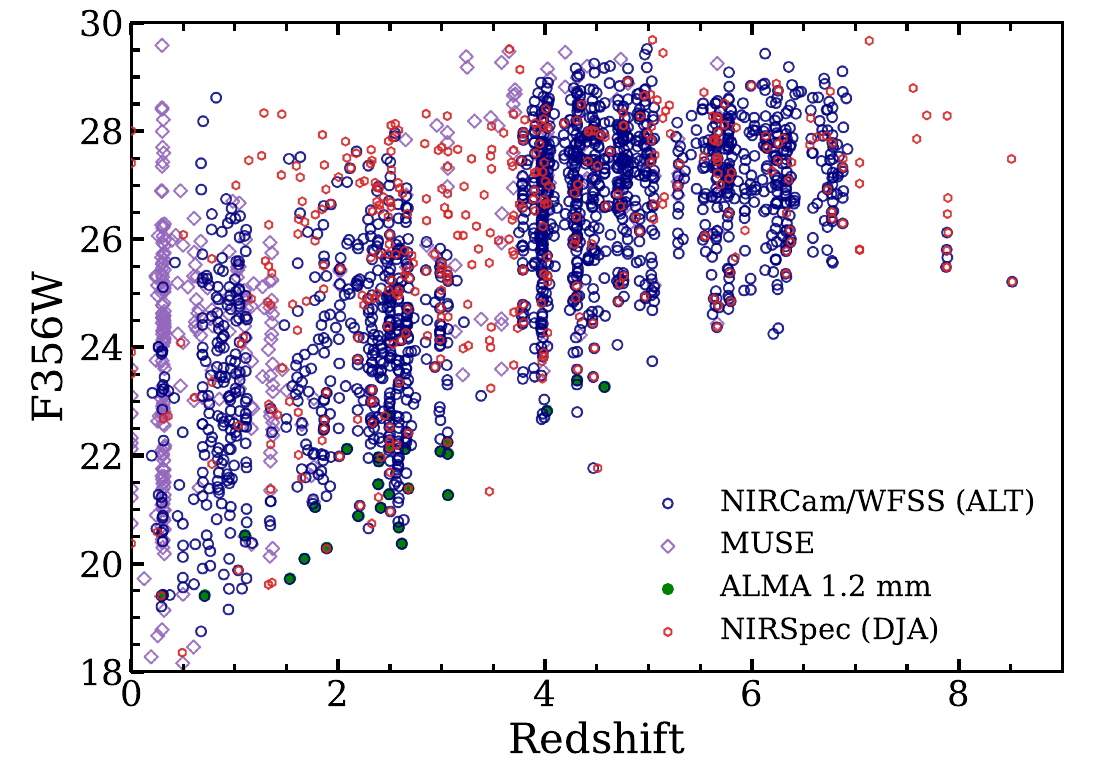}
    \caption{\textbf{The observed F356W magnitudes (not corrected for magnification) and spectroscopic redshifts of ALT DR1.} Blue circles show ALT galaxies ($N=1630$), while purple diamonds show sources with redshifts from VLT/MUSE ($N=328$; \citealt{Richard21}), red hexagons show redshifts from the DAWN JWST archive ($N=91$, \citealt{Heintz24}). Green filled circles show ALT galaxies that are detected in ALMA 1.2 mm continuum data ($N=27$; \citealt{DUALZ}).} 
    \label{fig:context}
\end{figure*}

Despite the increasing luminosity distance with redshift and the peak in the cosmic star formation rate density at $z\approx2$ \citep[e.g.][]{madau&dickinson}, we have detected roughly equal numbers of galaxies above and below $z=4$. This primarily originates from the fact that the H$\alpha$ and [OIII] luminosity produced per unit of star formation is much higher than the HeI, Paschen, Brackett, and [SIII] line luminosities. For example, under standard Case-B recombination assumptions, Pa$\alpha$ is expected to be $\approx9\times$ weaker than H$\alpha$ \citep[e.g.,][]{Reddy23}.

\begin{figure*}
    \centering
    \includegraphics[width=0.7\linewidth]{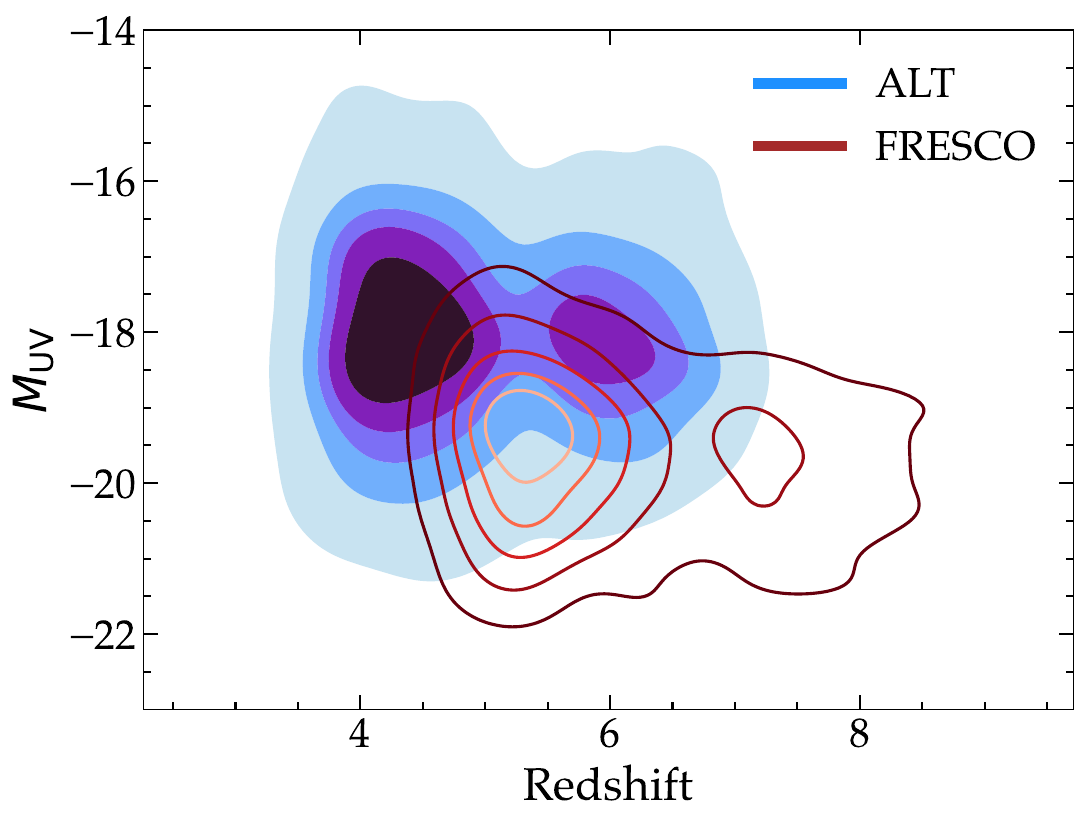}
    \caption{\textbf{The high-redshift parameter space probed by ALT compared to the FRESCO survey.} Comparison of the ALT ($z\approx3.5$-7; blue) and FRESCO ($z\approx5$-8.5; brown) H$\alpha$ and [OIII] samples \citep[][]{Meyer24, CoveloPaz24}. FRESCO uses the F444W grism and captures galaxies at higher redshifts. The bimodal distribution for each survey is due to the narrow redshift window where neither H$\alpha$ nor [OIII] are detected (see e.g., Figure \ref{fig:redshifts}). ALT pushes towards order-of-magnitude fainter galaxies ($\approx0.005 L^{*}$, $M_{\rm{UV}}\approx-15$) compared to existing grism surveys in blank extragalactic fields, of which FRESCO is the pioneering example ($\approx0.1 L^{*}$, $M_{\rm{UV}}\approx-18$). We present first results enabled by this new parameter space in \S\ref{sec:results}.} 
    \label{fig:contextFRESCO}
\end{figure*}

\begin{figure*}
    \centering
    \begin{tabular}{cc}
    \includegraphics[width=0.49\linewidth]{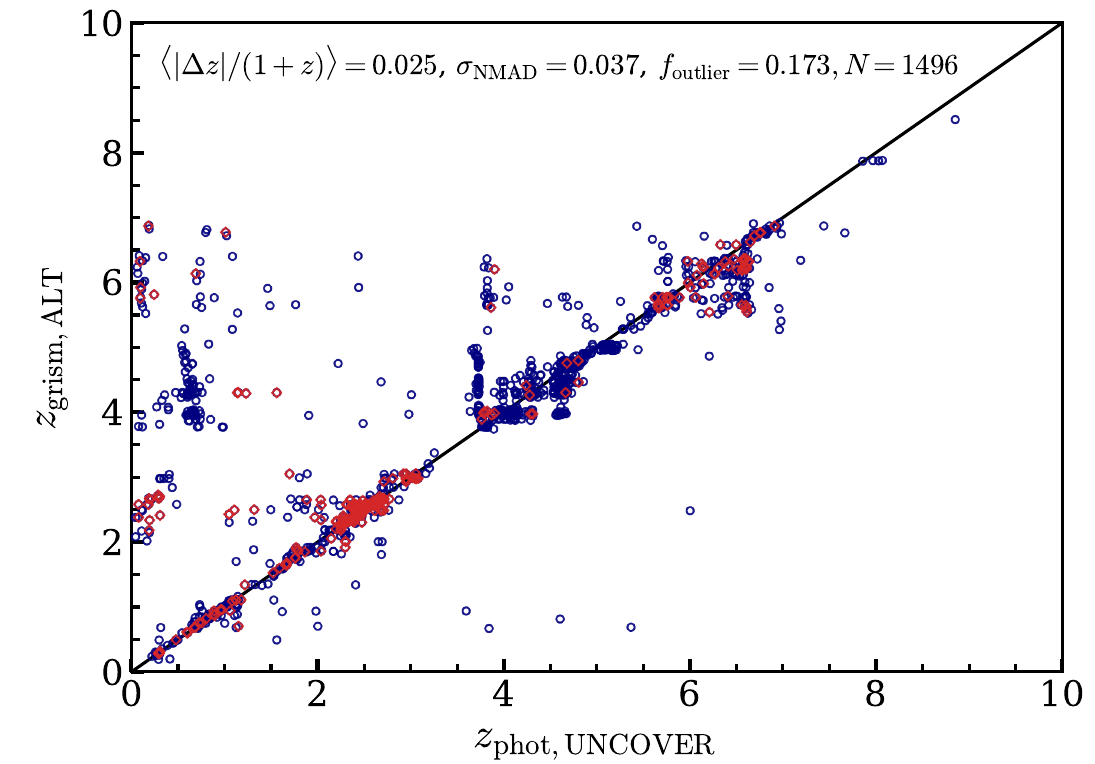} & 
     \includegraphics[width=0.49\linewidth]{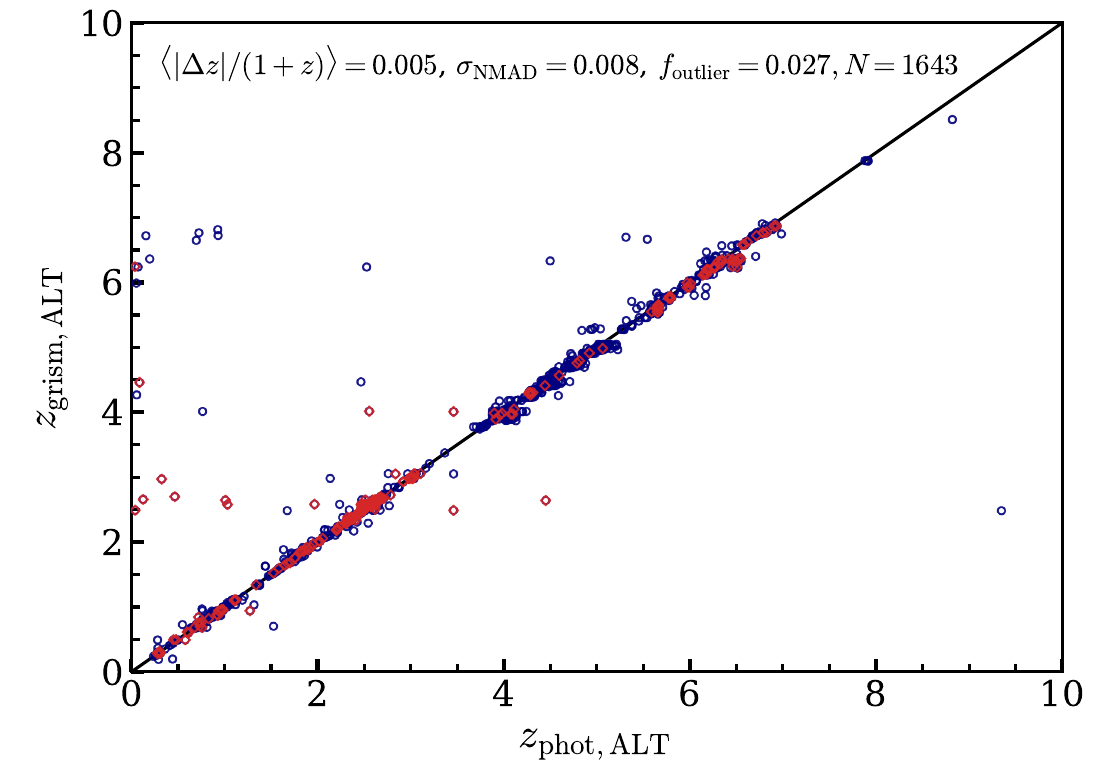} \\  
     \end{tabular}
    \caption{\textbf{Comparison between the photometric and spectroscopic redshifts of our ALT galaxies.} Blue circles show galaxies for which we have detected a single emission-line, red circles are galaxies for which we have detected multiple emission-lines. We note that the photometric redshifts are used in the line-identification of single line-emitters. The black solid line shows the 1:1 relation. In the left panel, we compare our redshifts with the broadband photometric redshifts from UNCOVER DR1 \citep{Wang24SPS,Weaver24}. In the right panel, we compare with our photometric redshifts that include NIRCam medium-bands and F070W and F090W photometry. In each panel we list the average absolute redshift difference, the normalised median standard deviation $\sigma_{\rm NMAD}$ (see \citealt{Brammer08}), the outlier fraction (defined as the fraction of galaxies with $|\Delta z|/(1+z) > 0.15$) and the total number of comparison galaxies. This number is slightly lower for the UNCOVER DR1 comparison because the position angle of the ALT data is slightly different. Both panels show good photometric redshift performance, in particular excellent performance for photo-$z$s with F070W/F090W and medium-bands. The majority of the outliers are objects whose photometry is blended with foreground galaxies, or, in some cases, show strong evidence of AGN emission.}
    \label{fig:zphots}
\end{figure*} 

Figure $\ref{fig:context}$ shows the spectroscopic redshifts and F356W magnitudes of the ALT sample, compared to other samples in the Abell 2744 field. The VLT/MUSE redshifts \citep{Richard21}, that originate from a $3\times3$ arcmin$^2$ dataset around the main cluster core, contain a larger number of cluster galaxies at $z\approx0.3$ and also more galaxies at $z\approx1.5$, where it benefits from the use of rest-frame optical lines as [OII]. The MUSE sample contains few galaxies at $z\approx2-3$, where our sample has a large number of galaxies with redshifts from HeI+Pa$\gamma$. MUSE detects Lyman-$\alpha$ emission at $z=3$-6.6, including in extremely UV-faint galaxies \citep[e.g.][]{Maseda18}. However, Ly$\alpha$ emitters are only sample a sub-set of the galaxy population \citep[e.g.][]{Cassata2015,Kusakabe20,Matthee21} and at $z>5$ the sensitivity is strongly impacted by various night sky emission lines, such that the majority of ALT galaxies are not detected in Ly$\alpha$ emission. 

We also show spectroscopic redshifts from the public DAWN {\it JWST} archive\footnote{\url{https://dawn-cph.github.io/dja/spectroscopy/NIRSpec/}}, which are compiled from JWST/NIRSpec observations as part of various follow-up programs \citep{Heintz24}. The selection function of these archival observations appears to be biased towards high redshifts. With the full 1-5 $\mu$m range covered by NIRSpec, it is easier to detect galaxies at $z>7$, such as the relatively large number of spectroscopic redshifts for the over-density at $z\approx7.8$ \citep{Morishita23}. The NIRSpec sample also slightly fills our redshift gap at $z\approx3.5$. Apart from that, the ALT spectroscopic redshifts vastly outnumber the NIRSpec redshifts, in particular at $z\approx2.5$. It is furthermore interesting to note that while NIRSpec is in principle more sensitive than the NIRCam grism, there are few galaxies with spectroscopic redshifts that have a fainter F356W magnitude than those that we have identified. As this is likely due to prioritisation of more luminous systems in the NIRSpec observations, it highlights that the source-density of objects for which JWST can in principle measure spectroscopic redshifts is higher than the typically used maximum number of open NIRSpec shutters.

Various scientific results about the properties of galaxies can already be discerned from a glance at Figure $\ref{fig:context}$. We show 27 galaxies detected in the ALMA 1.2 mm continuum map \citep{DUALZ} with green-filled circles. The majority of these objects are among the most luminous, highly star forming galaxies in our field at $z\approx2$, but there are also three luminous H$\alpha$ line-emitting galaxies at $z\approx4$ with a high amount of obscured star formation. These galaxies (of which some are shown in Section $\ref{sec:lineprofiles}$) typically have detections of the [NII] line as well, indicative of a relatively high gas-phase metallicity \citep[e.g.,][]{Sanders23}. The luminous galaxy at $z=4.47$ with F356W$\approx22$ that is {\it not} detected in the ALMA data is the most luminous broad-line H$\alpha$ emitter known at $z\approx4.47$ (\citealt{Greene24}, Labbe et al. in prep.), indicative of different physical processes causing the high F356W luminosity (such as AGN activity), compared to the other luminous galaxies at similar redshift.

Figure $\ref{fig:context}$ also highlights a remarkable difference in the F356W magnitude distributions of ALT galaxies at different redshifts. The $z\approx1$-2 samples peak at F356W magnitudes $\approx24$ and drops at magnitudes fainter than 27. The $z\approx4$-7 samples peak at much fainter magnitudes, F356W$\approx28$, but rapidly drop at fainter magnitudes. These differences can be understood as follows: our sample selection is emission-line flux limited, and the F356W magnitude is a combination of line and continuum flux. The H$\alpha$ and [OIII] emission lines that are used to identify galaxies at $z\approx4$-7 have much higher equivalent widths (EWs) than the [SIII], HeI and Paschen lines that are used to measure redshifts at $z\approx1$-2. Therefore, we are simply not detecting galaxies at $z\approx2$ with magnitudes as faint as $\approx28$ as it would require these lines to have unusually high equivalent widths. The same argument applies to the $z\approx8$ samples, where we rely on lines like [OII] and H$\gamma$, whose EWs are significantly lower than [OIII]. At first glance, a comparison of the H$\alpha$ and [OIII] emitter samples suggests a lack of faint [OIII] emitters which could possibly be ascribed to a decrease in the EW distribution at faint (continuum) magnitudes, e.g., due to a lower metallicity. However, a detailed comparison shows that the drop in F356W magnitudes is very similar when normalising the H$\alpha$ and [OIII] distributions to their peak number densities. A detailed investigation on the implications of the continuum-dependence of the EW distributions \citep[e.g.][]{Endsley23} will be pursued in a future paper. A clear difference is that the H$\alpha$ sample extends to brighter F356W magnitudes, probably owing to the build up of the bright end of the luminosity function towards lower redshift \citep[e.g.,][]{Sobral13}.

In terms of spectroscopic completeness with respect to our photometric parent catalog (limited to F356W$<29$ mag), we have measured spectroscopic redshifts for 6\% of \textit{all} the sources. This fraction depends strongly on the photometric redshift, given that intrinsically stronger lines are accessible to ALT towards higher redshift. ALT's completeness is 4\% for $z_{\rm phot}=0-4$, 30\% at $z_{\rm phot}=4-7$, and peaks at 50\% at $z_{\rm phot}=6-6.5$. For objects with F115W$<27$ mag, we have measured 20\% of their redshifts over the full redshift range, ranging from 10\% at $z_{\rm phot}<4$ and 60\% at $z_{\rm phot}=4-7$. For context, comparing against a typical NIRSpec prism survey in extragalactic legacy fields (e.g., the UNCOVER survey's prism component in Abell 2744, \citealt{Price24}), the completeness fractions for F356W$<29$ mag are $0.7\%$ at $z<4$, and $2.0\%$ at $z=4-7$.

To place the ALT sample in the context of existing grism datasets in legacy extragalactic fields, in Figure \ref{fig:contextFRESCO} we compare the \textit{intrinsic} UV luminosity (i.e., corrected for magnification and redshift) of the ALT H$\alpha$ and [OIII] emitters at $z\approx3.5-7$ with the corresponding sample from the FRESCO survey \citep[][]{Oesch23} that targeted the GOODS fields with the redder F444W filter and identified galaxies at higher redshifts of $z\approx5-8.5$ \citep[e.g.,][]{Herard-Demanche23,CoveloPaz24,Xiao23,Meyer24,Helton24}. The FRESCO SEDs are modeled exactly as described in \S\ref{sec:sed} for fair comparison. This figure illustrates how ALT pushes roughly an order of magnitude deeper at fixed redshift, returning spectra for faint galaxies (down to $M_{\rm{UV}}\approx-15$, $\approx 0.005 L^{*}$), extending FRESCO's reach ($M_{\rm{UV}}\approx-18$, $\approx 0.1 L^{*}$) to lower luminosities. ALT is assisted by gravitational lensing (typically $\approx2\times$) as well as deeper integrations ($\approx7-28\times$) in a more sensitive filter ($\approx1.5\times$). Conversely, the rare $M_{\rm{UV}}\approx-20$ systems are sparse in the ALT sample, likely due to the smaller volume probed on-sky, as well as when adjusted for lensing ($1/\mu$). Overall, ALT shows a clear path to spectroscopically probing the faint end of the UVLF in lensed fields.

\subsection{Redshift performance} \label{sec:redshift_performance}

There are 79 ALT galaxies with a previously published redshift from VLT/MUSE \citep{Richard21}. Only 2/79 galaxies have a large redshift discrepancy. These are ALT IDs 28423 and 45074, which are at $z=2.5824$ and $z=2.494$, respectively. Both their redshifts are confirmed by multiple lines (HeI, [SIII]) and their photometric redshifts are consistent with the ALT redshift. The reported MUSE redshifts are $z=0.951, 0.787$, respectively. In the case of 45074, the redshift difference can be explained by misidentification of real CIII]$_{1907,1909}$ lines as an [OII] doublet. For the other galaxies, we find very good redshift agreement with a median $\Delta z_{\rm MUSE}- z_{\rm ALT}/(1+z) =0.0004$ ($\approx100$ km s$^{-1}$). This difference is driven primarily by the use of Lyman-$\alpha$ redshifts at $z>3$, which are known to be typically redshifted with respect to the systemic redshift \citep[e.g.][]{Verhamme18}. Indeed, when limiting our sample to $z<3$, we find a median offset of $10\pm60$ km s$^{-1}$. At $z>3$, we find a median offset of 220 km s$^{-1}$, which is a very common Ly$\alpha$ offset \citep{Trainor16,Muzahid20,Matthee21}. Overall, these comparisons are a confirmation of the absolute wavelength calibration of the NIRCam grism data, yielding a redshift precision of $\approx60$ km s$^{-1}$ (consistent with \citealt{Torralba-Torregrosa24, Bordoloi24}). 

In Figure $\ref{fig:zphots}$ we compare our spectroscopic redshifts to photometric redshifts. In the left panel, we use published UNCOVER DR1 photometric redshifts that use HST/ACS, WFC3 and (broad-band) JWST/NIRCam data \citep{Weaver24}, while we also additionally include our F070W and F090W data, medium-band imaging \citep{Suess24}, and all other literature data (\S\ref{sec:publicdata}) in the photometric redshift measurements (\S\ref{sec:photozs}) that are displayed in the right panel. In each panel, we show the median redshift differences $\langle |\Delta z|/(1+z) \rangle$. In particular the photometric redshifts with medium-band data are extremely precise with $\langle |\Delta z|/(1+z) \rangle = 0.005$ ($\approx1500$ km s$^{-1}$). We also quantify the overall quality of the photometric redshifts using the normalised median deviation, $\sigma_{\rm NMAD}$ \citep{Brammer08}, which is 0.037 for the UNCOVER photo-$z$s and reduces to values as low as $\sigma_{\rm NMAD}=0.008$ for our photo-$z$s. Outliers are defined as objects for which $|\Delta z|/(1+z) \geq 0.15$ and constitute 17.3 \% and 2.7 \% of the sample, respectively. The majority of these 2.7 \% outliers are objects for which the photometry is blended with a foreground galaxy (for example, object \#3 in Figure $\ref{fig:rolls}$) or AGNs. 

We note that the ALT photometric redshift is used in the line-identification of single line-emitters, which could lead to an artificially precise agreement between the grism-$z$ and photo-$z$. However, for objects in which we detect multiple emission lines (and for which the photo-$z$ was not necessarily used), we find that the photometric redshifts are even more precise, $\sigma_{\rm NMAD}=0.025, 0.005$, respectively. For the majority of objects that are single line-emitters, the photo-$z$s are used to distinguish whether lines are either H$\alpha$ at $z\approx4$, or lower redshift Paschen lines, which have vastly distinguishable SEDs.

Our photometric redshift precision has improved significantly relative to Cycle 1 data \citep[][]{Weaver24,Wang24SPS} owing to the use of F070W/F090W and medium-band data. One could question the added value of spectroscopic redshifts over photometric redshifts. We should however note that the photometric redshifts of our emission-line sample are relatively `easy' to get right thanks to the multitudes of emission-lines boosting various medium-bands. Indeed, the results from \cite{Suess24}, who report a slightly higher 7\% outlier fraction and $\sigma_{\rm NMAD}=0.015$, stem from a spectroscopic sample that is primarily based on JWST/NIRSpec observations which also includes objects with weaker emission-lines. We also discuss in \S\ref{sec:whygrism}, that despite this wealth of data there exist critical blind-spots in the  photo-$z$s (yielding ``phantom overdensities", possibly due to the template set included in \texttt{eazy}) and science applications that are only possible with the grism redshifts (e.g., clustering). %eazy

%%%%%%%%%%%

\section{Results}
\label{sec:results}

In this section we demonstrate the discovery space opened by a deep grism survey over a lensing cluster. The first results presented here are made possible by the unique combination of ALT's depth, statistics, all-inclusive selection function, and spectral and spatial resolution. To begin this exploration, we present some general findings in this paper that also characterize our dataset and its potential for mapping early galaxies in unprecedented ways. We will present more specific results in dedicated forthcoming publications that will address the main science goals of the survey outlined in \S\ref{sec:intro}.

\subsection{Unlocking the Potential of Lensing Clusters I: Highly Magnified Arcs}
\label{sec:arcs}

\begin{figure*}
\centering
\includegraphics[width=0.8\linewidth]{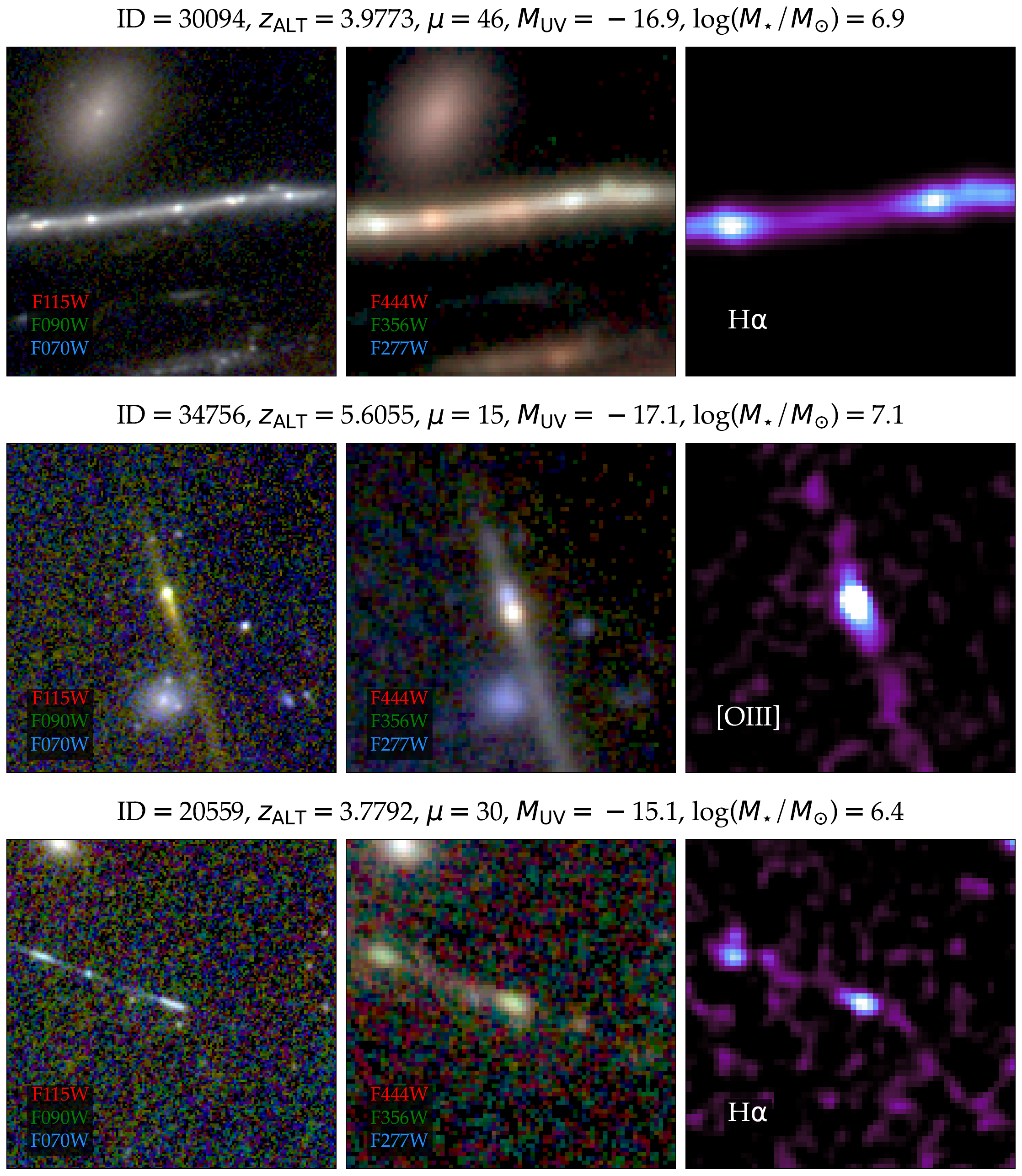}
\caption{{\bf Spatially resolved spectra of highly magnified arcs.} 3\farcs0$\times$3\farcs0 RGB cutouts of NIRCam images (left, center) are shown for comparison with ALT emission line maps (right). Discovering such highly lensed sources was our key motivation for surveying a cluster field, and has allowed us to probe sources as faint as $M_{\rm{UV}}\approx-15$ (bottom row). The grism spectra capture the complex morphology of the emission lines that are challenging for slit spectroscopy. The high spectral resolution separates [OIII] from H$\beta$ and [NII] from H$\alpha$ to help isolate metal-poor clumps that display relatively weak metal lines compared to the Balmer lines.
}
\label{fig:arcs}
\end{figure*}

Sources with significant gravitational magnification represent the key reason to survey lensing cluster fields to complement blank fields. These highest-value targets are often strongly sheared into arcs and have complex, extended morphology (see Figure \ref{fig:arcs}). They present the rare opportunity of resolving early galaxies into individual star-clusters and probing galaxy evolution on sub-kpc scales \citep[e.g.,][]{Claeyssens23,Adamo24, Fujimoto24, Mowla24}. However, spectroscopy of such sources comes with the challenge of charting the strong spatial variation of line emission across the extended arcs. It is difficult to measure the variations with slit spectroscopy, it is time consuming to build statistical samples with IFUs, and it is unwieldy to extract emission line ratios of interest (e.g., [OIII]/H$\beta$) that are blended in photometry, even with medium-bands.

In Figure \ref{fig:arcs} we show spatially resolved spectra of lensed arcs surveyed by ALT. The galaxy in the top-row of Figure \ref{fig:arcs} is an illustrative case-study (ID = 30094 at $z_{\rm{ALT}}=3.9773$; for prior work on this source see \citealt{Mahler18, Vanzella22, Bergamini23, Lin23jellybean, Katz24}). The SW image covering the rest-UV shows half a dozen clumps embedded in diffuse light, whereas the line map shows only a fraction of them are bright in H$\alpha$. Capturing this kind of spatial variation is key to the quest for metal-poor clumps (e.g., by identifying [NII]-weak and H$\alpha$-strong or [OIII]-weak and H$\beta$-strong regions). Furthermore, such spectra are essential for measuring the total line luminosity across all clumps to e.g., evaluate the contribution to cosmic reionization of such faint systems. Overall, ALT DR1 includes 72 sources with $\mu>5$ and 27 sources with $\mu>10$, with the faintest reaching $M_{\rm{UV}}\approx-15$.

Another challenge is that the most magnified arcs are often found in extremely crowded regions, amidst cluster members, their globular clusters, and embedded in the intra-cluster light. Indeed, as seen in Figure \ref{fig:mumap}, the critical lines crisscross the cluster centers. In Figure \ref{fig:rolls}, with Object \#3, we illustrate how the butterfly mosaic strategy is able to overcome the challenges of crowding. In particular, this ``peekaboo galaxy" is almost entirely hidden behind a cluster member. However, we are able to extract a spectrum by finding the [OIII] doublet first in the grism image, and then associating it with a source location that is invariant across the multiple rolls and modules.

\subsection{Unlocking the Potential of Lensing Clusters II: Multiply Imaged Systems}
\label{sec:lensing}

\begin{figure*}
\centering

\includegraphics[width=\linewidth]{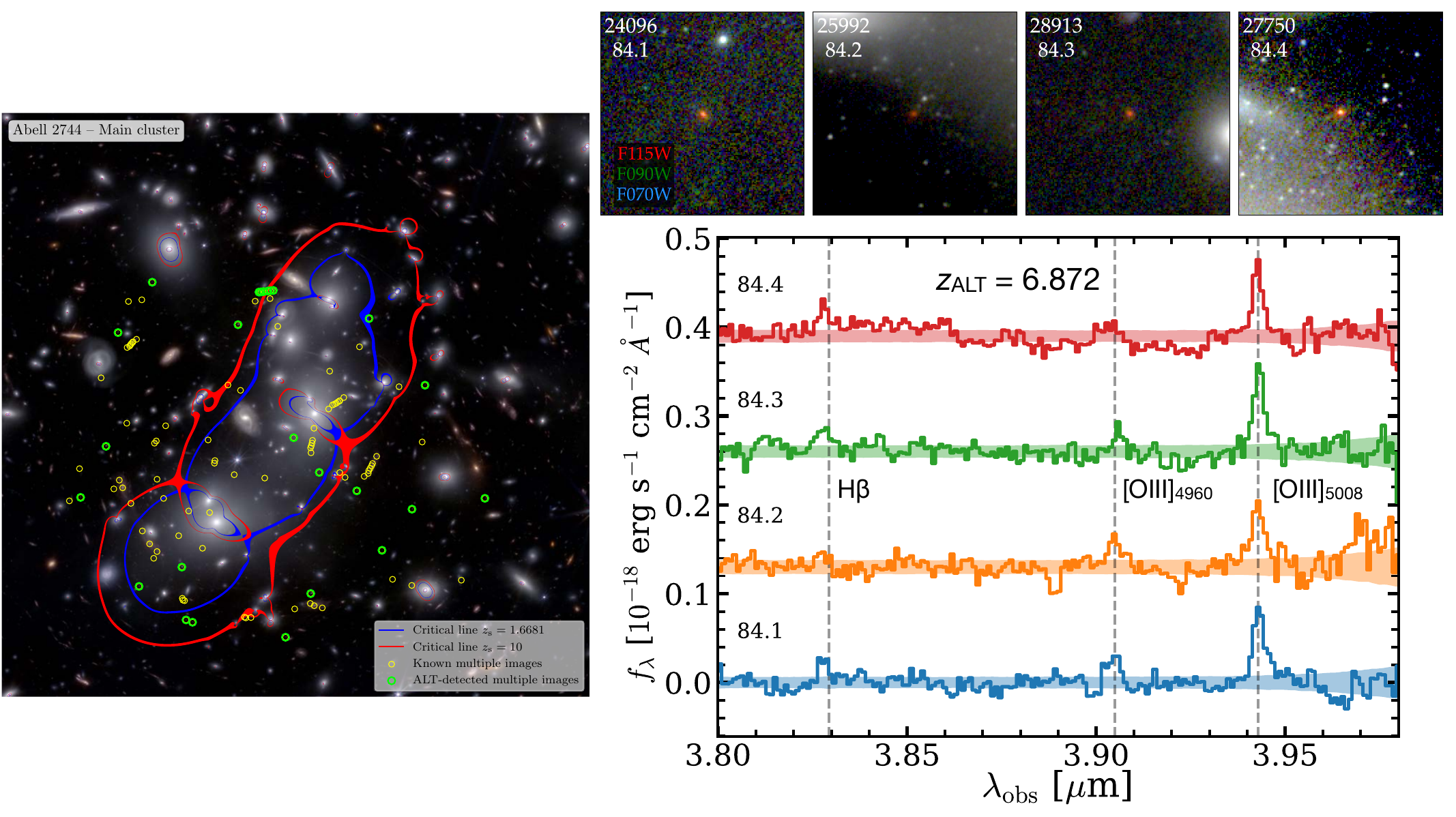}
\caption{\textbf{ALT efficiently characterizes multiply imaged systems}. Critical curves from the \citet{Furtak23,Price24} model are overlaid on the central core of the Abell 2744 cluster (left). All known spectroscopic multiple image systems are shown in yellow, with those found in ALT ($\approx40\%$ of all known systems) highlighted in green. The sources missed by ALT almost all lie at $z\lesssim3.5$ -- in redshift gaps (see Figure \ref{fig:redshifts}), or at redshifts where ground-based surveys push deeper using luminous emission lines such as [OIII] as opposed to e.g., Pa$\alpha$ in ALT. Conversely, ALT is highly complete at $z\approx3.5$-7, and detects systems such as the one highlighted in the right panel -- four images of a $z=6.873$ source, including images buried deep in the intra-cluster light (such as 84.4, top-right).}
\label{fig:multiple}
\label{fig:mumap}
\end{figure*}

A robust lensing model is critical to reconstructing the intrinsic properties of magnified sources. With well-constrained models, the scientific output from a cluster field is greatly amplified. As an example, with the Hubble Frontier Fields program \citep[e.g.,][]{Lotz17}, it was not only the ambitious investment of telescope time, but also the reliable models developed by multiple teams (compiled in \citealt[][]{Meneghetti17}) which helped UV LF studies push as faint as $M_{\rm{UV}}\gtrsim-15$ \citep[e.g.,][]{Bouwens17b, Bouwens22HFFII, Livermore17, Atek18}. 

Precise spectroscopic redshifts of multiply-imaged sources are the gold standard inputs for lens modeling (see \citealt[][]{Natarajan24} for a recent review). The high spectroscopic completeness, wide redshift coverage, and statistics of surveys like ALT are tailor-made for this task. For example, MUSE redshifts have played a crucial role in the modeling of several clusters \citep[e.g.,][]{Richard21} -- ALT adds sources in MUSE's ``redshift desert" at $z\approx2-3$ and extends coverage well into the reionization epoch, where Ly$\alpha$ that suffers from absorption in the ISM, CGM and IGM was previously the primary tracer available. 

ALT redshifts independently confirm twelve multiple image systems of the thirty-one currently known in A2744 \citep{Price24}, with three being new discoveries -- systems \#71 at $z_{\rm{ALT}}=2.582$, \#84 at $z_{\rm{ALT}}=6.873$, and \#85 at $z_{\rm{ALT}}=4.753$ with IDs as per \citet{Furtak23} and updated in \citet{Price24}. In Figure \ref{fig:multiple} we plot the locations of all the multiple image systems around the main cluster core along with the lensing model that is now fit to these systems. Here we highlight System \#84 whose redshifts were measured for the first time by ALT, including for images that were deeply buried in intra-cluster light. 

Given that A2744 is one of the most well-studied clusters with a wealth of prior redshifts, ALT's relative contribution to the census of multiple images is seemingly modest. Nevertheless, the results presented here demonstrate that an ALT-like survey would be highly efficient at characterizing clusters without JWST coverage -- e.g., in a single shot it would return lensed arcs as well as multiple images without requiring any JWST pre-imaging or pre-existing photometry to select targets from.

\subsection{Environments on Large Scales: The $z=0$-7 Galaxy Distribution Behind A2744} 
\label{sec:envlarge}

\begin{figure*}[htb]
    \centering
    \includegraphics[width=0.99\linewidth]{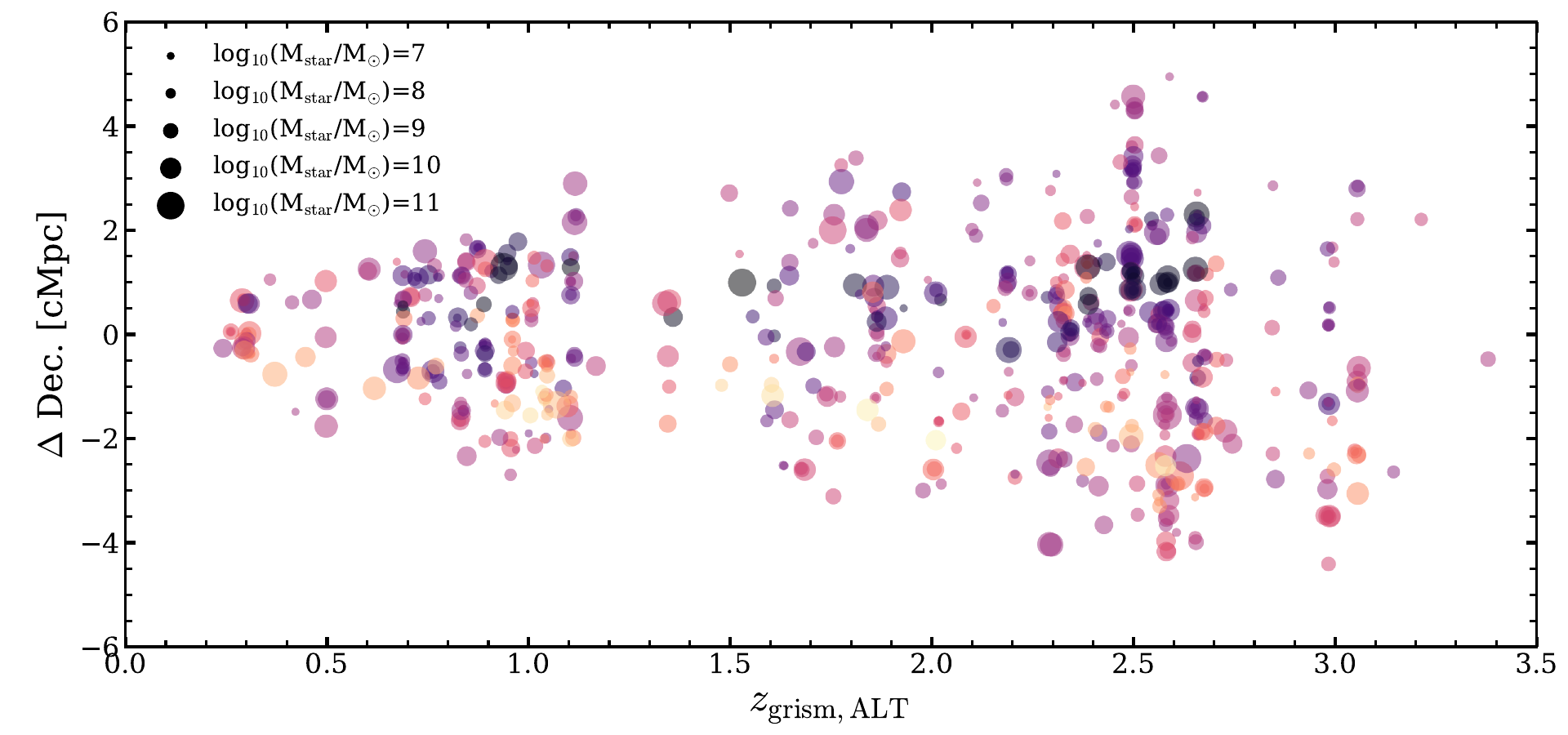} \\
    \includegraphics[width=0.99\linewidth]{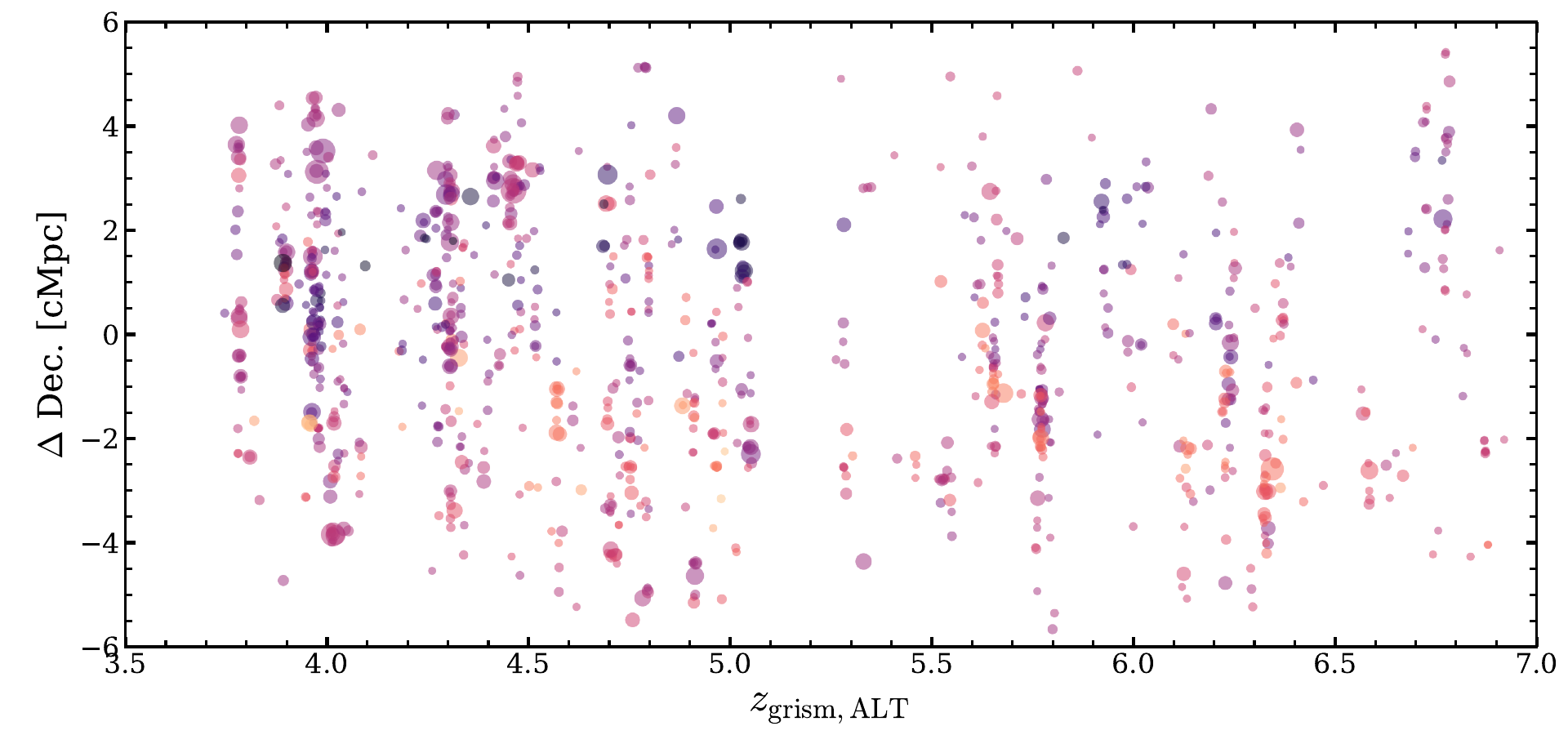} \\
    \caption{Large scale galaxy distribution behind A2744 as measured by ALT. The projected source plane distribution of the galaxies is illustrated on the y-axis (Dec.) and through the color scaling (RA). The sizes of the data-points correlate with stellar mass. Rich structure is readily apparent, with cold vertical spikes that span the plot identifying large-scale overdensities whereas clustered points identify more local overdensities. The largest points representing the most massive galaxies tend to fall in these regions.}
    \label{fig:overdensities}
\end{figure*}

\begin{figure*}
    \centering
    \includegraphics[width=0.80\linewidth]{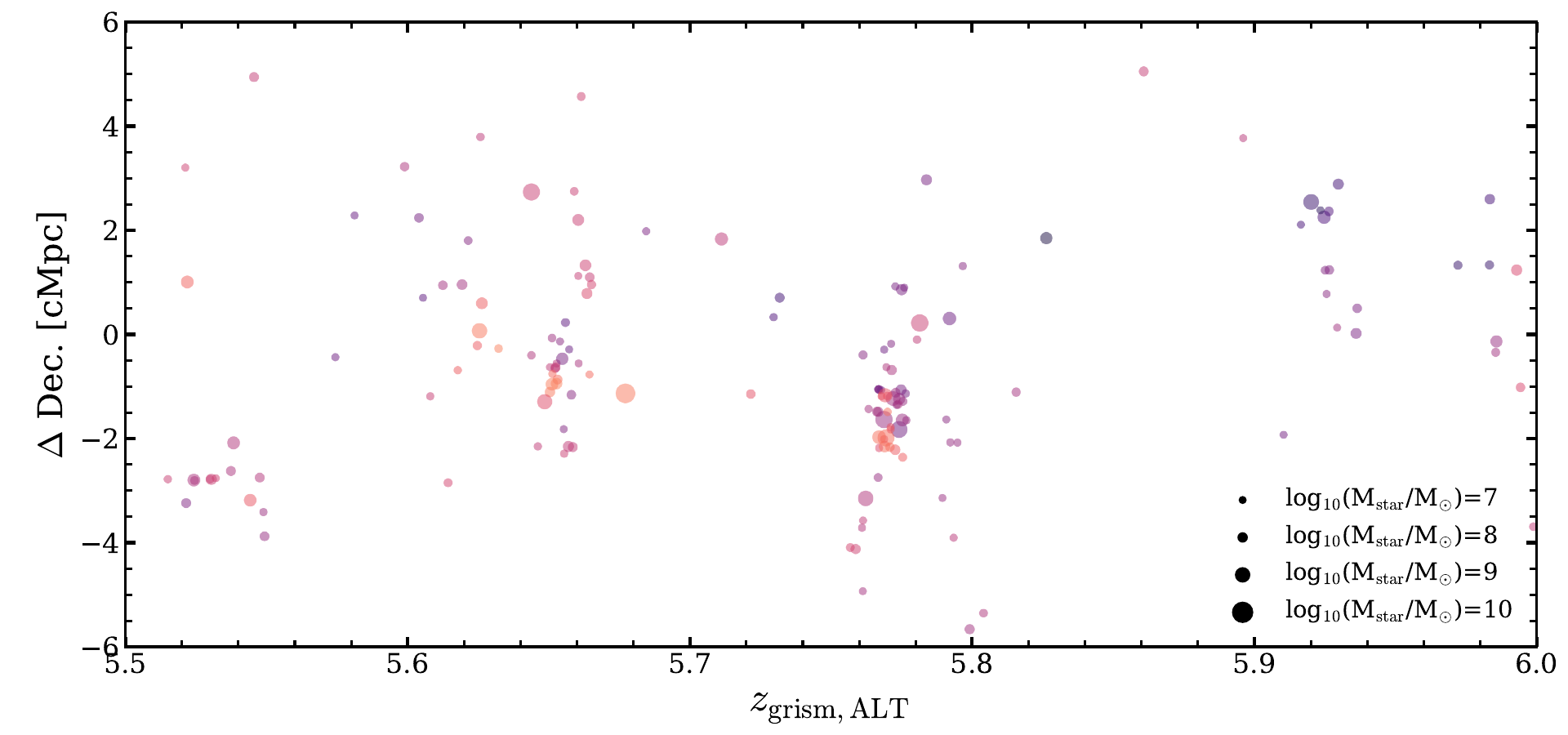} \\
    \begin{tabular}{cc}
    \includegraphics[width=0.49\linewidth]{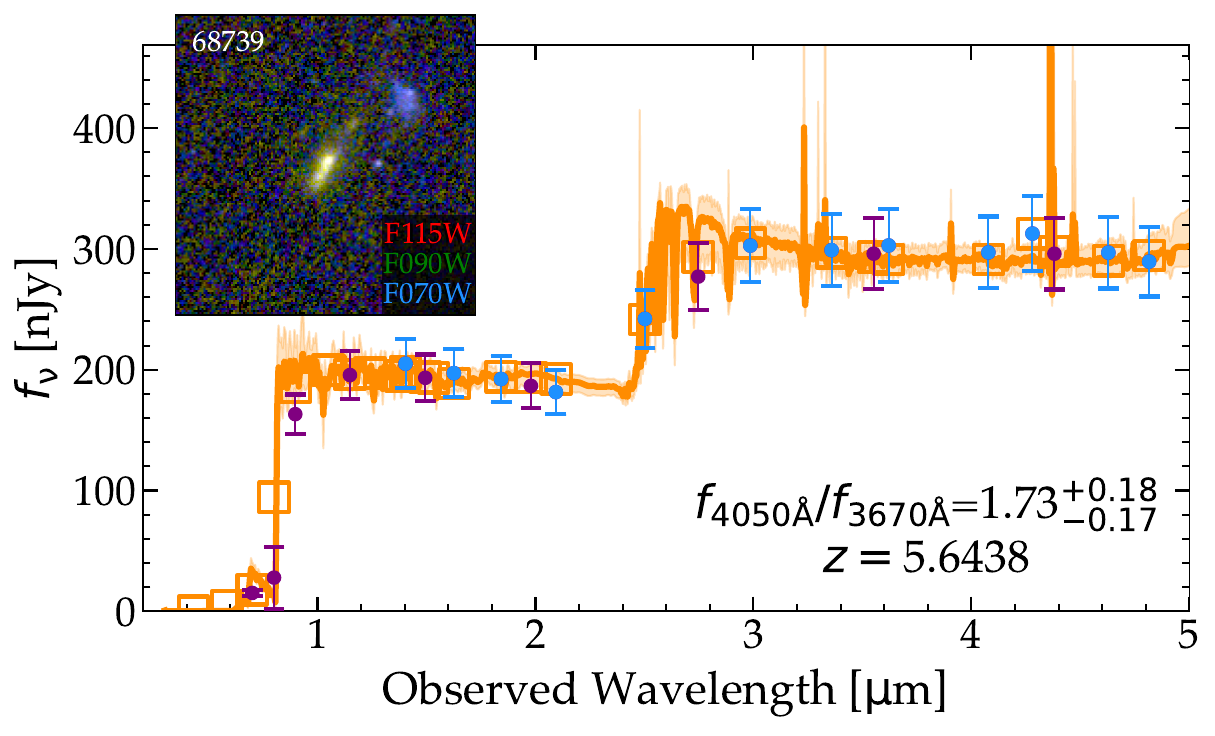} & \includegraphics[width=0.49\linewidth]{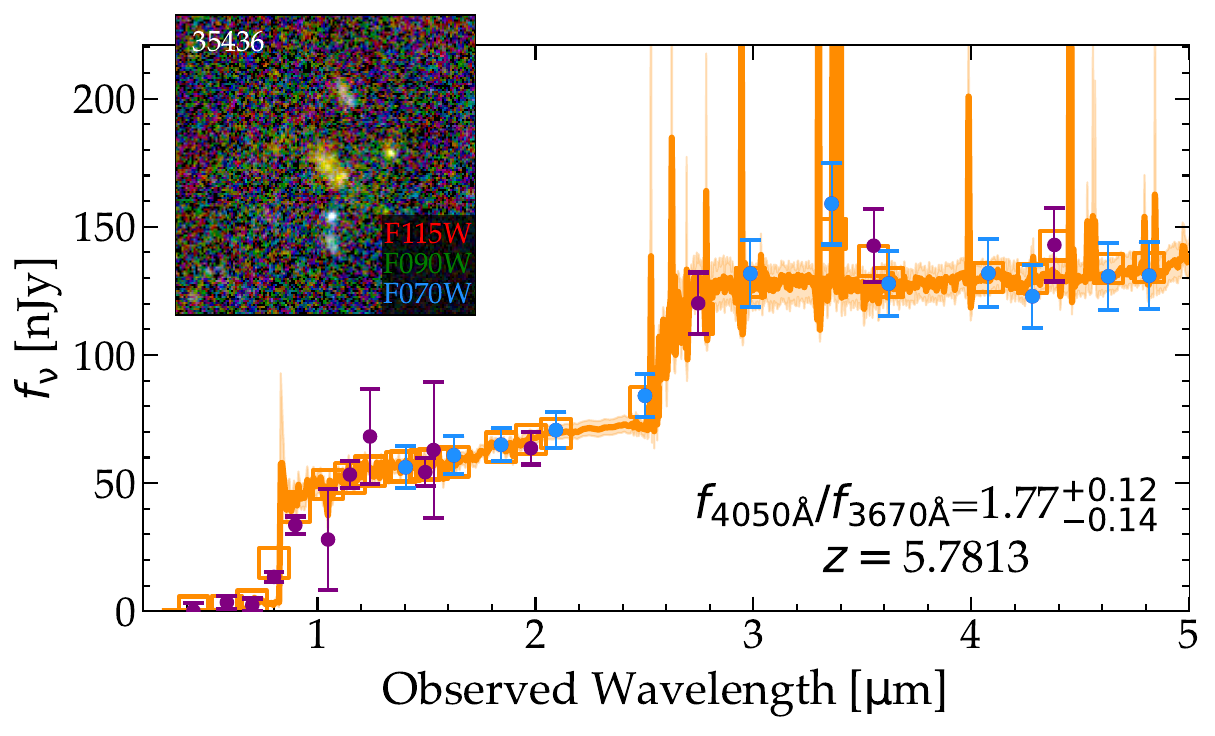} \\  \includegraphics[width=0.49\linewidth]{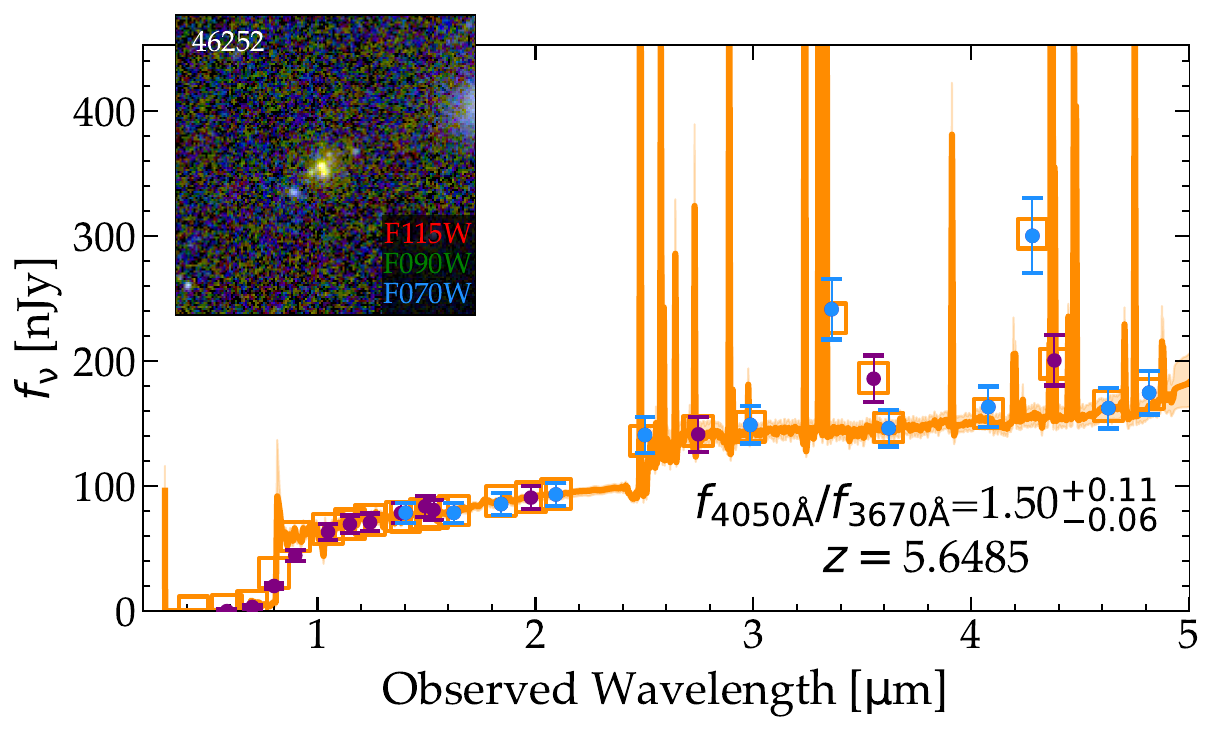} & \includegraphics[width=0.49\linewidth]{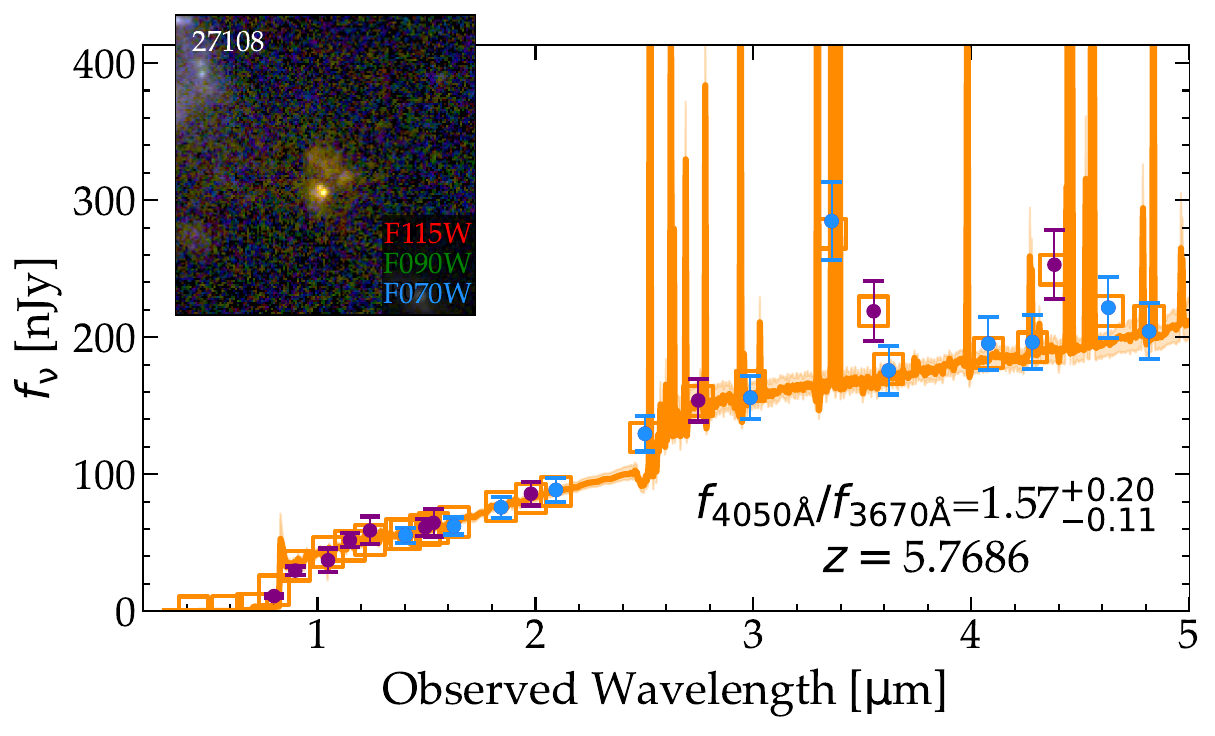} \\           
     \end{tabular}
    \caption{{\bf ALT demonstrates the connection between environment and galaxy properties.} In $\Lambda$CDM, the most overdense regions are predicted to collapse first, and are hence expected to host ancient galaxies. Indeed, two of the most over-dense regions in ALT at $z=5.66$ and $z=5.77$ (top panel, zoomed-in version of Figure \ref{fig:overdensities}) contain several galaxies with strong Balmer breaks, which are a signature of older stellar populations. The Balmer break strength is indicated in each panel ($f_{\rm{4050{\AA}}}/f_{\rm{3670{\AA}}}$) and denotes the ratio of $f_{\rm{\nu}}$ in two feature-less windows on either side of the break. The ALT sample as a whole has $f_{\rm{4050\AA}}/f_{\rm{3670\AA}}\approx1$, in agreement with typical values found at these redshifts \citep[e.g.,][]{Roberts-Borsani24}.}
    \label{fig:bbreaks}
\end{figure*}

\begin{deluxetable*}{cclc}[htb]
\tabletypesize{\footnotesize}
\tablecaption{The Largest Overdensities Found in ALT}
\tablehead{
\colhead{Redshift} & \colhead{Members} & \colhead{Notable Members}
}
\startdata
\vspace{-0.3cm}\\
2.496 & 51 & 26222, LRD with HeI absorption (\S\ref{sec:lineprofiles}, Figure \ref{fig:lineprofiles2})\\
 &  & 72835, elliptical galaxy with \mstar$=10.9$\\\\
 \hline\\
2.582 & 36 & 26974, broad-absorption line AGN (\S\ref{sec:lineprofiles}, Figure \ref{fig:lineprofiles2})\\
 & & 19866, most massive galaxy in ALT with \mstar$=11.2$\\\\
 \hline\\
3.970 & 84 & 65118, most massive H$\alpha$ emitter with \mstar$=10.4$\\
 & & 48598, quenched galaxy from \citet[][]{Setton24}\\\\
  \hline\\
4.304 & 59 & 39921, \mstar$=9.3$, H$\alpha$+[NII] and ALMA 1.2mm continuum detection\\
 & & 71962, massive galaxy with \mstar$=9.6$\\\\
 \hline\\
5.655 & 33 & 46252, 37715, 68739\\
& & massive galaxies with \mstar$>9$ and strong Balmer breaks\\\\
 \hline\\
5.769 & 50 & 35436, 27108, 26124, 30041, 18849\\  
& & massive galaxies with \mstar$>9$ and strong Balmer breaks\\\\
% & & 27108, 26124, 30041, 18849 with \mstar$>9$ and strong Balmer breaks\\
 \hline\\
6.328 & 29 & 26902, LRD with log($M_{\rm{BH}}/M_{\rm{\star}})=8.1$, ID = 13821 in \citet{Greene24}
\label{table:overdensity}
\enddata
\end{deluxetable*}

In Figure $\ref{fig:overdensities}$, we show the full distribution of galaxies in our DR1 dataset. Besides the redshift, we also illustrate the projected spatial distribution of the sources (in comoving coordinates) and their inferred stellar mass. Objects are shown at the locations in the source plane as predicted from our lensing model. Our data map the galaxy distribution over $z\approx0$-7, with small gaps (e.g., around $z\sim1.3$, $z\sim3.3$, $z\sim5.1$) where no strong emission lines fall in F356W. At $z<1$ the covered comoving volume is relatively small and no galaxy is detected in front of the cluster itself ($z\approx0.3$). The typical stellar mass of our sample at $z<4$ is higher (\mstar = 8.8) than at $z>4$ (\mstar \,= 7.7) owing to the weaker emission lines accessible to ALT at lower redshifts. Large-scale clustering over $\approx10$ cMpc scales is clearly visible in the form of sharp vertical structures at fixed redshift owing to the pencil-beam nature of our survey. More localized overdensities that are not part of these structures are also apparent (e.g., at $z=4.46$ and $z=5.04$).

To approximately identify the most prominent overdensities, we project our dataset into comoving coordinates and apply the DBSCAN algorithm \citep[][]{Ester96}. The seven overdensities found with at least 20 members are listed in Table \ref{table:overdensity} and correspond to structures apparent ``by eye" in Figure $\ref{fig:overdensities}$. DBSCAN is an unsupervised clustering technique that excels at finding overdense regions of arbitrary shape such as the filaments seen in protoclusters \citep[e.g.,][]{Herard-Demanche23, Sun24}. We set the ``epsilon" parameter that defines the size of the ``neighborhood" around each galaxy to 5 cMpc and require at least 5 galaxies to define a dense region. To control for inhomogeneity, we split the data based on the emission line used to identify the sources and impose a stellar mass limit equal to the 20$^{\rm{th}}$ percentile of each split. We stress that this is a very rough treatment to pinpoint the most striking structures to enable further investigations, and we do not account for e.g., completeness corrections and lensing distortions.

These overdensities host some of the most spectacular objects in the ALT sample (see Table \ref{table:overdensity}). Furthermore, they provide simple demonstrations of the dark matter paradigm at high redshift. For example, we observe in Figure \ref{fig:overdensities} that the most massive galaxies in our sample are typically found in the most overdense regions, exactly as expected in hierarchical structure formation \citep[e.g.,][]{Wechsler18}. In a companion paper (Matthee et al. in prep), we show that galaxy stellar mass indeed strongly correlates with the degree of overdensity across the whole \mstar=7-10 range at $z=4$-5.

A fundamental prediction of structure formation is that clustered regions generally collapse at relatively earlier epochs making them likelier to host ancient galaxies (see e.g., discussions of ``assembly bias" \citealt{Desjacques18}). Therefore, high-redshift overdensities are precious sites for galactic archaeology as they host some of the first galaxies that formed in the universe. Indeed, using the $z=5.655$ and $z=5.769$ structures as a case study, in Figure $\ref{fig:bbreaks}$, we demonstrate that a preponderance of galaxies in these two overdensities display strong Balmer breaks that are a hallmark of old stellar populations (see also \citealt{Morishita24, Witten24}). Detailed study of the old stellar populations in these Balmer break sources is a powerful, complementary way to approach the questions surrounding the surprisingly early and vigorous onset of cosmic dawn that JWST is revealing \citep[e.g.,][]{Oesch16, Bunker23, Naidu22, Castellano22, Castellano24, Robertson24, Carniani24}.

\subsection{Environments on Small Scales: A Case Study of a ``Local Group" at $z=5.7$ and High-$z$ Tests of $\Lambda$CDM}
\label{sec:envsmall}

\begin{figure*}
\centering
\includegraphics[width=0.8\linewidth]{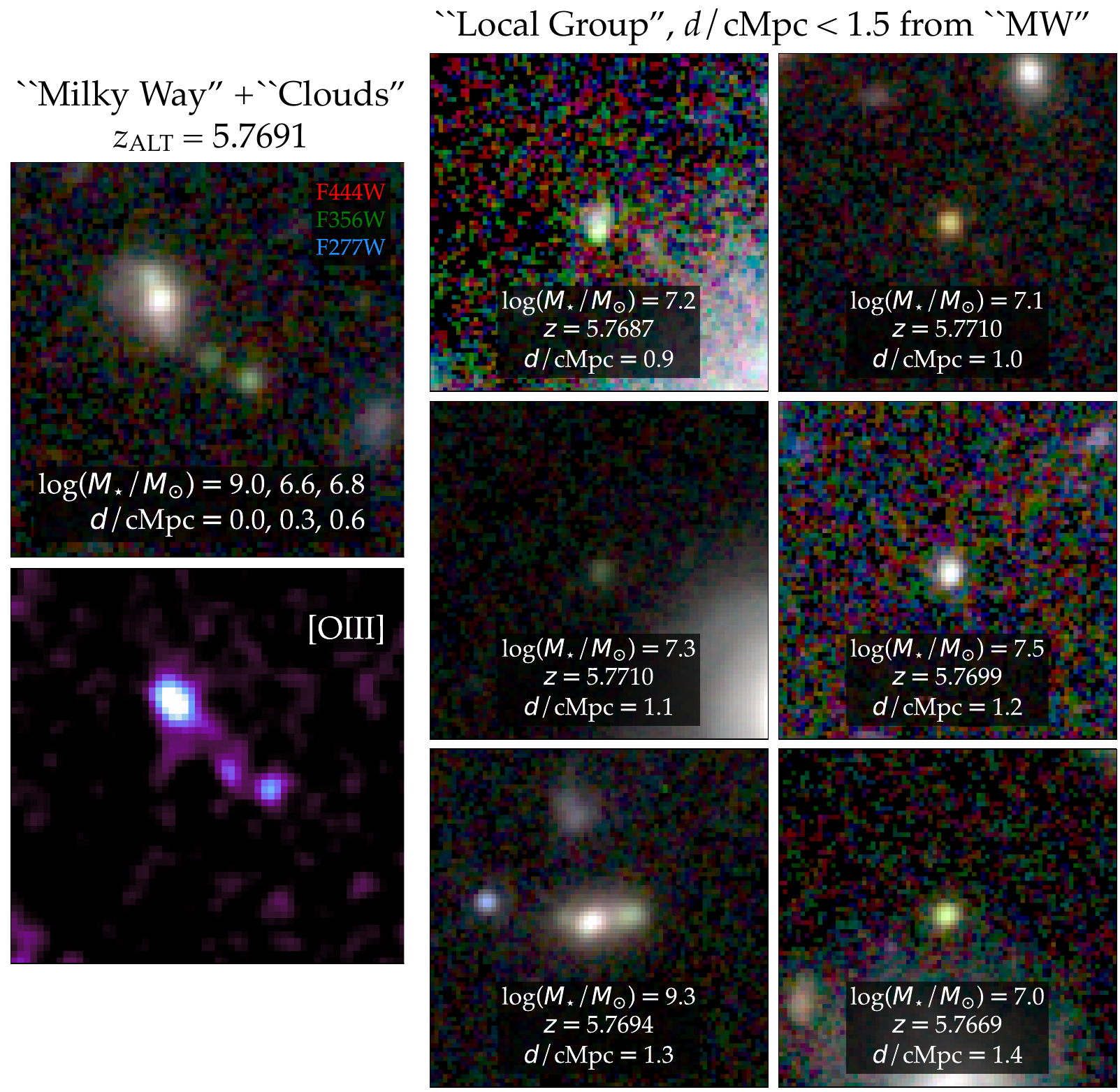}
\caption{{\bf Approximate analogs of the Milky Way, Magellanic Clouds, and the Local Group at $z=5.77$ identified owing to the depth, completeness, and redshift precision of ALT.} Left: 3\farcs0$\times$3\farcs0 RGB image (top) and [OIII] line map (bottom) of a galaxy whose stellar mass is comparable to that of the Milky Way at $z\approx6$. The two neighboring galaxies, reminiscent of the Magellanic Clouds, are only a few 100 cKpc ($<100$ pKpc) away as borne out by the appearance of all three sources in the [OIII] line map. Right: All sources within 1.5 cMpc (approximately the scale of the Local Group) of the Milky Way-like galaxy ordered by separation. These objects include several dwarf galaxies as well as an Andromeda analog with similar mass as the Milky Way analog (bottom-left). All these galaxies are only moderately magnified ($\mu=1.7-2.3$), which is the typical magnification found across the ALT footprint. Such an exhaustive census of satellites, the likes of which have posed challenges to $\Lambda$CDM in the local Universe, may now be extended to higher redshifts with ALT-like surveys.}
\label{fig:MW}
\end{figure*}

The detailed properties of the Milky Way's satellite galaxies have posed a succession of challenges to the $\Lambda$CDM paradigm (e.g., the ``missing satellites" problem; \citealt{Bullock17,Kim18, Perivolaropoulos22}). While these challenges are being addressed to various degrees, it is clear that the dwarf sector remains a fertile ground to stress-test and refine the current paradigm \citep[e.g.,][]{Drlica-Wagner22}. A key advance in recent years has been the extension of the census of satellites around the Milky Way to other galaxies of comparable mass out to $\approx40 $ Mpc \citep[e.g.,][]{Geha17, Carlsten22, Mao24}. These surveys have provided critical statistical context to address the Milky Way's uniqueness vis-a-vis the various proposed challenges for $\Lambda$CDM.

With ALT, we now have the opportunity to extend such surveys of Milky Way-like galaxies to higher redshift. In particular, with increasingly precise knowledge of the Milky Way's assembly timeline emerging from surveys exploiting \textit{Gaia} \citep{GaiaDR3}, we are now able to define ``Milky Way-like" not only at $z=0$, but also out to $z\approx6$ \citep[e.g.,][]{Bonaca20, Naidu21, Conroy22, Belokurov24, Chandra24}. What new challenges might censuses of satellites at high-redshift reveal? Already at these epochs, the remarkable efficiency of structure formation as seen in the early emergence of massive \citep[e.g.,][]{Xiao23, Akins23, Baggen23, Baggen24}, quenched \citep[e.g.,][]{Carnall24,degraaff24,Weibel24QG}, and bright galaxies \citep[e.g.,][]{Conselice24,Akins24, Casey24, Harikane24} is at odds with standard models of galaxy formation and perhaps even $\Lambda$CDM \citep[e.g.,][]{Shen24}.

Here we demonstrate that a detailed census of satellites around Milky Way-like galaxies at high redshift is viable with ALT-like surveys. In Figure \ref{fig:MW}, we present a galaxy (ID: 45621, $z_{\rm{ALT}}=5.7691$) with stellar mass comparable to that inferred for the Milky Way at $z\approx5$-6 (\mstar$\,\approx7$-9; Chandra et al. in prep., \citealt{deGraaff24MW, Buch24,Belokurov24, Semenov24}). Remarkably, the environment of this galaxy has a number of features reminiscent of the present-day Local Group -- e.g., a pair of satellites akin to the Magellanic Clouds hovering just outside the virial radius, a slightly more massive Andromeda-like companion $\approx1$ cMpc away, and several other dwarf galaxies comparable to Fornax within 1.5 cMpc. Of course, this comparison must not be taken too literally, and the fate of this ``Local Group" is likely to have diverged from ours. For example, the Clouds in the Milky Way likely entered its virial radius only in the last 1 Gyr, and were at $\approx1$-3 cMpc at $z\approx6$ \citep[e.g.,][]{Besla07,Conroy21, Chiti24, Chandra24reflex}. It is also possible that all these galaxies pictured in Figure \ref{fig:MW} merge rapidly with the ``Milky Way". Nonetheless, it is remarkable that such systems of satellites may be studied in detail, at such high redshift, owing to the depth and statistics of the survey. 

Beyond this one example, in a more statistical sense, we argue in a companion paper via clustering metrics that about 20\% of the galaxies in our sample at $z=4$-5 (where our survey is most sensitive) are likely satellites of other galaxies in our sample (Matthee et al., in prep.). Overall, it is clear that ALT-like surveys are well-suited to deliver galleries of high-redshift ``Local Group"s such as what is shown in Figure \ref{fig:MW} that are ready for detailed analysis akin to, e.g., the SAGA survey's gallery of 100 $z\approx0$ Milky Way-like galaxies (matched on stellar mass) and their diverse satellite systems (\citealt{Mao24}, see their Fig. 7).

\subsection{Environments on All Scales: Grism Redshifts Uniquely Enable Clustering}
\label{sec:whygrism}
\begin{figure*}
    \centering
\includegraphics[width=16cm]{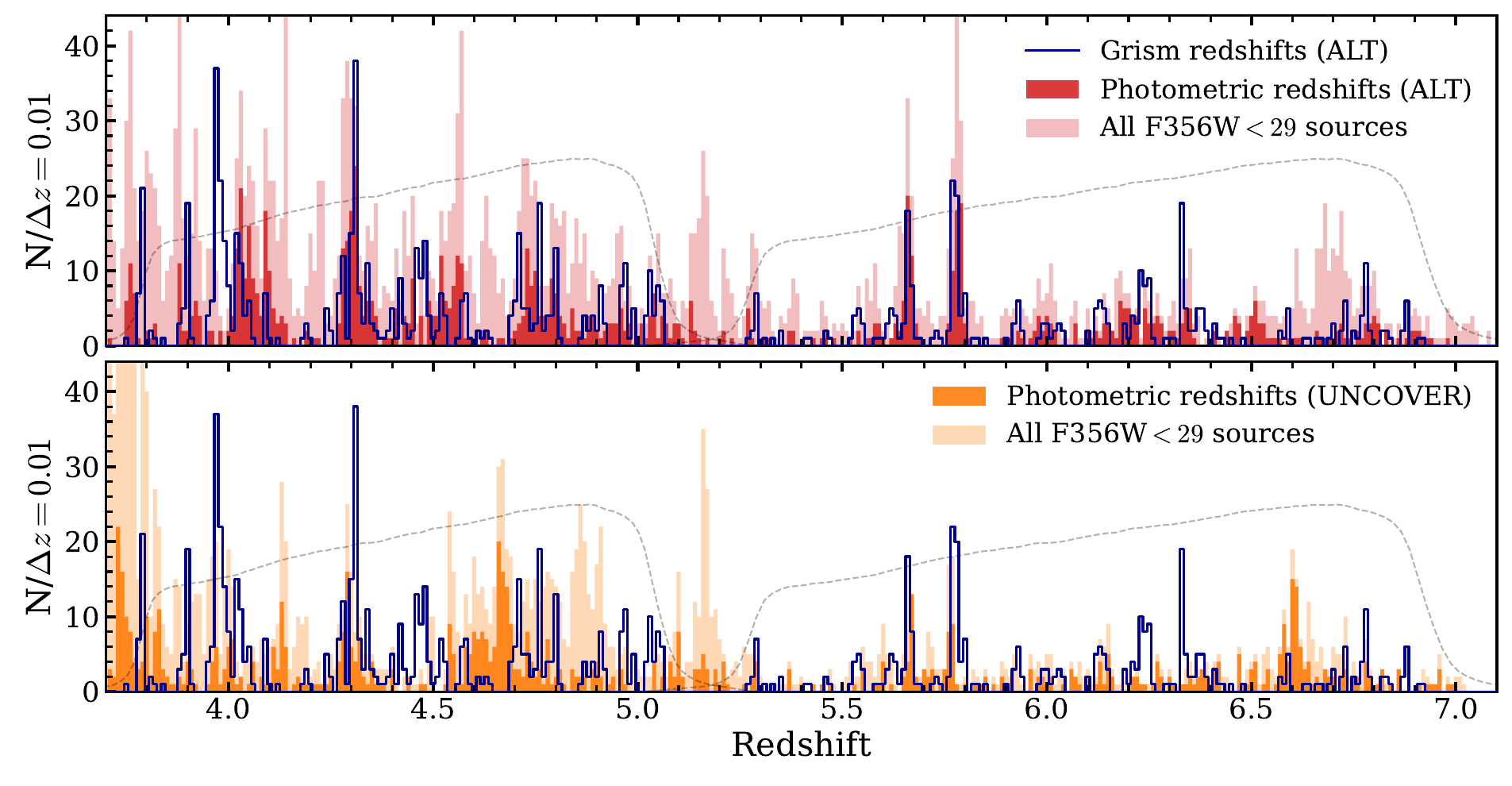}
    \caption{{\bf Detailed comparison of grism redshifts (blue) and photometric redshifts (red, orange).} Darker colors represent photo-$z$s of objects in ALT DR1, while lighter colors show all objects in the field of view with F356W$<29$ mag. Grey dashed lines show the F356W filter curve to highlight the gap in our data at $z=5.0$-5.2 between H$\alpha$ and [OIII] coverage. While the photo-$z$s agree well generally, ``phantom overdensities" (e.g., $z=4.5$, $z=6.6$) that show up in the photo-$z$s are a key failure mode. Further, extremely prominent structures such as the spikes at $z=4.0$ are missed both with the UNCOVER imaging (bottom) as well as when the full NIRCam filter-set is included (top).}
    \label{fig:redshift_phantoms}
\end{figure*} 

\begin{figure}
    \centering
\includegraphics[width=8cm]{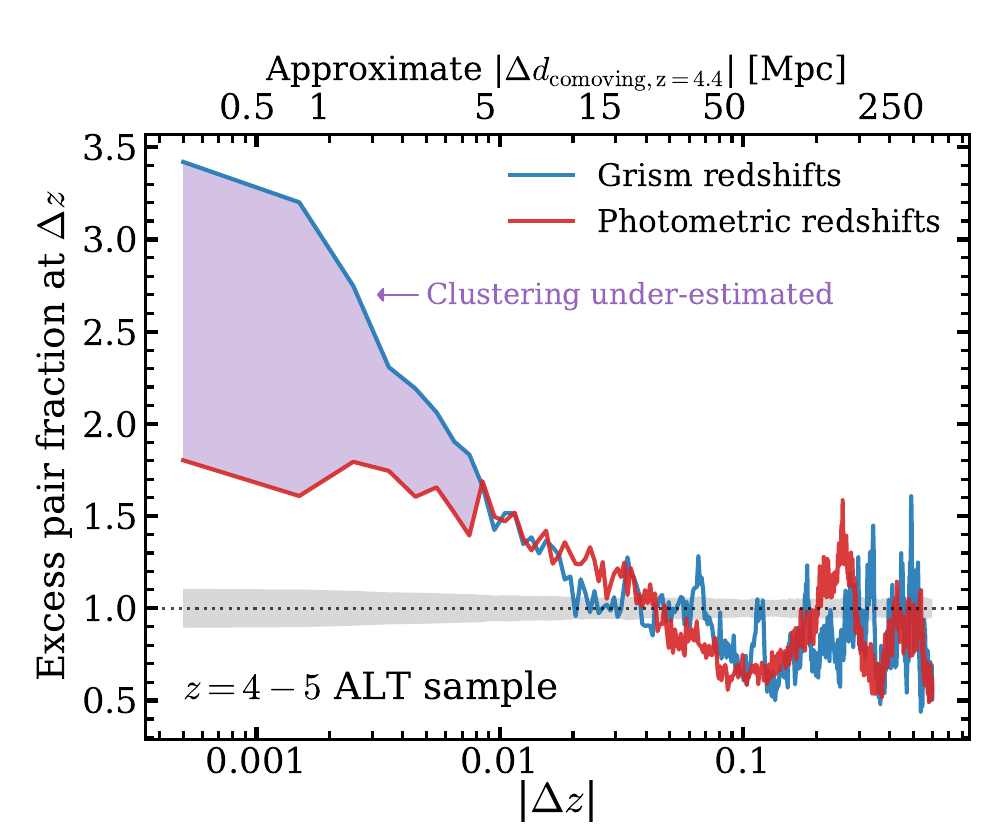}
    \caption{{\bf The excess pair fraction as a function of the redshift difference of galaxy pairs.} While the excess pair fractions derived from photo-$z$s broadly follow those based on grism redshifts on large scales, indicating galaxy clustering, the clustering signal is severely underestimated on 5 Mpc scales. This demonstrates that even with the most optimal possible filter coverage, photo-$z$s are too coarse to probe environments required to, e.g., situate sources in dark matter halos of appropriate mass or to find Milky Way-like local groups.}
    \label{fig:redshift_clustering}
\end{figure}

One could question: why do we even need grism redshifts when the photometric redshifts are as accurate as we find ($\langle |\Delta z|/(1+z) \rangle = 0.005, \sigma_{\rm NMAD}=0.008$, with only 3$\%$ catastrophic outliers; Figure $\ref{fig:zphots}$)? The answer to this question depends on the required precision of the redshifts. 

Figure $\ref{fig:redshift_phantoms}$ zooms in on the H$\alpha$ and [OIII] redshift distributions. The photo-$z$s derived with the full NIRCam filter-set are clearly better than the Cycle 1 UNCOVER photometry at finding redshift spikes, such as the major ones at $z=4.3$ and $z=5.6,5.8$. However, even these updated photo-$z$s still miss major structures such as the one at $z=4.0$. Another striking feature in this figure is the appearance of ``phantom" photo-$z$ spikes, which imply the presence of massive overdensities, but that are completely absent from the grism sample. There are several such phantom spikes, e.g., at $z\approx4.6$ and $z\approx6.6$. These are spurious solutions that coincide with photometric redshift solutions where no strong-line falls in an MB. There are perhaps also subtle issues with the theoretical filter curves, such that the uncertainties around the edges of the curves translate to spikes at particular redshifts. Even the older broad-band photo-$z$s from UNCOVER miss several prominent spikes, and also have similarly prominent phantom overdensities.

In the context of measuring galaxy clustering, in Figure $\ref{fig:redshift_clustering}$ we compare the fraction of galaxy-pairs as a function of the redshift difference using either grism or photometric redshifts (in excess to the random expectation assuming a flat redshift distribution). Here we use the subset of ALT sources with redshifts between $z=4$-5. A clustering signal is detected below separations of $\Delta z < 0.04$ using both grism and photometric redshifts. Note that at large scales (i.e., $\Delta z >0.1$), the results in Figure $\ref{fig:redshift_clustering}$ are cosmic variance dominated as we are mostly counting the separations to the most prominent redshift spikes in our data, and the numbers are lower than the random expectation due to the integral constraint (as we normalised the number of random pairs to the total number of pairs). The photometric redshifts roughly match the excess pair fraction measurements from the grism redshifts around $\Delta z=0.01-0.05$, but they strongly under-estimate the clustering on scales below $\Delta z < 0.005$, which correspond to line of sight separations of about 5 cMpc ignoring peculiar velocities. This implies that any clustering analysis below 5 cMpc / $\Delta z < 0.01$ requires grism redshifts, whereas clustering on somewhat larger scales could be performed with photometric redshifts in the ALT data, although the caveat of phantom redshift spikes discussed above is challenging to quantify.

\subsection{Revealing Obscured Star-Formation across Cosmic Time}
\label{sec:paschen}

\begin{figure*}
\centering
\includegraphics[width=0.8\linewidth]{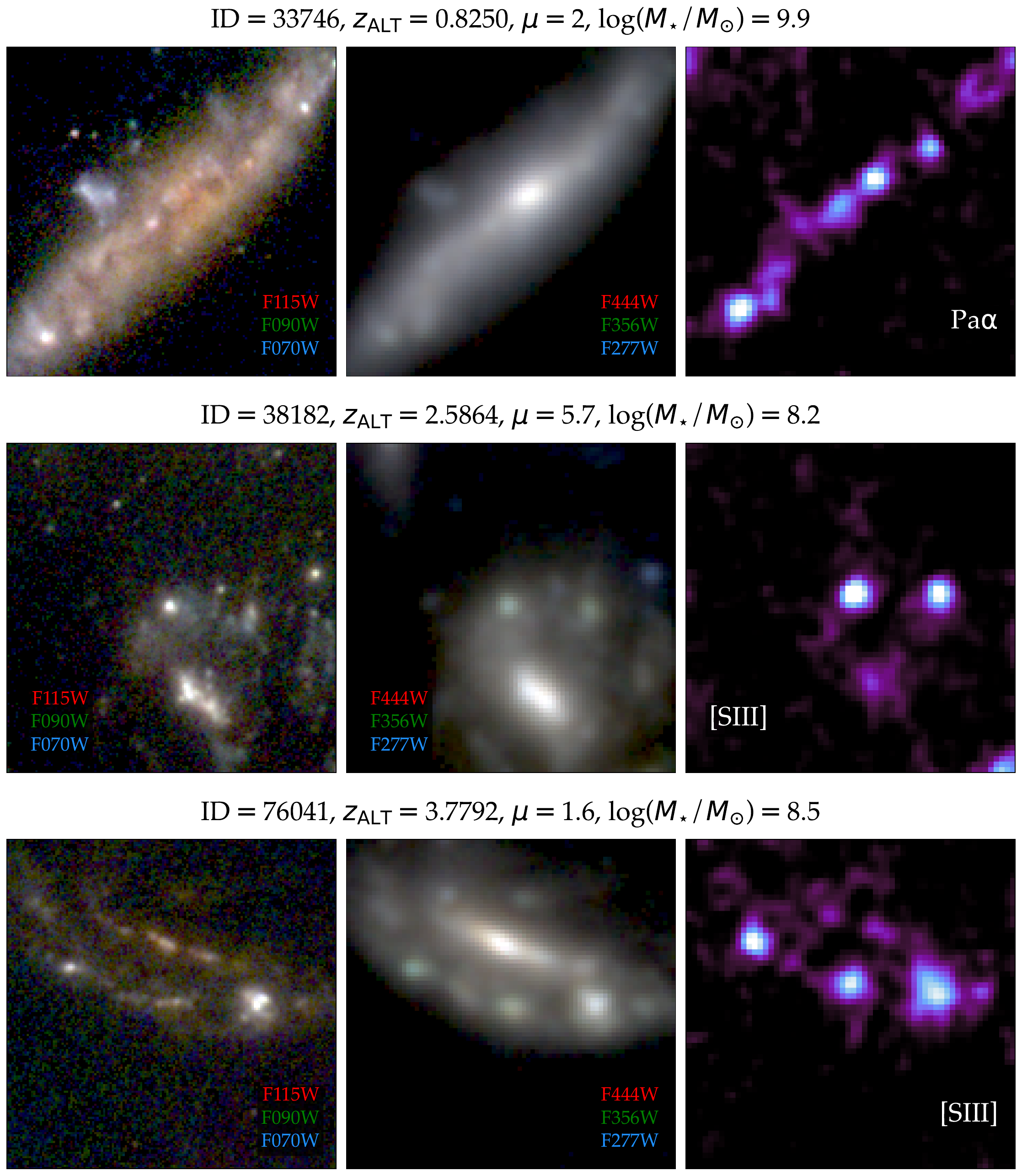}
\caption{\textbf{ALT provides a resolved unobscured view of star-formation at cosmic noon.} With ALT's depth, we are able to measure weak rest-infrared emission lines such as Pa$\alpha$ and [SIII] that are effectively dust insensitive compared to classic rest-optical tracers such as H$\alpha$. Further, we are able to produce high-SNR line maps (right column) that paint a striking picture of clumpy clustered complexes dominating the overall star-formation in these galaxies. The dramatic contrast between the line maps and NIRCam images (left columns) demonstrates the power of these data -- they reveal how relatively nondescript, faint clumps sitting behind veils of dust in the RGB images dominate the line maps and therefore the star-formation budget in these galaxies.}
\label{fig:paschen}
\end{figure*}

\begin{figure*}
    \centering
    \begin{tabular}{cc}
     \includegraphics[width=8.2cm]{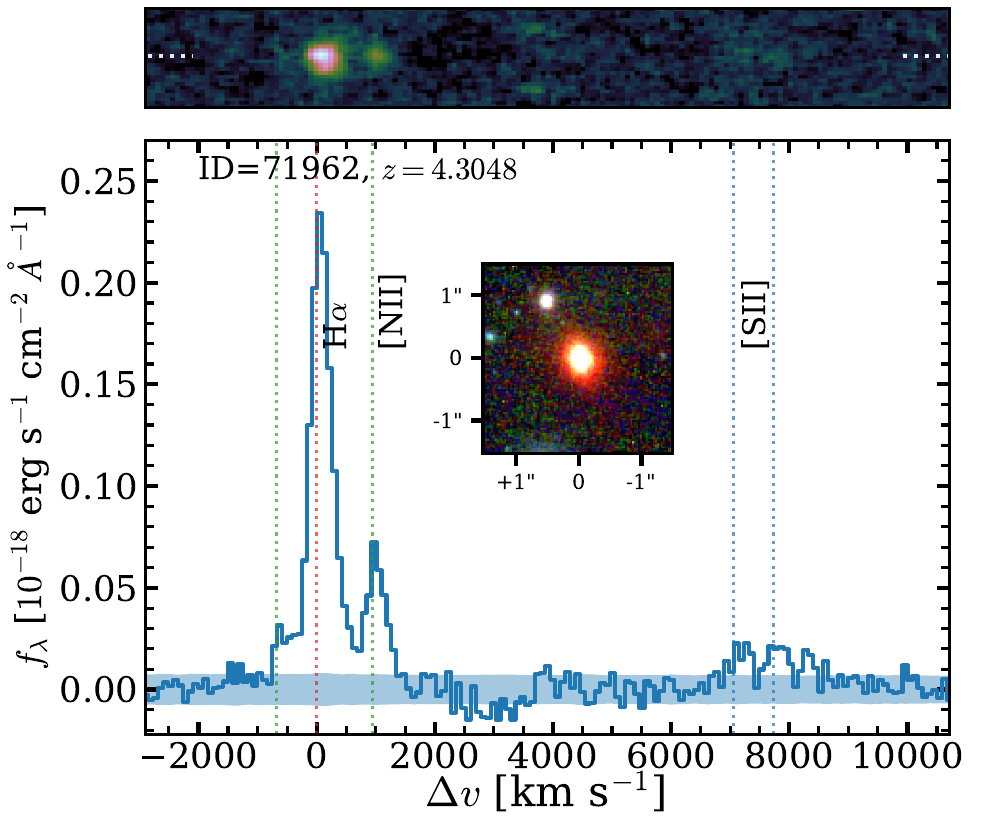} &
    \includegraphics[width=8.2cm]{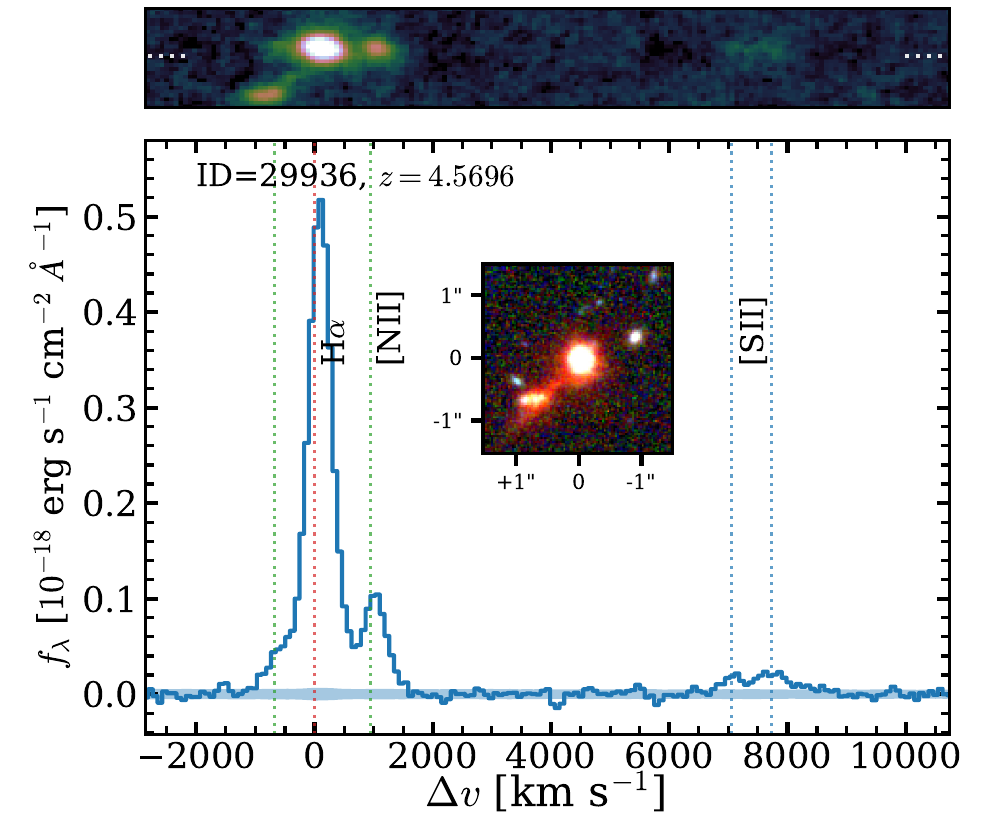} \\
     \end{tabular}
    \caption{{\bf The H$\alpha$, [NII] and [SII] spectra of two luminous, massive galaxies at redshifts $z\approx4-5$.} These galaxies have masses log($M_{\rm{\star}}/M_{\rm{\odot}}$)$\approx9.3$ and $\approx10$ respectively. These spectra illustrate that H$\alpha$ and [NII] are well-resolved by the NIRCam grism. The 2D spectrum of ID 29936 further shows the diffuse H$\alpha$ emission bridging two clumps of the galaxy.}
    \label{fig:lineprofiles}
\end{figure*}

\begin{figure*}
    \centering
    \begin{tabular}{cc}
    \includegraphics[width=8.2cm]{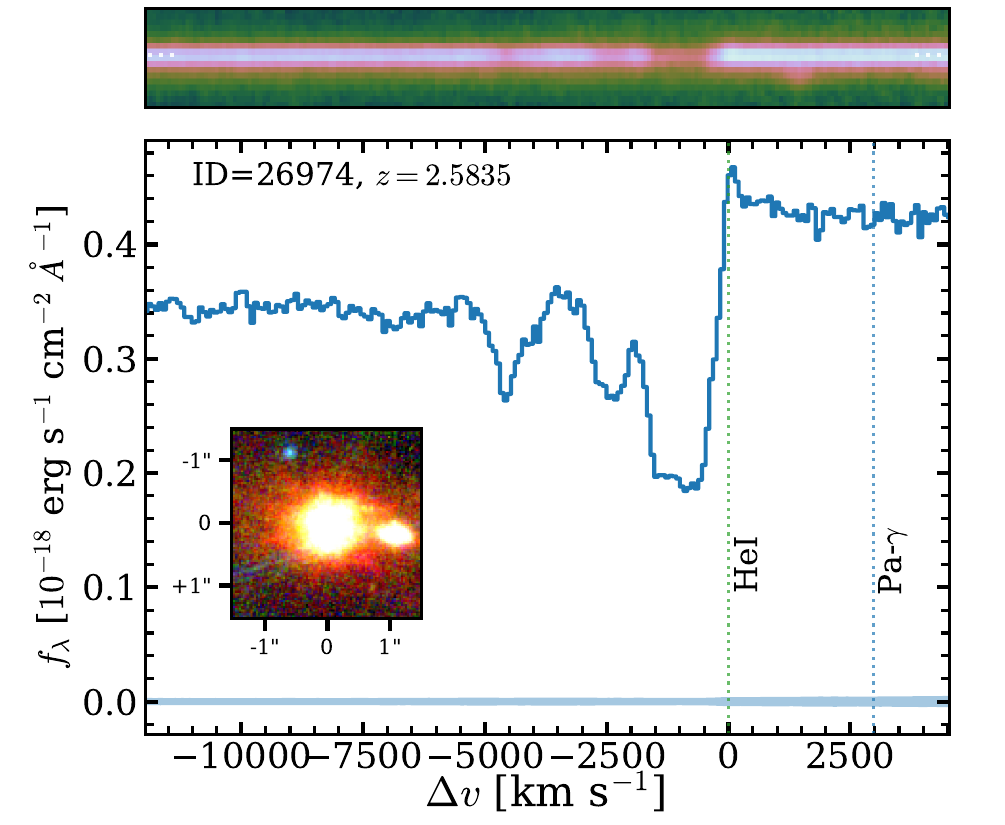} & 
     \includegraphics[width=8.2cm]{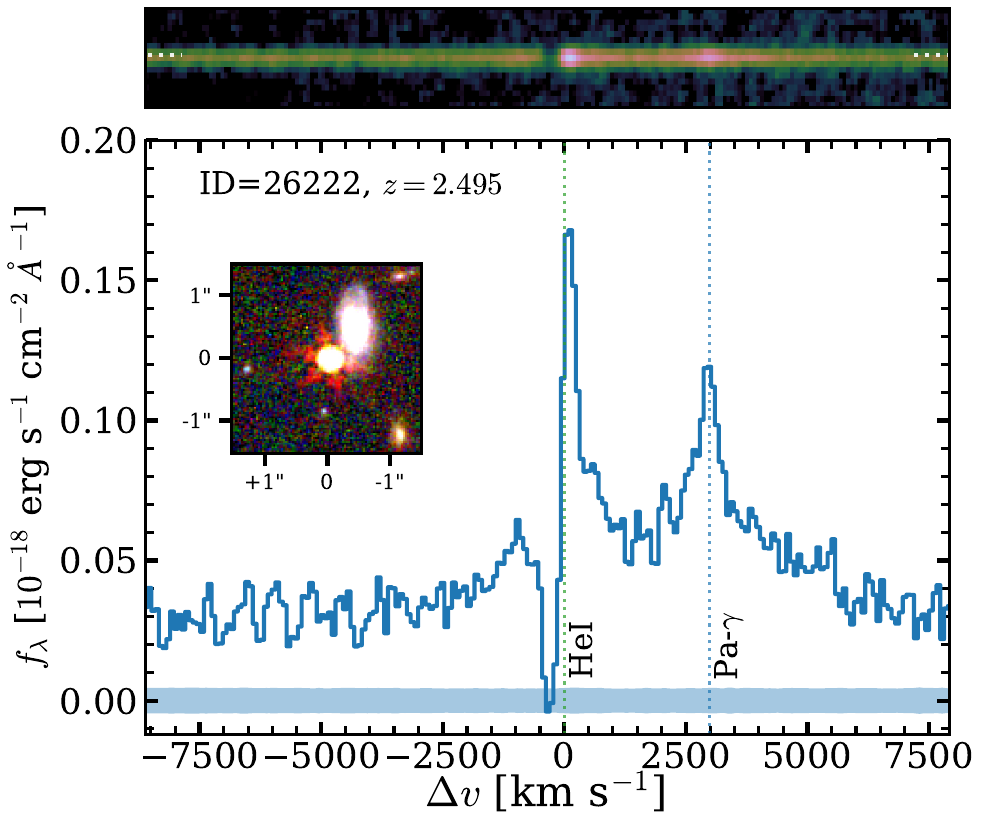} \\  
     \end{tabular}
    \caption{{\bf The HeI + Pa$\gamma$ spectra of two galaxies with AGN emission at $z\approx2.5$.} Unlike in the rest of the paper, these spectra are the original grism spectra without removing the continuum (fortunately available without contamination in subsets of the data). The x-ray detected ID 26974 shows various broad, blue-shifted HeI absorption systems indicative of ionized outflows. HeI absorption is also detected in the compact AGN ID 26222, but this is very narrow and close to the systemic redshift. ID 26222 further shows broad Pa$\gamma$ emission and resembles the $z>4$ Little Red Dots in various ways.  }
    \label{fig:lineprofiles2}
\end{figure*}

Some of the key unresolved questions in galaxy evolution are: How is star formation regulated and quenched in galaxies? Which physics establishes the star-forming main sequence of galaxies? How do galaxies grow and build-up their structure, including their bulges and disks? To address these questions, we need to map star formation on spatially-resolved scales within galaxies across cosmic time \citep[e.g.,][]{Tacchella15, Nelson16, Nelson21, Baker23, Matharu24}. The fundamental challenge is that light emitted from stars is obscured by dust, which severely complicates empirical measurements of star formation rate in galaxies \citep[e.g.,][]{Shivaei15, Nelson16dust,Tacchella18dust, Popping22, Miller22, Zhang23, LeBail24}.

With JWST, for the first time, relatively dust-insensitive infrared emission-line tracers (e.g., Paschen lines, [SIII]) are now accessible out to $z\approx4$ \citep[e.g.,][]{Reddy23, Neufeld24, Liu24, Shapley24}. These lines, however, are weak -- e.g., Pa$\alpha$ is typically $\approx9\times$ weaker than H$\alpha$ at fixed SFR \citep[e.g.,][]{Reddy23}. ALT's depth, aided by magnification, not only allows us to capture an enormous statistical sample of these lines (Figure \ref{fig:redshifts}), but also allows us to create robust line maps that reveal the spatial structure of star-formation (Figure \ref{fig:paschen}).

The most striking aspect of the Paschen and [SIII] line maps in Figure \ref{fig:paschen} is that the star-forming clumps are barely visible in the rest-UV and rest-optical wavelengths are, in fact, the dominant centers of star-formation (compare the NIRCam SW RGBs in the left column with the line maps on the right). Even comparing photometry at rest-near-infrared wavelengths (center column) vs. the line maps shows dramatic differences owing to the weak EWs of these lines, underscoring the unique value of these line maps. Consider the galaxy in the bottom-row (ID: 76041, $z_{\rm{ALT}}=3.7792$). The [SIII] map resolves the star-formation in this galaxy into three major complexes with comparable luminosity, of which one is entirely obscured in the rest-UV. Furthermore, parts of the galaxy that have comparable luminosity to that of the clumps in the NIRCam LW image (e.g., the core) disappear when viewed in [SIII], implying weak star-formation activity and pointing to inside-out growth. We anticipate such rich, resolved line maps delivered by ALT-like surveys will help usher an era of high-redshift star-formation studies that boast a richness of detail comparable to studies set in the nearby universe \citep[e.g.,][]{Krumholz19, Kruijssen19, Tacconi20, Chevance20, Schinnerer23}.

\subsection{Maximizing Serendipity: Pure Spectroscopic Discovery of Rare Sources} \label{sec:lineprofiles}
The spectral resolution of the NIRCam grism is relatively high ($R\sim1600$), i.e., about a factor 10 higher than the resolution of the Prism mode on NIRSpec (at comparable wavelength). Combined with the unbiased spectroscopic coverage (i.e., coverage without any dependence on previous interpretations of photometric data) of all sources in the field of view, this provides a significant serendipitous discovery space for galaxies with unusual line-profiles or closely separated emission-lines. An example of this is the discovery of broad line H$\alpha$ emitters (``Little Red Dots") in the first NIRCam grism surveys \citep{Matthee24} (see also \citealt{Lin24}). In fact, grism surveys have yielded more LRD spectra than Cycle 1 NIRSpec programs \citep[e.g.][]{Harikane23,Maiolino23}, perhaps due to their peculiar colors eluding typical selection criteria used for selecting follow-up candidates. Indeed, in the ALT data, in addition to verifying various previously published broad-line H$\alpha$ emitters \citep[e.g.][]{Greene24}, we have identified a further four such sources at $z=4-5$ (presented in Matthee et al. in prep), some of which were missed even in photometric colour selections targeting LRDs \citep[e.g.][]{Labbe23}.

In Figure $\ref{fig:lineprofiles}$, we further illustrate the H$\alpha$+[NII]+[SII] spectra of two luminous H$\alpha$ emitting galaxies that are among the most massive galaxies in the $z=4$-5 sample, with stellar masses $10^{10}$ M$_{\odot}$ and $2\times10^9$ M$_{\odot}$ for ID 71962 and ID 29936, respectively. While [NII] is faint or even undetected in grating spectra of low mass galaxies that typically dominate the observed number at high-redshift \citep[e.g.][]{Cameron23}, the [NII] emission-line doublet, especially the redder line at $\approx+1000$ km s$^{-1}$ from the H$\alpha$ line, can be identified relatively clearly in these rare massive sources. The [SII] doublet, that is sensitive to the electron density is further also detected and resolved in these data. ID 71962 is the most massive galaxy in a very large overdensity of galaxies (60 galaxies within $\pm1000$ km s$^{-1}$). ID 29936 is among the few galaxies in our data above $z>4$ that has an ALMA continuum detection (as illustrated in Figure $\ref{fig:context}$; \citealt{DUALZ}). This indicates that the galaxy has a high rate of obscured star formation \citep[e.g.][]{Casey21}, consistent with its relatively red UV slope of $\beta=-1.1$, possibly associated with the merging activity that can be identified through diffuse H$\alpha$ emission in Figure $\ref{fig:lineprofiles}$, likely due to interactions with a nearby galaxy. 

In Figure $\ref{fig:lineprofiles2}$, we show two original grism spectra (i.e., spectra that have not been continuum-subtracted) of two galaxies that likely show AGN activity at $z\approx2.5$. These emission/absorption line features were first found in the filtered data, but the unfiltered spectra were then obtained by using only the subsets of the data where we found that (by fortunate coincidence) the trace of these galaxies was not significantly contaminated by spectra from other sources. These spectra illustrate that absorption-line science can be performed with the NIRCam grism for brighter sources (these have F356W=20.7 and 22.3, respectively). In ID 26974, which is an x-ray detected, very massive galaxy with a stellar mass of $10^{11}$ M$_{\odot}$, we detect multiple broad, high velocity HeI absorbers at velocities $\approx -1000$ to $\approx -5000$ km s$^{-1}$ indicative of multiple bursts of outflow events similar to broad absorption line quasars \citep[e.g.][]{Liu15,Wildy21}. The rest-frame optical SED of ID 26974 is well fit with stellar population models, indicating a stellar mass of $10^{11}$ M$_{\odot}$, and it is closely located to the mass-weighted center of a large overdensity at $z=2.58$ (of which it is the most massive member). 

The galaxy ID 26222 (right panel of Figure $\ref{fig:lineprofiles2}$),  resembles a lower redshift version of an LRD \citep{Matthee24}, i.e., very compact, broad hydrogen emission-lines (in this case Pa$\gamma$), no detection in the x-rays, and narrow absorption close to the systemic redshift (possibly allowing us to study how outflows develop, e.g., \citealt{Leighly11,Zhang18}). Similarly to the source studied in \cite{Wang24}, the SED of ID 26222 is not well explained by stellar population models alone, in particular due to a steep and red continuum bump from $\lambda_0\approx0.4$-1\,$\mu$m, above which wavelengths the continuum is relatively flat. While ID 26222 is at the edge of a large overdensity, it is not in its center nor is it the most massive galaxy associated to it (which is ID 34715, a passive galaxy with HeI emission likely excited by shocks; \citealt{Brinchmann23}). These properties are strikingly similar to the $z\gtrsim4$ little red dots (\citealt{Wang24}; Matthee et al. in prep), making this a rare and useful lower redshift version of such systems.

\section{Summary}

This paper provided an overview of the design, implementation, and yield of ALT, which is JWST's deepest NIRCam grism survey yet. ALT sought to survey faint dwarf galaxies out to the Epoch of Reionization, leveraging the magnification of the Abell 2744 cluster to further amplify the power of the NIRCam grism: high resolution ($R\approx1600$), coverage of every source in the field of view, and spatially resolved spectroscopy. ALT acquired sensitive $3-4\mu$m spectra reaching average 5$\sigma$ depths of $8\times10^{-19}$ erg s$^{-1}$ cm$^{-2}$, with simultaneous F070W and F090W imaging over the same 30 arcmin$^{2}$ reaching depths of $\approx30$ mag. To execute and analyze this first-of-a-kind survey, we deployed:
\begin{itemize}
    \item A ``butterfly" mosaic in which each source is dispersed at two slightly different roll angles. Galaxies in the ``wings" have the most divergent traces, whereas towards the center, we optimize for overlap of the NIRCam A+B modules that disperse sources in opposite directions. We demonstrate how this strategy solves spectral confusion and contamination, and validate it as an effective template for future deep grism surveys. [Figs. \ref{fig:RADEC}, \ref{fig:rolls}, \S\ref{sec:butterfly}]

    \item Photometric redshifts of exquisite quality: $\langle |\Delta z|/(1+z) \rangle = 0.005$, a standard deviation of 0.16, and 3\% catastrophic failures. The ALT photometric catalog incorporates all existing JWST and HST imaging in Abell 2744, which now includes the entire set of NIRCam broad-bands and medium-bands. The SEDs of galaxies are thoroughly constrained by 27 JWST+HST bands, and display subtle features such as Balmer jumps and breaks. [Figs. \ref{fig:seds}, \ref{fig:zphots}, \S\ref{sec:photozs}]

    \item Two independent spectroscopic catalog construction methods -- \texttt{Allegro}, which takes a ``line-first" approach and then seeks source associations assisted by photometric redshifts, and \texttt{grizli}, which takes a ``source-first" approach and searches for emission lines in windows around the source photometric redshift. [Fig. \ref{fig:flowchart}, \S\ref{sec:grism}]
    
\end{itemize}

Along with this paper, we release a catalog of 1630 sources with ALT grism redshifts that double the number of JWST spectra in this critical legacy field. Some notable features of this catalog are:

\begin{itemize}
    
    \item Pan-redshift coverage spanning A2744 cluster members at $z\approx0.3$ detected in Br$\beta$ to $z>8$ sources detected in H$\delta$, with ``deserts" in a handful of redshift slices (e.g., $z=3.3-3.8$) where no strong feature occurs at $3-4\mu$m. 
    
    \item Roughly equal numbers at $z<4$ and $z>4$ owing to the relative strength of the tracers (e.g., Paschen lines are much weaker than the Balmer lines at higher redshift at fixed SFR). [Fig. \ref{fig:redshifts}]

    \item Order-of-magnitude fainter galaxies (0.05 $L^{*}$, $M_{\rm{UV}}\approx-15$) compared to existing grism surveys that push to (0.1 $L^{*}$, $M_{\rm{UV}}\approx-18$). [Fig. \ref{fig:contextFRESCO}]

    \item Redshift precision of $\approx60$ km s$^{-1}$ validated against a compilation of ground-based (e.g., MUSE) and JWST/NIRSpec redshifts. [\S\ref{sec:redshift_performance}]
    
\end{itemize}

We present first science results and demonstrations based on the ALT sample to provide a tangible taste of the discovery space opened by this survey [\S\ref{sec:results}]:
\begin{itemize}
    \item Highly magnified arcs are among the highest value targets in cluster fields. Spatially resolved ALT spectra capture the complex morphology of these sources, resolve the clumps required to identify low-metallicity pockets, and collect the total flux crucial to the reionization budget. Such sources are challenging for slit spectroscopy and inefficient to survey using IFUs. [Fig. \ref{fig:arcs}, \S\ref{sec:arcs}]

    \item Multiply imaged sources are the key input for lensing models that underpin all science in cluster fields. ALT has independently measured redshifts in twelve multiple image systems of the thirty one currently known in A2744, with three new discoveries. [Fig. \ref{fig:multiple}, \S\ref{sec:lensing}]
    
    \item The completeness and precision of grism redshifts uniquely enable environmental clustering measurements. Photo-$z$s, even in this field where every HST+JWST broad and medium-band filter is available, severely underestimate clustering on small scales ($<5$ cMpc) and are limited by systematics such as ``phantom overdensities" on larger scales. [Figs. \ref{fig:redshift_phantoms}, \ref{fig:redshift_clustering}, \S\ref{sec:whygrism}]
    
    \item Galaxies are fundamentally shaped by their environments. In Fig. \ref{fig:overdensities} we present the density distribution of all galaxies behind A2744, i.e., the environmental context of every galaxy in this field. Several striking overdensities on 5 cMpc scales with dozens of members are clearly visible as redshift spikes (listed in Table \ref{table:overdensity}).
    
    \item The relationship between environment and galaxy properties is apparent in the ALT data. Some of the most massive galaxies, quenched galaxies, and galaxies with large Balmer breaks occur in the largest overdensities, as predicted for a hierarchical universe where overdensities collapse first [\S\ref{sec:envlarge}, Fig. \ref{fig:bbreaks}, Table \ref{table:overdensity}]. 
    
    \item On smaller scales ($<1$cMpc), a census of satellites around Milky Way progenitors is possible with the precision of grism redshifts even out to $z\approx6$. This opens the door to tests of cosmology with satellite statistics at higher redshifts analogous to tests performed in the Local Group. [Fig. \ref{fig:MW}, \S\ref{sec:envsmall}]

    \item ALT's rest-frame infrared tracers (e.g., Pa$\alpha$, [SIII]) paint remarkable portraits of obscured star-formation in sensitive spatially resolved emission line maps. These low-EW lines are difficult to infer from photometry, which makes their grism spectra particularly valuable. [Fig. \ref{fig:paschen}, \S\ref{sec:paschen}] 
    
    \item The unbiased spectroscopic coverage coupled with high resolution are particularly suited to the serendipitous discovery of rare sources  (e.g., the ``Little Red Dots" found in the EIGER+FRESCO surveys \citealt{Matthee24}). As examples, we present massive, metal-rich galaxies with [NII] emission at $z\approx4.5$, continuum spectra of an AGN with an ionized outflow, and an x-ray faint $z\approx2.5$ analog of the LRDs. [Figs. \ref{fig:lineprofiles}, \ref{fig:lineprofiles2}, \S\ref{sec:lineprofiles}]

\end{itemize}

Surveys of strong lensing clusters may be the most efficient way JWST can answer some of the fundamental questions it was built for, by peering deep into the farthest, faintest, first galaxies. Here, for the first time, we have demonstrated the pan-redshift power of the NIRCam grism in such strong lensing fields. The results in this paper are proof that an ALT-like strategy will be successful at surveying clusters that JWST is yet to look at by providing simultaneous imaging and spectroscopy, complete statistics, multiple images for lensing models, and the spatial and spectral resolution required to study highly magnified arcs -- all in one shot! We hope the rich returns of ALT inspire the community to design such surveys and complete the spectroscopic census of the smallest galaxies from the edge of the Milky Way to the brink of the Big Bang.

\section*{Acknowledgments}

ALT was developed in part at the International Space Science Institute (ISSI) in Bern, Switzerland during the February 2023 meeting of ISSI Team \#562 “First Light at Cosmic Dawn: Exploiting the James Webb Space Telescope Revolution” led by Pascal Oesch and Michael Maseda. RPN extends his heartfelt gratitude to the staff at the Swiss consulate in Vienna, particularly Marina Bont\`{a}, for their assistance with Schengen visa issues, and for their excitement for astronomy in general and ALT in particular.

ALT is but the latest link in a long chain of surveys that have transformed Abell 2744 into a premier, public extragalactic legacy field. We are grateful to all the teams that have observed this field with ground-based and space-based campaigns over the last few decades (see \S\ref{sec:publicdata}). We are particularly thankful to the following teams that developed JWST programs delivering immediately public data that we have incorporated in our analysis: \#1324 (GLASS, PI: Treu), \#2561 (UNCOVER, PIs: Labbe \& Bezanson), \#4111 (MegaScience, PI: Suess), \#2883 (MAGNIF, PI: Sun), \#3538 (PI: Iani), \#2756 (PI: Chen), \#3990 (PI: Morishita).

RPN acknowledges funding from {\it JWST} program GO-3516. Support for this work was provided by NASA through the NASA Hubble Fellowship grant HST-HF2-51515.001-A awarded by the Space Telescope Science Institute, which is operated by the Association of Universities for Research in Astronomy, Incorporated, under NASA contract NAS5-26555. Funded by the European Union (ERC, AGENTS, 101076224). Views and opinions expressed are however those of the author(s) only and do not necessarily reflect those of the European Union or the European Research Council. Neither the European Union nor the granting authority can be held responsible for them. This work has received funding from the Swiss State Secretariat for Education, Research and Innovation (SERI) under contract number MB22.00072, as well as from the Swiss National Science Foundation (SNSF) through project grant 200020\_207349. A.F. acknowledges support from NSF-AAG grant AST-2307436. The work of CCW is supported by NOIRLab, which is managed by the Association of Universities for Research in Astronomy (AURA) under a cooperative agreement with the National Science Foundation.

Computations supporting this paper were run on MIT's Engaging cluster. This publication made use of the NASA Astrophysical Data System for bibliographic information. Some of the data products presented herein were retrieved from the Dawn JWST Archive (DJA). DJA is an initiative of the Cosmic Dawn Center (DAWN), which is funded by the Danish National Research Foundation under grant DNRF140. Software used in developing this work includes: \texttt{matplotlib} \citep{matplotlib}, \texttt{jupyter} \citep{jupyter}, \texttt{IPython} \citep{ipython}, \texttt{numpy} \citep{numpy}, \texttt{scipy} \citep{scipy}, \texttt{TOPCAT} \citep{topcat}, and \texttt{Astropy} \citep{astropy}.

This work is based on observations made with the NASA/ESA/CSA James Webb Space Telescope. The data were obtained from the Mikulski Archive for Space Telescopes at the Space Telescope Science Institute, which is operated by the Association of Universities for Research in Astronomy, Inc., under NASA contract NAS 5-03127 for \textit{JWST}. These observations are associated with program \# 3516.

\bibliography{MasterBiblio}
\bibliographystyle{apj}

\end{document}